\documentclass[twocolumn,english,aps,pra,superscriptaddress]{revtex4-1}
\usepackage[T1]{fontenc}
\usepackage[latin9]{inputenc}
\usepackage{color}
\usepackage{array}
\usepackage{float}
\usepackage{multirow}
\usepackage{amsmath}
\usepackage{amssymb}
\usepackage{graphicx}
\usepackage{esint}

\makeatletter

\providecommand{\tabularnewline}{\\}
\newcommand{\lyxdot}{.}

\makeatother

\usepackage{babel}
\begin{document}

\title{Dynamics of transient cat-states in degenerate parametric oscillation
with and without nonlinear Kerr interactions}

\author{R. Y. Teh}

\affiliation{Centre for Quantum and Optical Science, Swinburne University of Technology,
Melbourne 3122, Australia}

\author{F.-X. Sun}

\affiliation{State Key Laboratory for Mesoscopic Physics and Collaborative Innovation
Center of Quantum Matter, School of Physics, Peking University, Beijing
100871, China}

\affiliation{Centre for Quantum and Optical Science, Swinburne University of Technology,
Melbourne 3122, Australia}

\author{R. E. S. Polkinghorne}

\affiliation{Centre for Quantum and Optical Science, Swinburne University of Technology,
Melbourne 3122, Australia}

\author{Q. Y. He}

\affiliation{State Key Laboratory for Mesoscopic Physics and Collaborative Innovation
Center of Quantum Matter, School of Physics, Peking University, Beijing
100871, China}

\author{Q. Gong}

\affiliation{State Key Laboratory for Mesoscopic Physics and Collaborative Innovation
Center of Quantum Matter, School of Physics, Peking University, Beijing
100871, China}

\author{P. D. Drummond}

\affiliation{Centre for Quantum and Optical Science, Swinburne University of Technology,
Melbourne 3122, Australia}

\affiliation{Institute of Theoretical Atomic, Molecular and Optical Physics (ITAMP),
Harvard University, Cambridge, Massachusetts 02138, USA}

\author{M. D. Reid}

\affiliation{Centre for Quantum and Optical Science, Swinburne University of Technology,
Melbourne 3122, Australia}

\affiliation{Institute of Theoretical Atomic, Molecular and Optical Physics (ITAMP),
Harvard University, Cambridge, Massachusetts 02138, USA}
\begin{abstract}
A cat-state is formed as the steady-state solution for the signal
mode of an ideal, degenerate parametric oscillator, in the limit of
negligible single-photon signal loss. In the presence of signal loss,
this is no longer true over timescales much longer than the damping
time. However, for sufficient parametric nonlinearity, a cat-state
can still exist as a transient state. In this paper, we study the
dynamics of the creation and decoherence of cat-states in degenerate
parametric oscillation, both with and without the Kerr nonlinearity
found in recent superconducting-circuit experiments that generate
cat-states in microwave cavities. We determine the time of formation
and the lifetime of a cat-state of fixed amplitude in terms of three
dimensionless parameters $\lambda$, $g$ and $\chi$. These relate
to the driving strength, the parametric nonlinearity relative to signal
damping, and the Kerr nonlinearity, respectively. We find that the
Kerr nonlinearity has little effect on the threshold parametric nonlinearity
($g>1$) required for the formation of cat-states, and does not
significantly alter the decoherence time of the cat-state, but can
reduce the time of formation. The quality of the cat-state increases
with the value of $g$. To verify the existence of the cat-state,
we consider several signatures, including interference fringes and
negativity. We emphasize the importance of taking into account more
than one of these signatures. We simulate a superconducting-circuit
experiment using published experimental parameters and find good agreement
with experimental results, indicating that a nonclassical cat-like
state with a small Wigner negativity is generated in the experiment.
Interference fringes however are absent, requiring higher $g$ values.
 Finally, we explore the feasibility of creating large cat-states
with a coherent amplitude of $20$, corresponding to 400 photons,
and study finite temperature reservoir effects.
\end{abstract}
\maketitle

\section{introduction}

After Schr\"odinger's famous paradox, a ``cat-state'' is a quantum
superposition of two macroscopically distinguishable states, often
taken to be coherent states \citep{schrodinger_1935}. The cat-state
plays a fundamental role in motivating experiments probing the validity
of quantum mechanics for macroscopic systems \citep{Frowis_RMP2018}.
More recently, it has been recognized that cat-states are a useful
resource for quantum information processing and metrology \citep{Chuang_PRA1997,Zurek_metrology_Nature2001,vanEnk_PRA2001,Toscano_metrology_PRA2006,caves_metrology_OptsCom2010,Joo_PRL2011,Mirrahimi_NJP2014,Wang_Science2016,Marios_PRX2016}.
There has been success in creating mesoscopic superposition states,
including in optical cavities, ion traps, and for Rydberg atoms \citep{Brune_PRL1996,Monroe_Science1996,Friedman:2000aa,Greiner:2002aa,Walther:2004aa,Mitchell:2004aa,Leibfried:2005aa,Ourjoumtsev_nature2007,Palacios-Laloy:2010aa,Afek879,Monz_PRL2011,Kirchmair_Nature2013Kerr,Kovachy:2015aa,omran2019generation}.
In microwave experiments that utilise superconducting circuits to
enhance nonlinearities, cat-states with up to $80$ photons \citep{Wang_Science2016}
and $100$ photons \citep{Vlastakis_Science2013} have been reported.

Recently, a two-photon driven dissipative process based on superconducting
circuits has been used to generate cat-like states in a microwave
cavity \citep{Leghtas_Science2015,Safavi-Naeini2019QuantumDynamics}.
Following the proposal by Mirrahimi et al. \citep{Mirrahimi_NJP2014},
the experiment demonstrates confinement of a state to a manifold mostly
spanned by two coherent states with opposite phases. The creation
of a cat-like state in a superposition of the two coherent states
$\pi$ out of phase is made possible by the strong nonlinearity due
to a Josephson junction and a comparatively low single-photon damping
of the signal \citep{Wallraff_nature2004,Leghtas_PRA2013,Leghtas_PRL2013,Vlastakis_Science2013,Mirrahimi_NJP2014,Leghtas_Science2015,Wang_Science2016}.
This process is an example of degenerate parametric oscillation (DPO).
In an optical DPO, current setups give a much smaller nonlinearity
and cat-states are not generated. Rather, the system evolves to a
bistable situation, being in a classical mixture of the two coherent
amplitudes, with quantum tunneling possible between the two states
\citep{Drummond_PRA1989,Kinsler_PRA1991}.

In this paper, we study the generation, dynamics and eventual decoherence
of a cat-state in a degenerate parametric oscillator (DPO). We extend
previous quantum treatments of the DPO to include the additional Kerr
nonlinearities that arise in the recent superconducting experiments.
In both the standard DPO (without Kerr nonlinearity) and the DPO
with Kerr nonlinearity, we demonstrate the possibility of the formation
of cat-states in a transient regime if the two-photon effective nonlinear
driving is sufficiently strong. We fully characterize the parameter
regimes necessary for the formation of the cat-states, determining
the threshold nonlinearity required, and the time-scales over which
the cat-states are generated. In the presence of signal-photon losses
from the cavity, the cat-states eventually decohere. We determine
the lifetime of the cat-states for the full parameter regime. To fully
evaluate the dynamics of cat-state formation, we consider several
signatures of cat-states, including the negativity of the Wigner function
and interference fringes.

Understanding the dynamics of the formation of cat-states in degenerate
parametric oscillation is motivated by applications in quantum information,
and by the development of the Coherent Ising Machine (CIM), an optimization
device capable of solving NP-hard problems \citep{Marandi_CIM_Nature2014,McMahon_CIM_science2019,Inagaki_CIM_science2019}.
Although current realizations of the CIM use the equivalent of a network
of optical DPOs which do not operate in a cat-state regime, the regime
of cat-states may be of interest in future devices. There already
exist proposals \citep{Nigg_PRL2012,Goto_PRA2019} and experiments
\citep{Pfaff_Nature2017} generating itinerant cat-states in a DPO,
which will be useful in a DPO network for adiabatic quantum computation
\citep{Goto_SciRep2016,Goto_PRA2016,Puri_Nature2017} to solve these
NP-hard problems.

A DPO consists of pump and signal modes resonant in a cavity at frequencies
$2\omega$ and $\omega$ respectively, and resembles a laser in exhibiting
a threshold behavior for the intensity of the signal mode \citep{Byer_1979,Heidmann_PRL1987,Villar_PRL2005,Laurat_OptLett2005,Keller_Opt_Express2008}.
The signal photons leave the cavity with a fixed cavity decay rate
$\gamma_{1}$. Unlike the laser, however, the steady-state solutions
for the amplitude of the signal field above threshold have a fixed
phase relation. A quantum analysis of the DPO was given by Drummond,
McNeil and Walls \citep{Drummond_OpActa1981}, who gave exact steady-state
solutions in the limit of a fast-decaying pump mode, which acts to
generate photon pairs at the signal frequency. The possibility of
generating a cat-state as a superposition of the two coherent steady-state
solutions $\pi$ out of phase was proposed by Wolinsky and Carmichael
\citep{Wolinsky_PRL1988}. While it was realized that the steady-state
solution forming over times much longer than $\gamma_{1}^{-1}$ would
not be a cat-state \citep{Reid_PRA1992}, it became clear that in
the limit of zero signal losses, a cat-state would form dynamically
from a vacuum state as a result of the two-photon driving process
\citep{Gilles_PRA1994,Hach_PRA1994,Krippner_PRA1994}. Cat-states
can be generated as a transient over suitable timescales even in the
presence of signal losses (which give decoherence) provided the nonlinearity
is sufficiently dominant \citep{Krippner_PRA1994}.

In this paper, we provide a complete analysis of the dynamics of the
cat-states in terms of three parameters that define the system. The
parameters are the driving strength $\lambda$, the parametric nonlinearity
$g$ (scaled relative to cavity and pump decay rates), and the time
of evolution $\tau$ (scaled relative to the signal cavity decay rate
$\gamma_{1}$). Our study assumes that the pump field decays much
more rapidly than the intracavity signal field. Whether a cat-state
or a mixture is formed depends on the competition between how fast
one can generate a cat-state and how fast one loses it, due to decoherence
from signal-photon loss. A minimum $g>1$ is required for the formation
of a cat-like state. We find that the value of $g$ also determines
the lifetime and quality of the cat state, in the presence of the
signal damping. We analyze the limit as $g\rightarrow\infty$, showing
that the cat-state becomes increasingly stable, consistent with the
analysis of Gilles et al \citep{Gilles_PRA1994}.

The Hamiltonian describing the superconducting cat-system is that
of the DPO, but with an additional term due to a Kerr nonlinearity.
This introduces a fourth scaled parameter $\chi$. Recent works by
Sun et al. \citep{sun2019discrete,sunschrodinger} have revealed
that the cat-states can form in the presence of the Kerr terms, in
the limit of zero signal loss, but that the final steady-state solution
where signal loss is present cannot be a cat-state. The analysis
presented in this paper determines the threshold condition for the
formation of transient cat-states including the Kerr nonlinearity.
The Kerr nonlinearity has little effect on the threshold parametric
nonlinearity required for the formation of cat-states. We also predict
how fast a cat-state can be generated for a given Kerr nonlinearity,
and how fast the cat-state decays. For cat-states of a fixed size,
the time of formation can be reduced for a fixed parametric nonlinearity,
provided the driving field or Kerr nonlinearity can be increased and
that the parametric nonlinearity satisfies $g>1$.

The paper is organized as follows. In Section \ref{sec:The-System},
we introduce the Hamiltonian modeling the degenerate parametric oscillator.
In this work, we solve using the master equation expanded in the
number-state basis, which provides a set of partial differential equations
for all matrix elements of a density operator (up to a cutoff number).
The master equation and the corresponding steady states in certain
limits are described in Section \ref{sec:steady_state_number_state_expansion}.
In Appendix \ref{sec:Cat signature}, we consider cat-state signatures,
including both interference fringes in the quadrature probability
distribution \citep{PhysRevLett.57.13,Yurke_physicalB1988} and the
photon-number distribution. The Q and Wigner functions \citep{Wigner_PhysRev1932,Cahill_PhysRev1969,HILLERY_PhysRep1984}
are also considered. For cat-states, the Wigner function becomes
negative, and the corresponding Wigner negativity \citep{Kenfack_2004}
can be computed from the Wigner function, as a signature of the cat-state.
The zeros of a Q function for a pure state \citep{Lutkenhaus_PRA1995}
serve the same purpose. Technical issues are also mentioned in this
section as some of the signatures are numerically hard to compute.

In Sections \ref{sec:Results}-\ref{sec:Large-transient-cat}, we
present the results for different DPO parameters. In Section \ref{sec:Results},
we compute the dynamics of a degenerate parametric oscillator at zero
temperature without detuning and Kerr nonlinearity, and give a full
study the corresponding time evolution and decoherence of the cat-state
signatures. The effect of detuning and Kerr nonlinearity are examined
in Sections \ref{sec:Detuning} and \ref{sec:Degenerate-parametric-oscillatio}.
In Section \ref{sec:Large-transient-cat}, we simulate an experiment
using published superconducting circuit experimental parameters, and
find that our numerical results agree well with the experimental observations.
Based on these realistic parameters, we explore the feasibility of
generating large transient cat-states, and study the effects of finite
temperatures. We conclude in Section \ref{sec:conclusion}.

\section{The Hamiltonian \label{sec:The-System}}

\subsection{Degenerate parametric oscillation}

The Hamiltonian for a degenerate parametric oscillator (DPO) is given
by \citep{Drummond_OpActa1981}
\begin{align}
H_{1}\! & =\!\hbar\omega_{1}a_{1}^{\dagger}a_{1}+\hbar\omega_{2}a_{2}^{\dagger}a_{2}+\frac{i\hbar}{2}\!\left(\bar{g}a_{2}a_{1}^{\dagger2}\!-\!\bar{g}^{*}a_{2}^{\dagger}a_{1}^{2}\right)\!\nonumber \\
 & +\!i\hbar\epsilon\left(a_{2}^{\dagger}e^{-i\omega_{p}t}\!-\!a_{2}e^{i\omega_{p}t}\right)\!+\!\sum_{i=1}^{2}\!\left(a_{i}^{\dagger}\Gamma_{i}\!+\!a_{i}\Gamma_{i}^{\dagger}\right)\,.\label{eq:DPO_hamiltonian}
\end{align}
Here $a_{i}$ are boson operators for the optical cavity modes at
frequencies $\omega_{i}$, with $\omega_{2}\approx2\omega_{1}$. The
modes with frequency $\omega_{2}$ and $\omega_{1}$ are the pump
and signal modes respectively. The pump mode is driven by an external,
classical light field of amplitude $\epsilon$ with frequency $\omega_{p}$,
and $\bar{g}$ is the coupling strength between the pump and signal
modes. The last term represents the couplings of the cavity modes
to the external environment and hence describes the single-photon
losses of pump and signal from the cavity to the environment \citep{Yurke_PRA1984,Yurke_Denker_PRA1984,Collett_PRA1984,PhysRevA.31.3761}.
We ignore thermal noise in the pump, but will include the thermal
noise in the signal, if necessary.

In this work, we set the driving laser frequency $\omega_{p}$ to
be on resonance with the pump mode frequency $\omega_{2}$, and transform
the system into the rotating frame of the driving frequency. The resulting
Hamiltonian is then given by
\begin{align}
H_{2}\! & =\hbar\bar{\Delta}a_{1}^{\dagger}a_{1}+\!\frac{i\hbar}{2}\!\left(\bar{g}a_{2}a_{1}^{\dagger2}\!-\!\bar{g}^{*}a_{2}^{\dagger}a_{1}^{2}\right)\!\nonumber \\
 & +\!i\hbar\epsilon\left(a_{2}^{\dagger}\!-\!a_{2}\right)\!+\!\sum_{i=1}^{2}\!\left(a_{i}^{\dagger}\Gamma_{i}\!+\!a_{i}\Gamma_{i}^{\dagger}\right)\,\label{eq:rotating_frame_Hamiltonian}
\end{align}
where $\bar{\Delta}=\omega_{1}-\omega_{p}/2$. A nonzero $\bar{\Delta}$
implies that the signal mode frequency $\omega_{1}$ is not exactly
half the pump mode frequency $\omega_{2}$.

When the pump mode single-photon decay rate is much larger than the
signal mode decay rate, i.e. $\gamma_{2}\gg\gamma_{1}$, the pump
mode can be adiabatically eliminated \citep{Drummond_OpActa1981}.
In this case, the pump mode amplitude has a steady state $\alpha_{2}^{0}=\left(\epsilon-\bar{g}\alpha_{1}^{2}/2\right)/\gamma_{2}$,
which is determined by the signal mode amplitude expectation value
$\alpha_{1}$ \citep{Drummond_OpActa1981}. The signal-mode amplitude
evolves in time according to a simpler Hamiltonian involving only
the signal mode \citep{sunschrodinger}:
\begin{align}
H & =\hbar\bar{\Delta}a_{1}^{\dagger}a_{1}+i\hbar\left(\frac{\bar{g}\epsilon}{\gamma_{2}}a_{1}^{\dagger2}-\frac{\bar{g}^{*}\epsilon^{*}}{\gamma_{2}}a_{1}^{2}\right)\nonumber \\
 & +a_{1}^{\dagger}\Gamma_{1}+a_{1}\Gamma_{1}^{\dagger}+\frac{\left|\bar{g}\right|^{2}}{4\gamma_{2}}\left(a_{1}^{2}\Gamma_{2}^{\dagger}+a_{1}^{\dagger2}\Gamma_{2}\right)\,.\label{eq:simplified_hamiltonian}
\end{align}

A simple semi-classical analysis (in which noise terms are ignored)
indicates that this system undergoes a threshold when $\epsilon=\epsilon_{c}=\frac{\gamma_{1}\gamma_{2}}{\bar{g}}$
\citep{Drummond_OpActa1980,Drummond_OpActa1981} i.e. when
\begin{equation}
\lambda=\left|\bar{g}\epsilon\right|/\left(\gamma_{1}\gamma_{2}\right)=1\,.\label{eq:pump_threshold}
\end{equation}
Below this threshold ($\lambda<1$), the semi-classical mean signal
amplitude is zero. Above threshold ($\lambda>1$), the intensity of
the signal field increases with increasing driving field.

In certain regimes of parameters above threshold, the two-photon driven
dissipative process (\ref{eq:simplified_hamiltonian}) generates cat-states
of type \citep{Wolinsky_PRL1988,Hach_PRA1994,Gilles_PRA1994,Krippner_PRA1994,Munro_PRA1995}
\begin{align}
|\psi_{\text{even}}\rangle & =\mathcal{N}_{+}\left(|\alpha_{0}\rangle+|-\alpha_{0}\rangle\right)\nonumber \\
|\psi_{\text{odd}}\rangle & =\mathcal{N}_{-}\left(|\alpha_{0}\rangle-|-\alpha_{0}\rangle\right)\,\label{eq:even_odd_cat}
\end{align}
where $\mathcal{N}_{\pm}=\left[2\left(1\pm e^{-2\left|\alpha_{0}\right|^{2}}\right)\right]^{-1/2}$
and $|\pm\alpha_{0}\rangle$ are coherent states with amplitudes $\alpha_{0}=\pm\sqrt{2\epsilon/\bar{g}}$
respectively. Here, thermal noise is ignored. The $|\psi_{\text{even}}\rangle$
and $|\psi_{\text{odd}}\rangle$ are cat-states with even and odd
photon number respectively \citep{Buzek_PRA1992,Hach_PRA1994,Gilles_PRA1993}.
In particular, Hach and Gerry \citep{Hach_PRA1994} and Gilles et
al. \citep{Gilles_PRA1994} show that cat-states survive in this two-photon
driven dissipative process provided the single-photon losses for the
signal $a_{1}$ are neglected. Reid and Yurke showed that the single-photon
signal losses eventually destroy the cat-state \citep{Reid_PRA1992}.
They calculated the Wigner function of the steady state formed including
signal losses, showing that this function was positive and therefore
could not be a cat-state. For sufficiently strong coupling $\bar{g}$,
a cat-state can form in a transient regime \citep{Krippner_PRA1994}.
In Sections \ref{sec:Results} and \ref{sec:Detuning}, we extend
this earlier work, by examining the full dynamics of the formation
and decoherence of the cat-states over the complete parameter range.

\subsection{Degenerate parametric oscillation with a Kerr medium}

A promising system where single-photon signal damping can be small
relative to the nonlinearity is the superconducting circuit involving
a Josephson junction \citep{Kirchmair_Nature2013Kerr,Vlastakis_Science2013,Wang_Science2016}.
However, the implementation of the two-photon driven dissipative process
in Eq. (\ref{eq:simplified_hamiltonian}) in a superconducting circuit
leads to an additional Kerr-type nonlinear interaction. The resulting
Hamiltonian for this system (after the adiabatic elimination process)
is given by \citep{sunschrodinger,sun2019discrete}
\begin{align}
H & =\hbar\bar{\Delta}a_{1}^{\dagger}a_{1}+i\hbar\left(\frac{\bar{g}\epsilon}{\gamma_{2}}a_{1}^{\dagger2}-\frac{\bar{g}^{*}\epsilon^{*}}{\gamma_{2}}a_{1}^{2}\right)+\frac{\hbar\bar{\chi}}{2}a_{1}^{\dagger2}a_{1}^{2}\nonumber \\
 & +a_{1}^{\dagger}\Gamma_{1}+a_{1}\Gamma_{1}^{\dagger}+\frac{\left|\bar{g}\right|^{2}}{4\gamma_{2}}\left(a_{1}^{2}\Gamma_{2}^{\dagger}+a_{1}^{\dagger2}\Gamma_{2}\right)\,.\label{eq:simplified_full_hamiltonian}
\end{align}

It has been shown that the two-photon driven dissipative process
(\ref{eq:simplified_full_hamiltonian}) including $\bar{\chi}$ also
gives the threshold Eq. (\ref{eq:pump_threshold}) \citep{sunschrodinger}.
Here thermal noise is ignored. Above threshold, the process in the
absence of single-photon loss generates cat-states of type Eq. (\ref{eq:even_odd_cat})
\citep{Mirrahimi_NJP2014,sunschrodinger} but where  $|\pm\alpha_{0}\rangle$
are coherent states with amplitude $\alpha_{0}$ given by \citep{sunschrodinger}
\begin{align}
\alpha_{0} & =\sqrt{\frac{\epsilon}{\frac{\bar{g}}{2}\left(1+i\frac{2\gamma_{2}}{\bar{g}^{2}}\bar{\chi}\right)}}\,.\label{eq:cat-amplitude-chi}
\end{align}
As with the DPO, Sun et al. have shown that the cat-states are destroyed
in the limit of the steady-state if signal loss is nonzero \citep{sunschrodinger}.
In Sections \ref{sec:Degenerate-parametric-oscillatio} and \ref{sec:Large-transient-cat},
we examine the dynamics of the signal mode as it evolves from the
vacuum, identifying the parameter regimes which show the feasibility
of the formation of transient cat-states.

\section{Master equation and steady state solutions \label{sec:steady_state_number_state_expansion}}

\subsection{Master equation \label{subsec:Master-equation-and}}

A master equation takes into account the damping and quantum noise
fluctuations as well as the dynamics due to the system Hamiltonian,
in the Markovian approximation. The Hamiltonian in the previous section
has a corresponding master equation that describes the time evolution
of the signal mode $a\equiv a_{1}$. The full master equation corresponding
to Eq. (\ref{eq:simplified_full_hamiltonian}) including the effect
of thermal reservoirs is given by \begin{widetext}
\begin{align}
\frac{\partial}{\partial t}\rho & =-i\bar{\Delta}\left[a^{\dagger}a,\rho\right]+\frac{\left|\bar{g}\epsilon\right|}{2\gamma_{2}}\left[a^{\dagger2}-a^{2},\rho\right]-i\frac{\bar{\chi}}{2}\left[a^{\dagger2}a^{2},\rho\right]+\frac{1}{2}\left(\frac{\bar{g}^{2}}{2\gamma_{2}}\right)\left(2a^{2}\rho a^{\dagger2}-a^{\dagger2}a^{2}\rho-\rho a^{\dagger2}a^{2}\right)\nonumber \\
 & +\left(N+1\right)\gamma_{1}\left[2a\rho a^{\dagger}-a^{\dagger}a\rho-\rho a^{\dagger}a\right]+N\gamma_{1}\left[2a^{\dagger}\rho a-aa^{\dagger}\rho-\rho aa^{\dagger}\right]\,.\label{eq:master_eqn}
\end{align}
\end{widetext}  With no loss of generality, we can choose the phase
of $\bar{g}$ such that $\bar{g}\epsilon=\bar{g}^{*}\epsilon^{*}$
\citep{carmichael_QO4,Kinsler_PRA1991}. Here, $\rho$ is the density
operator of the signal mode. The first term on the right side of Eq.
(\ref{eq:master_eqn}) is due to the detuning between the driving
field and signal mode frequency. The second term describes the driving
of the signal mode by the pump. The third term arises from the Kerr-type
interaction, and the fourth term describes the two-photon loss process
where two signal-mode photons convert back to a pump mode photon,
which then subsequently leaks out of the system. The remaining terms
describe single-photon damping due to the interaction between the
system and its environment, where the parameter $N$ is the mean thermal
occupation number of the reservoir.

\subsection{Steady-state solutions}

The steady-state solution $\rho\left(\infty\right)$ that satisfies
$\partial\rho/\partial t=0$ is typically hard to obtain for driven
quantum systems out of thermal equilibrium. Using the generalized
P distribution \citep{Drummond_generalizedP1980}, the steady-state
solution in the quantum case where damping and parametric nonlinearity
are present can be obtained using the method of potentials \citep{Drummond_OpActa1981,Wolinsky_PRL1988,Kheruntsyan_OptCom1996}.
This was recently extended to the general quantum case where damping
and both Kerr and parametric nonlinearity are present \citep{sun2019discrete,sunschrodinger}.

\subsubsection{Two-photon dissipation and driving with no signal single-photon damping}

First, the steady-state solution in the absence of thermal noise where
the single-photon losses are neglected ($\gamma_{1}=0$), and where
the Kerr term ($\bar{\chi}=0$) is zero, has been shown to be of the
form \citep{Hach_PRA1994,Gilles_PRA1994}
\begin{align}
\rho\left(\infty\right) & =p_{++}|\psi_{\text{even}}\rangle\langle\psi_{\text{even}}|+p_{--}|\psi_{\text{odd}}\rangle\langle\psi_{\text{odd}}|\nonumber \\
 & +p_{+-}|\psi_{\text{even}}\rangle\langle\psi_{\text{odd}}|+p_{-+}|\psi_{\text{odd}}\rangle\langle\psi_{\text{even}}|\,\label{eq:steady_state}
\end{align}
This is a classical mixture of the even and odd cat-states $|\psi_{\text{even}}\rangle=\mathcal{N}_{+}\left(|\alpha_{0}\rangle+|-\alpha_{0}\rangle\right)$
and $|\psi_{\text{odd}}\rangle=\mathcal{N}_{-}\left(|\alpha_{0}\rangle-|-\alpha_{0}\rangle\right)$
given by Eq. (\ref{eq:even_odd_cat}). Here we assume no detuning
$\bar{\Delta}=0$. The coherent amplitude is found to be $\alpha_{0}=\pm\sqrt{2\epsilon/\bar{g}}$,
which can be given in terms of the pump parameter $\lambda$ (defined
in Eq. (\ref{eq:pump_threshold}) for the parametric oscillator with
signal damping)
\begin{equation}
\lambda\equiv\left|\bar{g}\epsilon\right|/\left(\gamma_{1}\gamma_{2}\right)\label{eq:lambda}
\end{equation}
 and a dimensionless two-photon dissipative rate
\begin{equation}
g\equiv\sqrt{\bar{g}^{2}/\left(2\gamma_{1}\gamma_{2}\right)}\label{eq:defn-g}
\end{equation}
 via
\begin{equation}
\alpha_{0}=\sqrt{\lambda}/g\,.\label{eq:dfn-alpha}
\end{equation}
This gives consistency with the work of Wolinsky and Carmichael who
had earlier pointed to the possibility of cat-states with amplitude
$\alpha_{0}=\sqrt{\lambda}/g$ in the limit of negligible signal damping
\citep{Wolinsky_PRL1988}. The amplitudes $\alpha_{0}=\pm\sqrt{\lambda}/g$
correspond to the steady-state solutions derived in a semi-classical
approach where quantum noise is ignored. The coefficients $p_{++},\,p_{--}$
can be interpreted as probabilities ($p_{++}+p_{--}=1$) and are obtained
from the initial state of the system where these coefficients are
the constants of motion. Following this, if the system has an initial
vacuum state, the steady state is an even cat-state $|\psi_{\text{even}}\rangle$.

The steady-state solution of Eq. (\ref{eq:master_eqn}) for the system
with an additional Kerr-type interaction $\bar{\chi}$ has recently
been analyzed by Sun et al. \citep{sunschrodinger}. The steady-state
is of the form (\ref{eq:steady_state}), except that the coherent
amplitude becomes
\begin{equation}
\alpha_{0}=\sqrt{\lambda/\left(g^{2}+i\chi'\right)}=\sqrt{\lambda/g^{2}\left(1+i\chi\right)}\label{eq:defn-alpha-kerr}
\end{equation}
 which is rotated in phase-space due to the nonlinear Kerr term $\bar{\chi}$.
Here, $\chi'=\bar{\chi}/\gamma_{1}$ is the scaled Kerr interaction
strength, and $\chi\equiv\chi'/g^{2}$ is the ratio of the Kerr strength
to the parametric gain, which will be used throughout Section \ref{sec:Degenerate-parametric-oscillatio}.

\subsubsection{Steady-solution in the presence of signal single-photon damping}

The steady-state solution for the general case where the single-photon
damping is taken into account is calculated using the complex P-representation
\citep{Drummond_generalizedP1980,Drummond_OpActa1981,sun2019discrete,sunschrodinger}.
Here, we ignore thermal noise. After adiabatic elimination of the
pump mode, a corresponding Fokker-Planck equation allows the analytical
steady-state potential solution to be obtained \citep{Drummond_OpActa1981}.
A steady-state solution in the positive P-representation was derived
by Wolinsky and Carmichael \citep{Wolinsky_PRL1988}, who pointed
out the potential to create cat-states in the large $g$ limit. However,
this approach is not valid for strong coupling and Kerr nonlinearities.

From the complex P solutions, a Wigner function can be derived which,
being positive, demonstrated that the steady-state solution itself
cannot be a cat-state \citep{Reid_PRA1992}. Beginning with an even
cat-state, for example, it is well-known that the loss of a signal
photon converts the system into an odd cat-state \citep{carmichael2009statistical}.
The presence of single-photon signal loss therefore leads to a mixture
of the odd and even cat-states being created. A 50/50 mixture of the
even and odd cat-states is equivalent to a 50/50 mixture of the two
coherent states $|\pm\alpha_{0}\rangle$. This gives the mechanism
by which ultimately the mesoscopic quantum coherence that gives the
cat-state is destroyed.

An analysis of the steady-state solution given by Sun et al. \citep{sunschrodinger}
yields that for the system where the signal mode is initially in a
vacuum state, the steady state solution for $g>1$ with an initial
vacuum state is given by a density operator in a mixture of the form
\citep{sunschrodinger}
\begin{align}
\rho_{ss} & =P_{ss}|\psi_{\text{even}}\rangle\langle\psi_{\text{even}}|+\left(1-P_{ss}\right)\rho_{\text{mix}}\,\label{eq:FX_steady_state-1}
\end{align}
where
\[
\rho_{mix}=\frac{1}{2}|\alpha_{0}\rangle\langle\alpha_{0}|+\frac{1}{2}|-\alpha_{0}\rangle\langle-\alpha_{0}|
\]
 and
\[
P_{ss}=[1+\text{exp}(-2\left|\alpha_{0}\right|^{2})]/[\text{exp}(2\left|\alpha_{0}\right|^{2})+\text{exp}(-2\left|\alpha_{0}\right|^{2})].
\]
where $\alpha_{0}$ is given by Eq. (\ref{eq:defn-alpha-kerr}).
The steady-state solution in Eq. (\ref{eq:steady_state}) is a good
approximation when the single-photon loss is low \citep{Krippner_PRA1994}.
There are proposals involving higher-order nonlinear interactions
that involve four-photon driven dissipation process which can reduce
the effect of single-photon losses \citep{Mirrahimi_NJP2014}. These
nonlinear interactions can be easily incorporated into our formalism,
but are not dealt with in this work.

\subsection{Number state expansion}

In the presence of damping and noise, a transient cat-state is nevertheless
possible for large $g$ \citep{Krippner_PRA1994,Munro_PRA1995}. In
order to fully capture the dynamics of the system, we give a numerical
solution of the master equation Eq. (\ref{eq:master_eqn}), by expanding
in the number state basis $\{|n\rangle\}$. This leads to time evolution
equations for each density operator matrix element $\rho_{nm}\equiv\langle n|\rho|m\rangle$:
\begin{widetext}
\begin{align}
\frac{\partial\rho_{nm}}{\partial\tau} & =-i\Delta\left(n-m\right)\rho_{n,m}+\frac{\lambda}{2}\left[\sqrt{n\left(n-1\right)}\rho_{n-2,m}\!+\!\sqrt{m\left(m-1\right)}\rho_{n,m-2}\!-\!\sqrt{\left(n+1\right)\left(n+2\right)}\rho_{n+2,m}\!-\!\sqrt{\left(m+1\right)\left(m+2\right)}\rho_{n,m+2}\right]\nonumber \\
 & -i\frac{\chi'}{2}\left[n\left(n-1\right)-m\left(m-1\right)\right]\rho_{n,m}+g^{2}\sqrt{\left(n+1\right)\left(n+2\right)\left(m+1\right)\left(m+2\right)}\rho_{n+1,m+2}-\frac{g^{2}}{2}\left[n\left(n-1\right)+m\left(m-1\right)\right]\rho_{n,m}\nonumber \\
 & +2\left(N+1\right)\sqrt{\left(n+1\right)\left(m+1\right)}\rho_{n+1,m+1}-\left(N+1\right)n\rho_{n,m}-\left(N+1\right)m\rho_{n,m}\nonumber \\
 & +2N\sqrt{nm}\rho_{n-1,m-1}-N\left(n+1\right)\rho_{n,m}-N\left(m+1\right)\rho_{n,m}\label{eq:number_state_expansion-1}
\end{align}
\end{widetext}where we introduce dimensionless parameters that are
scaled by $\gamma_{1}$: $\tau=\gamma_{1}t$, $\Delta=\bar{\Delta}/\gamma_{1}$,
$\lambda=\left|\bar{g}\epsilon\right|/\left(\gamma_{1}\gamma_{2}\right)$,
$\chi'=\bar{\chi}/\gamma_{1}$ and $g=\sqrt{\bar{g}^{2}/\left(2\gamma_{1}\gamma_{2}\right)}$.
For a given $n$ and $m$, the right side of Eq. (\ref{eq:number_state_expansion-1})
has contributions from indices other than $n$ and $m$. In other
words, we can express Eq. (\ref{eq:number_state_expansion-1}) as
follows:
\begin{align}
\frac{\partial}{\partial\tau}\rho_{n,m} & =\sum_{i}\sum_{j}\mathcal{L}_{nm}^{ij}\rho_{i,j}\,\label{eq:number_state_expansion_2-1-1}
\end{align}
where \begin{widetext}
\begin{align*}
\mathcal{L}_{nm}^{ij} & =\frac{\lambda}{2}\sqrt{n\left(n-1\right)}\delta_{n-2}^{i}\delta_{m}^{j}+\frac{\lambda}{2}\sqrt{m\left(m-1\right)}\delta_{n}^{i}\delta_{m-2}^{j}-\frac{\lambda}{2}\sqrt{\left(n+1\right)\left(n+2\right)}\delta_{n+2}^{i}\delta_{m}^{j}-\frac{\lambda}{2}\sqrt{\left(m+1\right)\left(m+2\right)}\delta_{n}^{i}\delta_{m+2}^{j}\\
 & -i\frac{\chi'}{2}\left[n\left(n-1\right)-m\left(m-1\right)\right]\delta_{n}^{i}\delta_{m}^{j}+g^{2}\sqrt{\left(n+1\right)\left(n+2\right)\left(m+1\right)\left(m+2\right)}\delta_{n+2}^{i}\delta_{m+2}^{j}-\frac{g^{2}}{2}\left[n\left(n-1\right)+m\left(m-1\right)\right]\delta_{n}^{i}\delta_{m}^{j}\\
 & +2\left(N+1\right)\sqrt{\left(n+1\right)\left(m+1\right)}\delta_{n+1}^{i}\delta_{m+1}^{j}-\left(N+1\right)n\delta_{n}^{i}\delta_{m}^{j}-\left(N+1\right)m\delta_{n}^{i}\delta_{m}^{j}\\
 & +2N\sqrt{nm}\delta_{n-1}^{i}\delta_{m-1}^{j}-N\left(n+1\right)\delta_{n}^{i}\delta_{m}^{j}-N\left(m+1\right)\delta_{n}^{i}\delta_{m}^{j}-i\Delta n\delta_{n}^{i}\delta_{m}^{j}+i\Delta m\delta_{n}^{i}\delta_{m}^{j}
\end{align*}
\end{widetext}Here, $\delta_{n}^{i}$ is a Kronecker delta function
with $\delta_{n}^{i}=1$ if $i=n$ and $\delta_{n}^{i}=0$ otherwise.

Eq. (\ref{eq:number_state_expansion_2-1-1}) is solved numerically
using the fourth-order Runge Kutta algorithm. Depending on the coherent
amplitude, a suitable photon-number cut-off is chosen. The validity
of this choice is checked by ensuring the diagonal matrix elements
with large photon number are not populated, and also by computing
the trace of the density operator, to ensure $Tr\rho=1$. Furthermore,
the convergence of the results is checked by increasing the cut-off
number. The time-step is chosen such that the time-step error is negligible.

\section{Transient cat-states with no Kerr nonlinearity \label{sec:Results}}

In this section, we analyze the dynamics of transient cat-states,
assuming zero detuning ($\bar{\Delta}=0$) and zero Kerr nonlinearity
($\bar{\chi}=0$). We ignore thermal noise. We solve the master equation
above numerically in the number state basis as explained in Section
\ref{sec:steady_state_number_state_expansion} and compute for the
quadrature probability distributions and their Wigner negativities.
These different cat-signatures are summarized in the Appendix \ref{sec:Cat signature},
and allow us to determine the onset of a cat-state.

We analyze for a complete range of parameters. In fact, three parameters
specify the transient behavior. These are $\lambda,$ $g$ given by
Eqns (\ref{eq:lambda}-\ref{eq:defn-g}) and defined earlier by Wolinsky
and Carmichael \citep{Wolinsky_PRL1988}, and the time $\tau=\gamma_{1}t$
scaled relative to the signal cavity decay time $1/\gamma_{1}$. In
fact, to analyze the strong coupling limit of large $g$, we find
it convenient to introduce a new set of parameters which completely
define the dynamics. These are: the pump strength scaled relative
to the oscillation threshold (as given in Eq. (\ref{eq:pump_threshold}))
\begin{equation}
\Lambda=\left|\bar{g}\epsilon\right|/\gamma_{2}=\gamma_{1}\lambda\,,\label{eq:capital-lambda}
\end{equation}
the scaled coupling strength
\begin{equation}
G=\sqrt{\bar{g}^{2}/\left(2\gamma_{2}\right)}=\sqrt{\gamma_{1}}g\,,\label{eq:capG}
\end{equation}
and the scaled time $T=G^{2}t.$ Using the parameters, the master
equation in Eq. (\ref{eq:master_eqn}) becomes
\begin{align}
\frac{\partial}{\partial T}\rho= & \frac{\Lambda}{2G^{2}}\left[a^{\dagger2}-a^{2},\rho\right]+\frac{1}{2}\left(2a^{2}\rho a^{\dagger2}-a^{\dagger2}a^{2}\rho-\rho a^{\dagger2}a^{2}\right)\nonumber \\
 & +\frac{\gamma_{1}}{G^{2}}\left(2a\rho a^{\dagger}-a^{\dagger}a\rho-\rho a^{\dagger}a\right).\label{eq:master_eqn_no_damping-1}
\end{align}
To make clear the relation with the case of signal damping $\gamma_{1}\neq0$,
we express $\Lambda$ and $G$ in terms of $\lambda$ and $g$
\begin{align}
\frac{\partial}{\partial T}\rho= & \frac{\lambda}{2g^{2}}\left[a^{\dagger2}-a^{2},\rho\right]+\frac{1}{2}\left(2a^{2}\rho a^{\dagger2}-a^{\dagger2}a^{2}\rho-\rho a^{\dagger2}a^{2}\right)\nonumber \\
 & +\frac{1}{g^{2}}\left(2a\rho a^{\dagger}-a^{\dagger}a\rho-\rho a^{\dagger}a\right).\label{eq:master_eqn_no_damping-1-1}
\end{align}
The first term is proportional to $\alpha_{0}^{2}$, which gives the
amplitudes $\pm\alpha_{0}=\pm\sqrt{\Lambda}/G=\pm\sqrt{\lambda}/g$
of the cat-state (that might be formed in the steady-state), as predicted
by Eq. (\ref{eq:dfn-alpha}). The last term in Eq. (\ref{eq:master_eqn_no_damping-1-1})
is zero in the case without single-photon damping ($\gamma_{1}=0$,
$g\rightarrow\infty$).

\subsection{{\normalsize{}Two-photon driving and dissipation with no single-photon
signal damping \label{subsec:G^2_scaling_lossless}}}

We would first like to understand the dynamics without single-photon
signal damping and at zero temperature. This corresponds to $\gamma_{1}=0$,
implying $g\rightarrow\infty$. Apart from the scaled time $T$ of
evolution, the master equation Eq. (\ref{eq:master_eqn_no_damping-1})
has only one free parameter which corresponds to the steady-state
coherent amplitude $\alpha_{0}=\sqrt{\Lambda}/G$. Here, we present
a determination of the interaction time $T$ required for the onset
of a cat-state, as a function of $\alpha_{0}=\sqrt{\Lambda}/G$, for
the full range of parameters, thus extending earlier work \citep{Gilles_PRA1994,Hach_PRA1994}.

\begin{figure}[H]
\begin{centering}
\includegraphics[width=0.51\columnwidth]{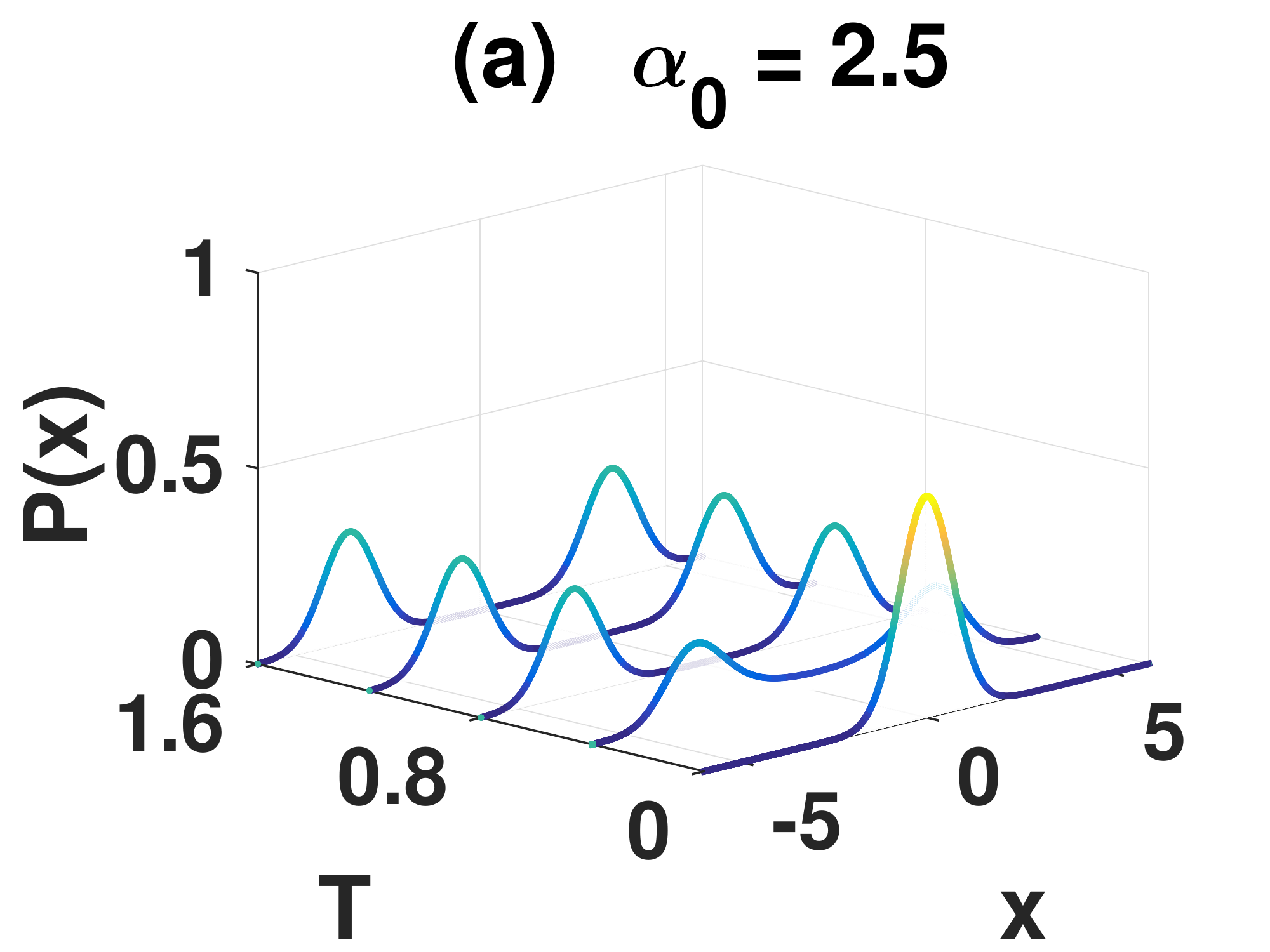}\includegraphics[width=0.51\columnwidth]{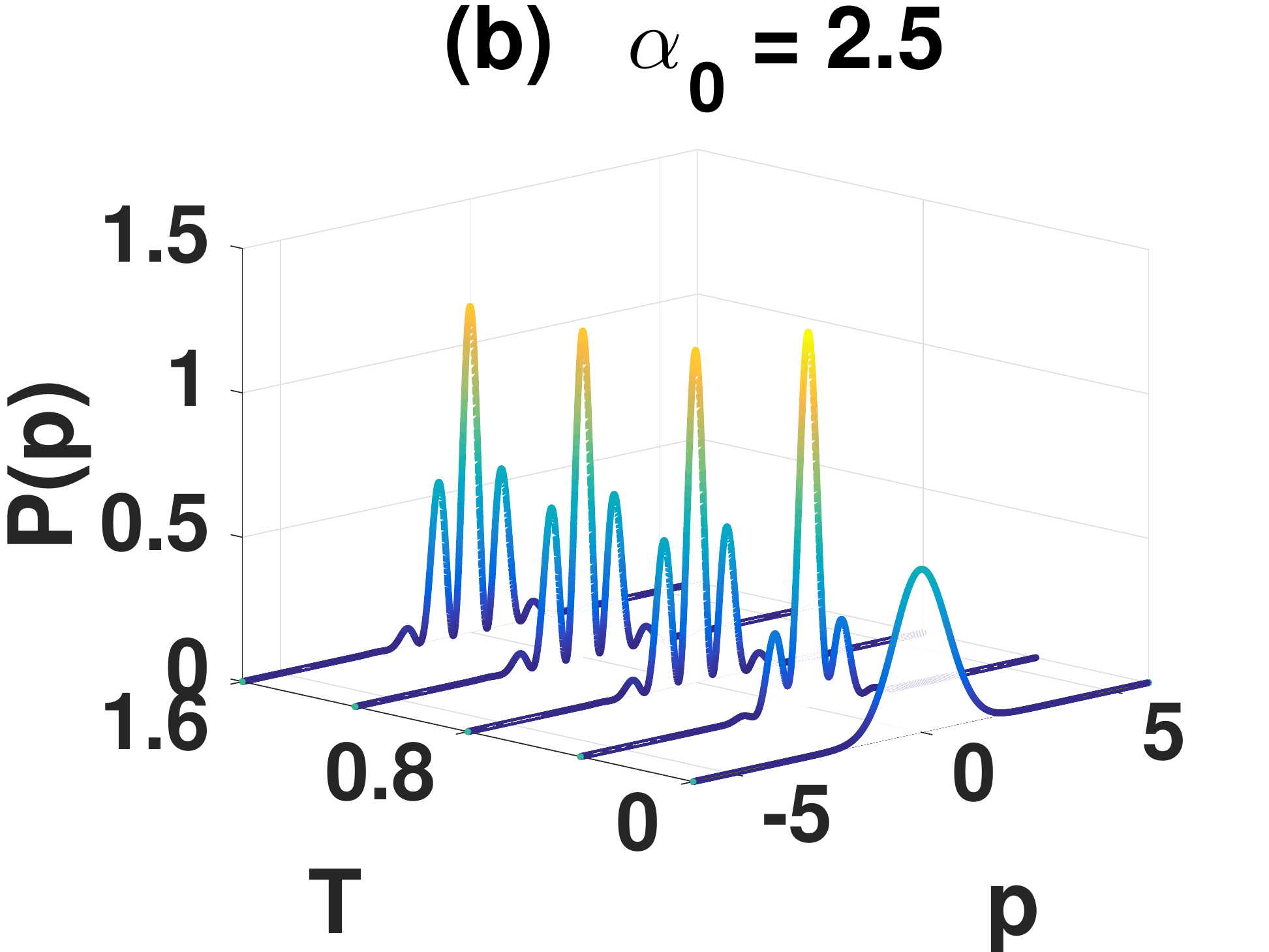}
\par\end{centering}
\begin{centering}
\bigskip{}
\includegraphics[width=0.51\columnwidth]{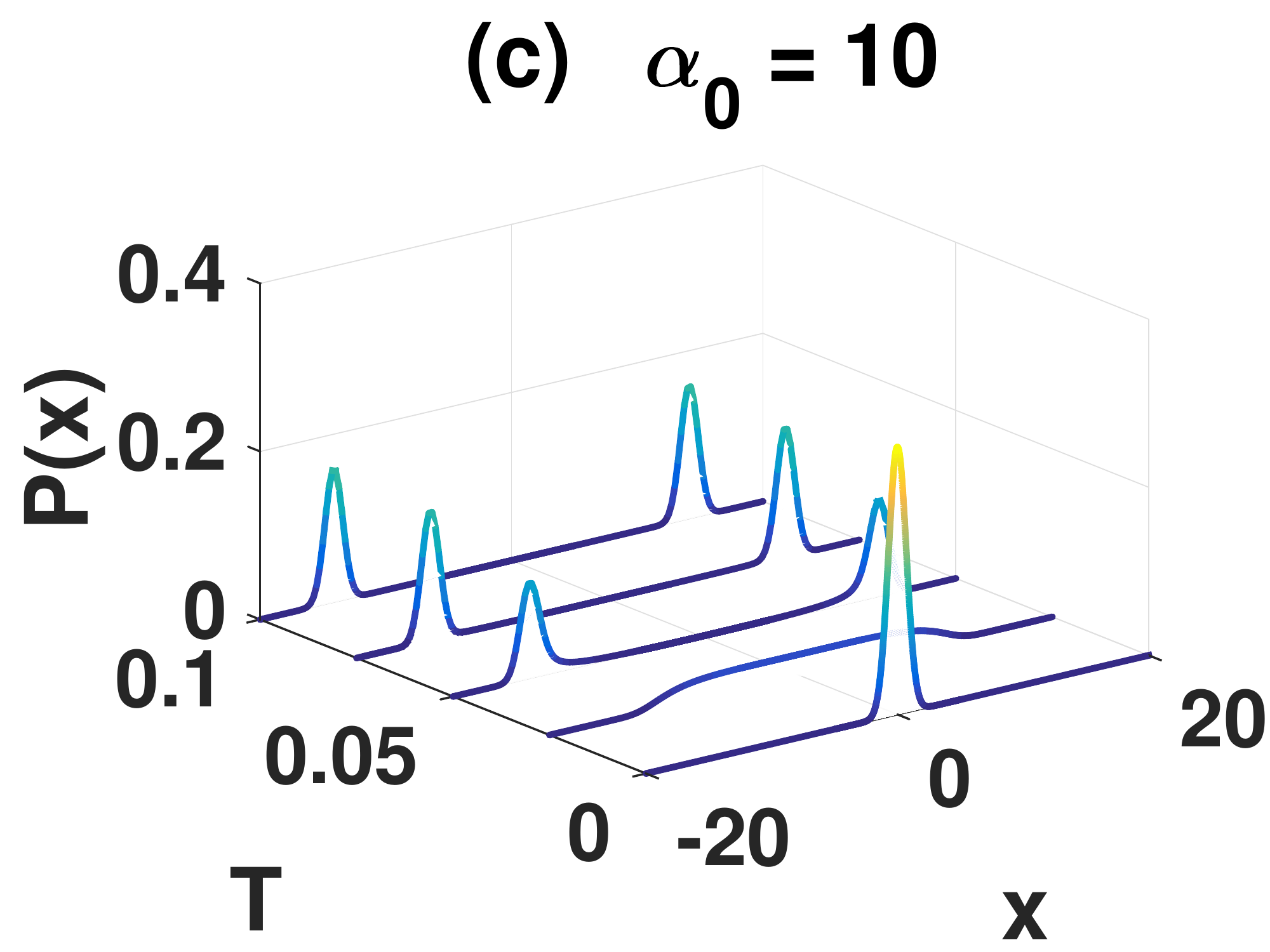}\includegraphics[width=0.51\columnwidth]{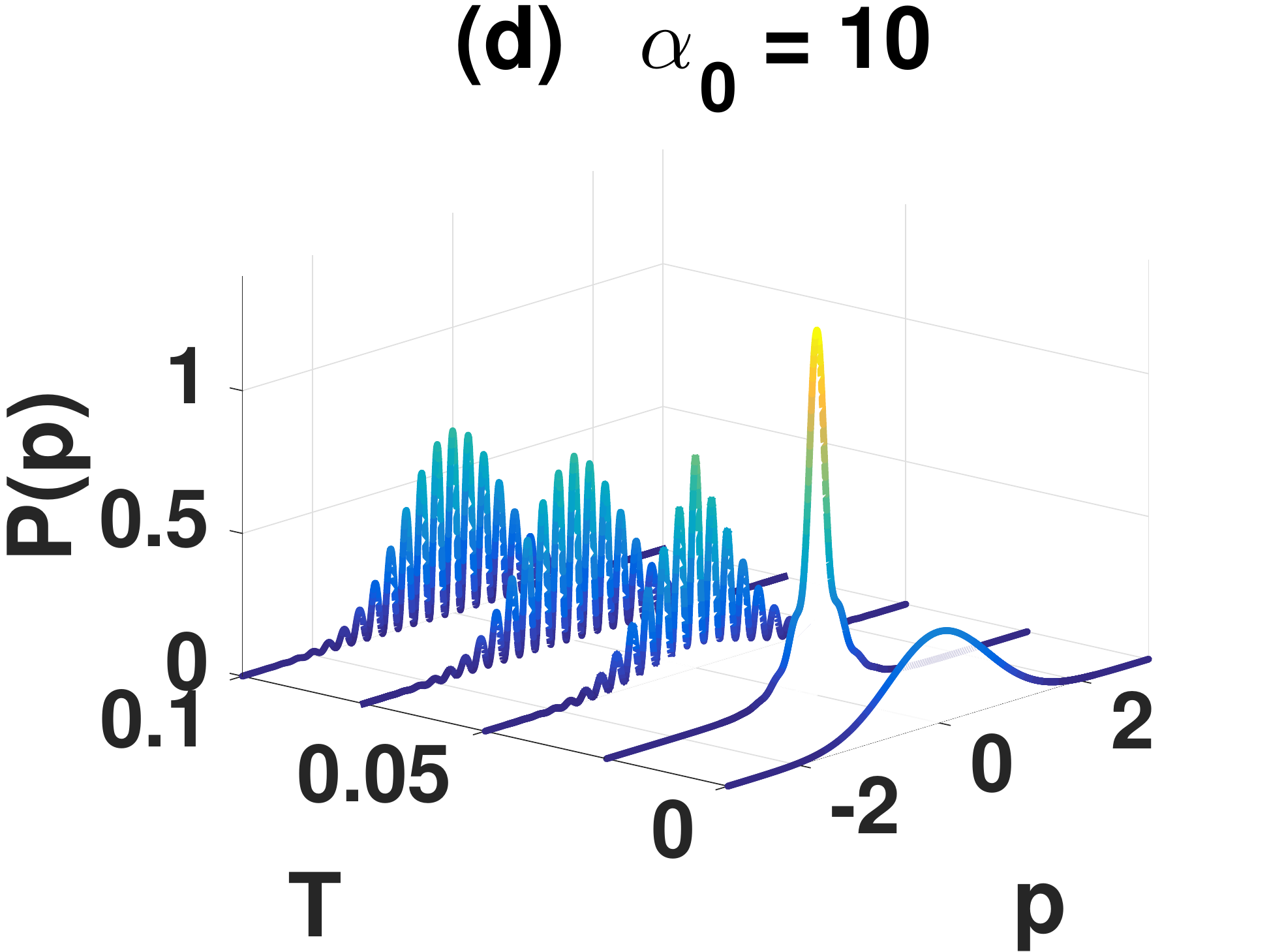}
\par\end{centering}
\caption{The $x$-quadrature probability distribution (a, c) and $p$-quadrature
probability distribution (b, d) as a function of scaled time $T=G^{2}t$
for $\alpha_{0}=2.5$ (a, b) and $\alpha_{0}=10$ (c, d). Here, $\gamma_{1}=0$.
\label{fig:lossless_Pp_a010}\textcolor{blue}{}}
\end{figure}

\begin{figure}[H]
\begin{centering}
\includegraphics[width=0.65\columnwidth]{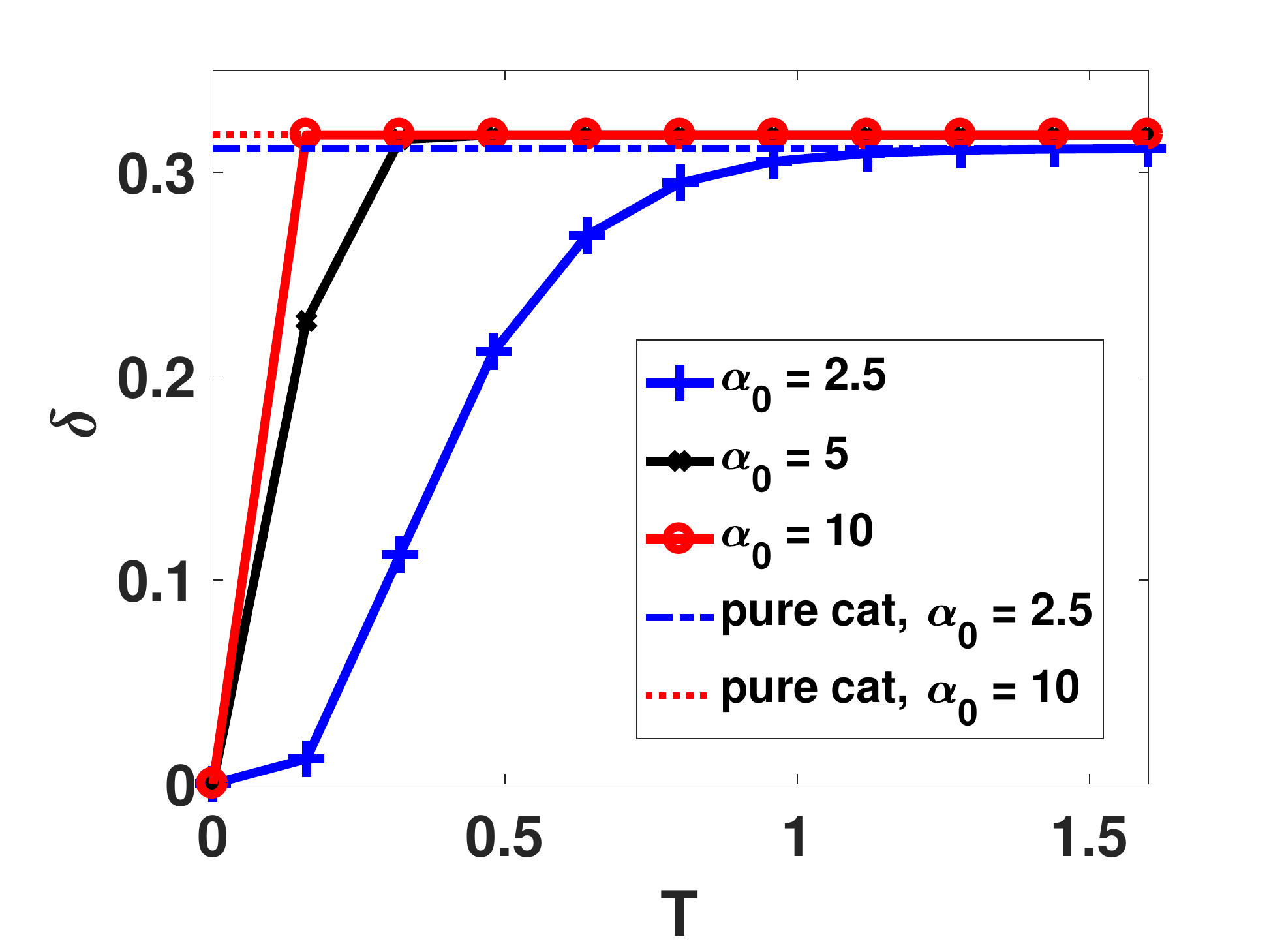}
\par\end{centering}
\caption{The time evolution of the Wigner negativity $\delta$ for $\alpha_{0}=2.5,\,5,$
and $10$, in terms of the scaled time $T=G^{2}t$. The blue (dashed),
black (solid) and red (dashed-dotted) lines correspond to $\alpha_{0}=2.5,\,5,$
and $10$ respectively. The blue dashed horizontal line shows the
Wigner negativity for a pure, even cat-state for $\alpha_{0}=2.5$
as calculated from the analytical Wigner function in Eq. (\ref{eq:Wigner_cat_state}).
The red dashed-dotted horizontal line corresponds to the same quantity
but for $\alpha_{0}=5$ and $10$, which have the same Wigner negativity.
The cat-formation time is calculated as the time taken for the Wigner
negativity to reach the analytical value associated with the pure
cat-state. \textcolor{black}{\label{fig:lossless_negativity_combine}
}\textcolor{red}{}}
\end{figure}

In Figures \ref{fig:lossless_Pp_a010} and \ref{fig:lossless_negativity_combine}
we fix $\alpha_{0}$ and determine the dimensionless time $T$ for
a transient cat state of amplitude $\alpha_{0}$ to appear, as measured
by the emergence of the fringes in $P(p)$ and the Wigner negativity
$\delta$. By comparing the numerical result of the Wigner negativity
time evolution with the Wigner negativity for a pure, even cat-state
of amplitude $\alpha_{0}$ in Eq. (\ref{eq:Wigner_cat_state}), the
dimensionless cat-formation time $T_{\text{cat}}$ is obtained when
the Wigner negativity from the numerical simulation agrees with the
analytical result of the ideal cat-state (refer Appendix) to four
significant figures.

These results demonstrate that larger cat-state amplitudes $\alpha_{0}$
have shorter scaled cat-state onset times $T_{\text{cat}}$. We next
discuss the cat-formation time $t_{\text{cat}}=T_{\text{cat}}/G^{2}$
for different cat-sizes $\alpha_{0}$ assuming $\gamma_{1}=0$. Recall
that a cat-state in the lossless case has an absolute coherent amplitude
$|\alpha_{0}|=|\sqrt{\Lambda}/G|$. In order to obtain $|\alpha_{0}|$
of a certain amplitude, one can either fix $G$ and change $\Lambda$
accordingly, or fix $\Lambda$ and change $G$, or change both. If
$G$ is fixed while $\Lambda$ is changed to obtain $|\alpha_{0}|$
of a certain amplitude, then $t_{\text{cat}}$ can indeed be shorter
for a larger cat-state (Table \ref{tab:cat_formation_time}). However,
$\Lambda$ scales as $\alpha_{0}^{2}G^{2}$ and this may quickly become
impractical for large $\alpha_{0}$.

\begin{table}[H]
\begin{centering}
\begin{tabular}{|c|c|c|c|}
\hline 
\multirow{2}{*}{$\alpha_{0}$} & \multirow{2}{*}{$T_{\text{cat}}$} & $t_{\text{cat}}=T_{\text{cat}}/G^{2},$ & $t_{\text{cat}}=T_{\text{cat}}\alpha_{0}^{2}/\Lambda$,\tabularnewline
 &  & (fixed $G$) & (fixed $\Lambda$)\tabularnewline
\hline 
\hline 
2.5 & $1.75\pm0.05$ & $\left(35.0\pm1.0\right)\text{\ensuremath{\mu s}}$ & $\left(35.0\pm1.0\right)\text{\ensuremath{\mu s}}$\tabularnewline
\hline 
5.0 & $0.45\pm0.035$ & $\left(9.0\pm0.7\right)\text{\ensuremath{\mu s}}$ & $\left(36.0\pm2.8\right)\text{\ensuremath{\mu s}}$\tabularnewline
\hline 
10.0 & $0.14\pm0.01$ & $\left(2.80\pm0.20\right)\text{\ensuremath{\mu}s}$ & $\left(44.8\pm3.2\right)\text{\ensuremath{\mu s}}$\tabularnewline
\hline 
\end{tabular}
\par\end{centering}
\caption{The cat-formation times for fixed $G$ and fixed $\Lambda$. \textcolor{black}{Here
}$T=G^{2}t$ \textcolor{black}{is the scaled time and $t_{\text{cat}}$
is the real time in seconds. In the third column, we use the estimated
value of $G=2.24\times10^{2}\sqrt{\text{Hz}}$ for the experiment
\citep{Leghtas_Science2015}. In the last column, we fix $\Lambda=3.13\times10^{5}\text{Hz}$.}\textcolor{blue}{}
\label{tab:cat_formation_time}}
\end{table}

To get a sense of the timescale in real times, we consider the parameters
from the experiment reported in \citep{Leghtas_Science2015}. The
nonlinear coupling strength is $\bar{g}/2\pi=225\text{kHz}$, and
the Kerr-type interaction strength is $\bar{\chi}/2\pi=4\text{kHz}$.
The single signal-photon damping rate is $\gamma_{1}/2\pi=3.98\text{kHz}$,
and single pump-photon damping rate $\gamma_{2}/2\pi=3.18\text{MHz}$.
In this section, we choose the pump field amplitude to be $\epsilon/2\pi=703\text{kHz}$
such that $|\alpha_{0}|=2.5$, without the Kerr term ($\bar{\chi}=0$),
according to Eq. (\ref{eq:cat-amplitude-chi}). These correspond
to parameter values $G=\sqrt{\bar{g}^{2}/\left(2\gamma_{2}\right)}=2.24\times10^{2}\sqrt{\text{Hz}}$
and $\Lambda=\left|\bar{g}\epsilon\right|/\gamma_{2}=3.13\times10^{5}\text{Hz}$.
In practice, it is better to modify both the parameters $G$ and
$\Lambda$ for different $\alpha_{0}$. For the sake of our discussion,
however, we consider the case where $\Lambda=3.13\times10^{5}\text{Hz}$
is fixed and we change $G$ accordingly, where $G$ scales as $\sqrt{\Lambda}/\alpha_{0}$.
Hence, $t_{\text{cat}}=T_{\text{cat}}/G^{2}=T_{\text{cat}}\alpha_{0}^{2}/\Lambda$.
The $t_{\text{cat}}$ for different cat sizes are shown in Table \ref{tab:cat_formation_time}.

\subsection{Single-photon signal damping}

Next, we include the effect of the signal damping ($\gamma_{1}\neq0$).
Apart from the time of evolution, the master equation Eq. (\ref{eq:master_eqn_no_damping-1-1})
has two free parameters $\alpha_{0}$ and $g$, which is the effective
ratio of the two-photon nonlinearity to the signal decay rate. For
sufficiently small $g$, cat-states cannot form. As mentioned previously,
the cat-size is given by the amplitude $\alpha_{0}=\sqrt{\lambda}/g$,
and we fix this for each Figure below. The parameter $g$ is changed
in order to find the threshold value of $g$ where interference fringes,
and hence a cat-state, emerge. We take $\alpha_{0}=2.5$ to be the
minimum value of $\alpha_{0}$ corresponding to a cat-state.

Figures 3, \ref{fig:lossy_Pp_ao2.5} and \ref{fig:lossy_Pp_ao10}
indicate that $g>1$ is the threshold for the emergence of fringes
(and hence of a cat-state), regardless of the amplitude $\alpha_{0}$
of the cat size. For $g>1$, the Figures show the interference fringes
to become more pronounced as $g$ increases. For long enough $T$,
the fringes vanish, as the system approaches a steady-state. The
steady state is not a cat-state, having a positive Wigner function
\citep{Reid_PRA1992}.

\begin{figure}[H]
\includegraphics[width=0.51\columnwidth]{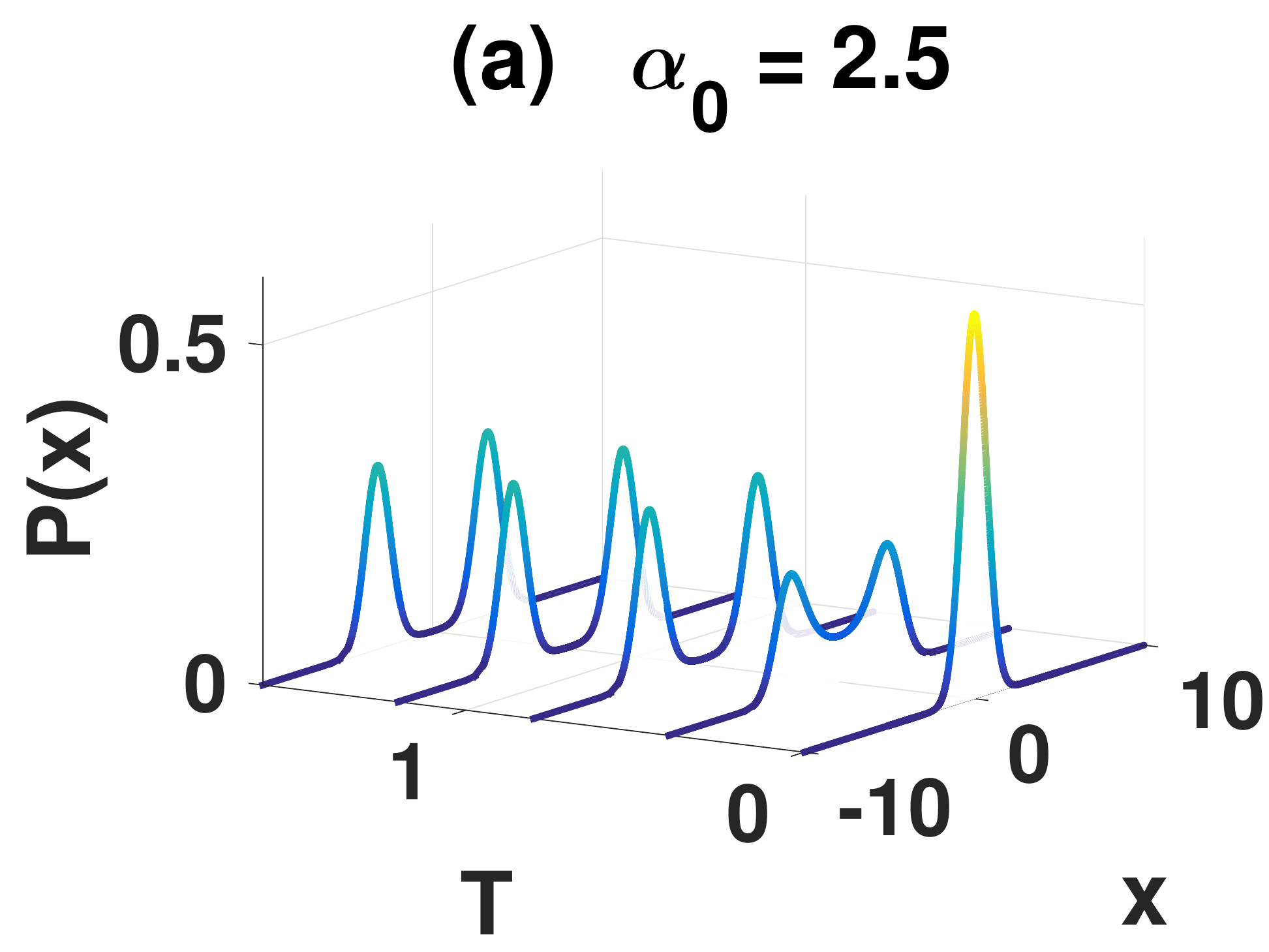}\includegraphics[width=0.51\columnwidth]{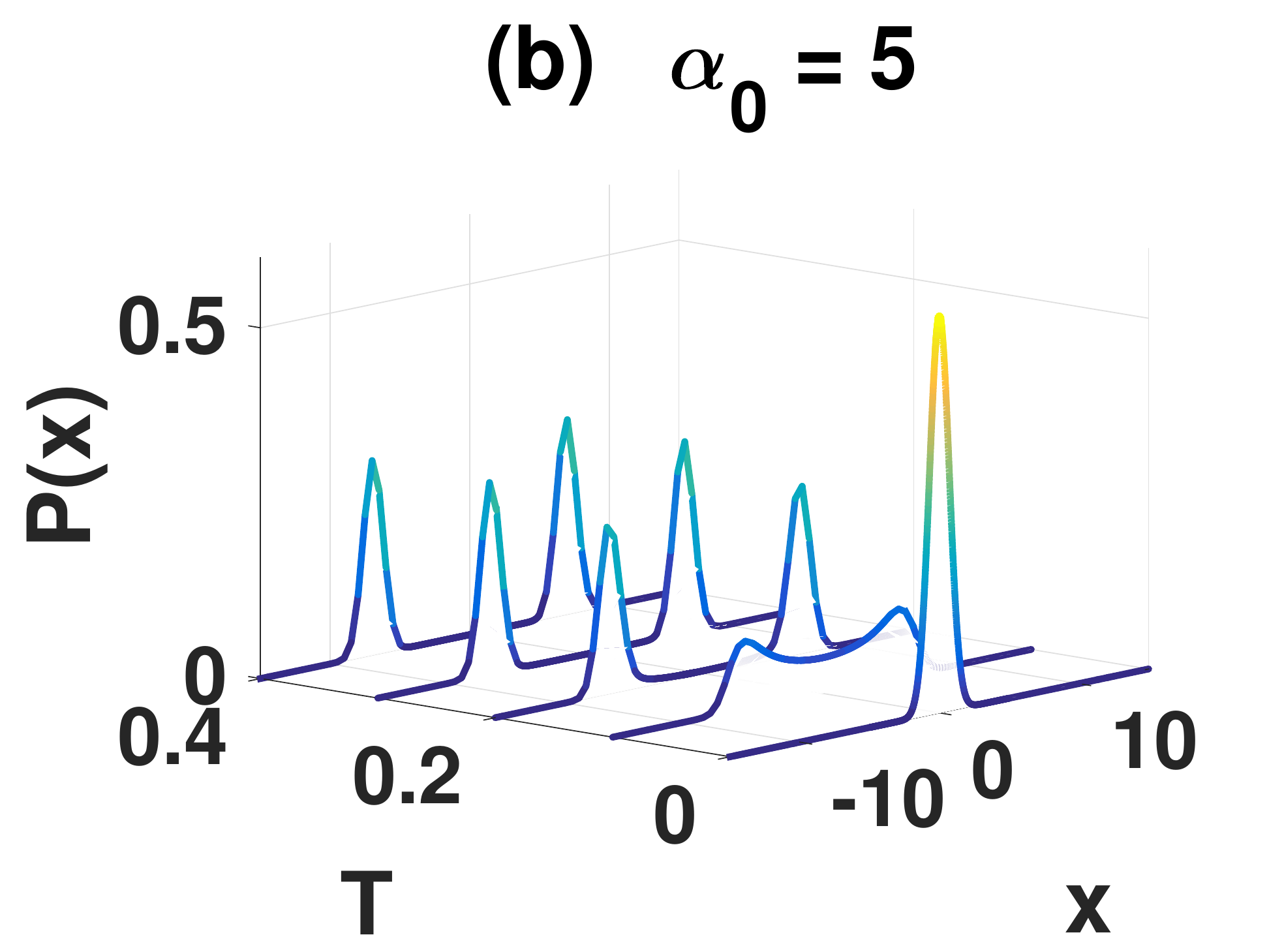}
\begin{centering}
\includegraphics[width=0.51\columnwidth]{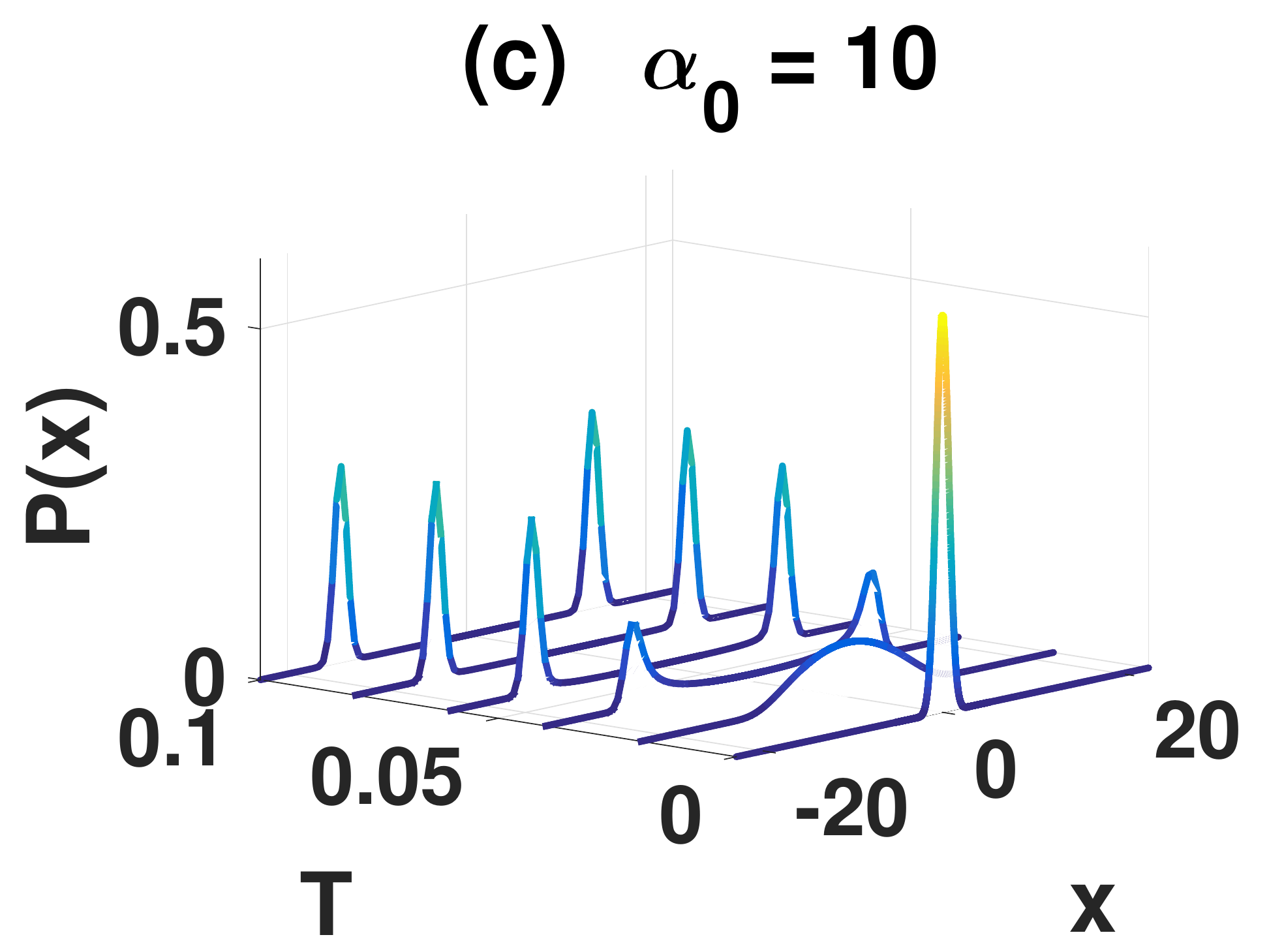}
\par\end{centering}
\caption{The $x$-quadrature probability distribution as a function of scaled
time $T=G^{2}t$ for $g=1$ for various $\alpha_{0}$. The distribution
is unchanged for $g=1.5,$ 2.0 and 2.5. We note that $g=G/\sqrt{\gamma_{1}}$.\label{fig:lossy_Pp_ao2.5-1}\textcolor{red}{}}
\end{figure}

\begin{figure}[H]
\includegraphics[width=0.51\columnwidth]{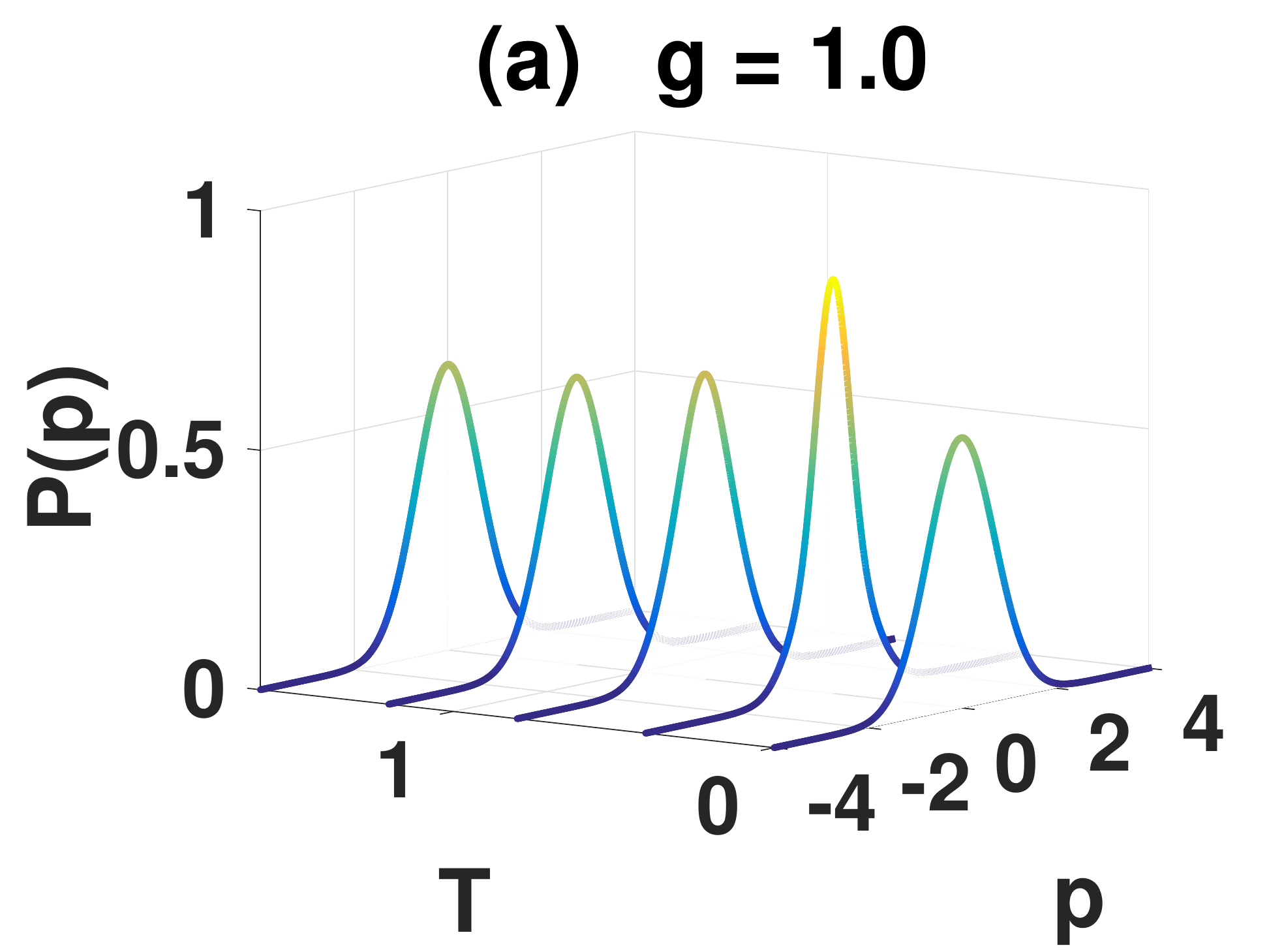}\includegraphics[width=0.51\columnwidth]{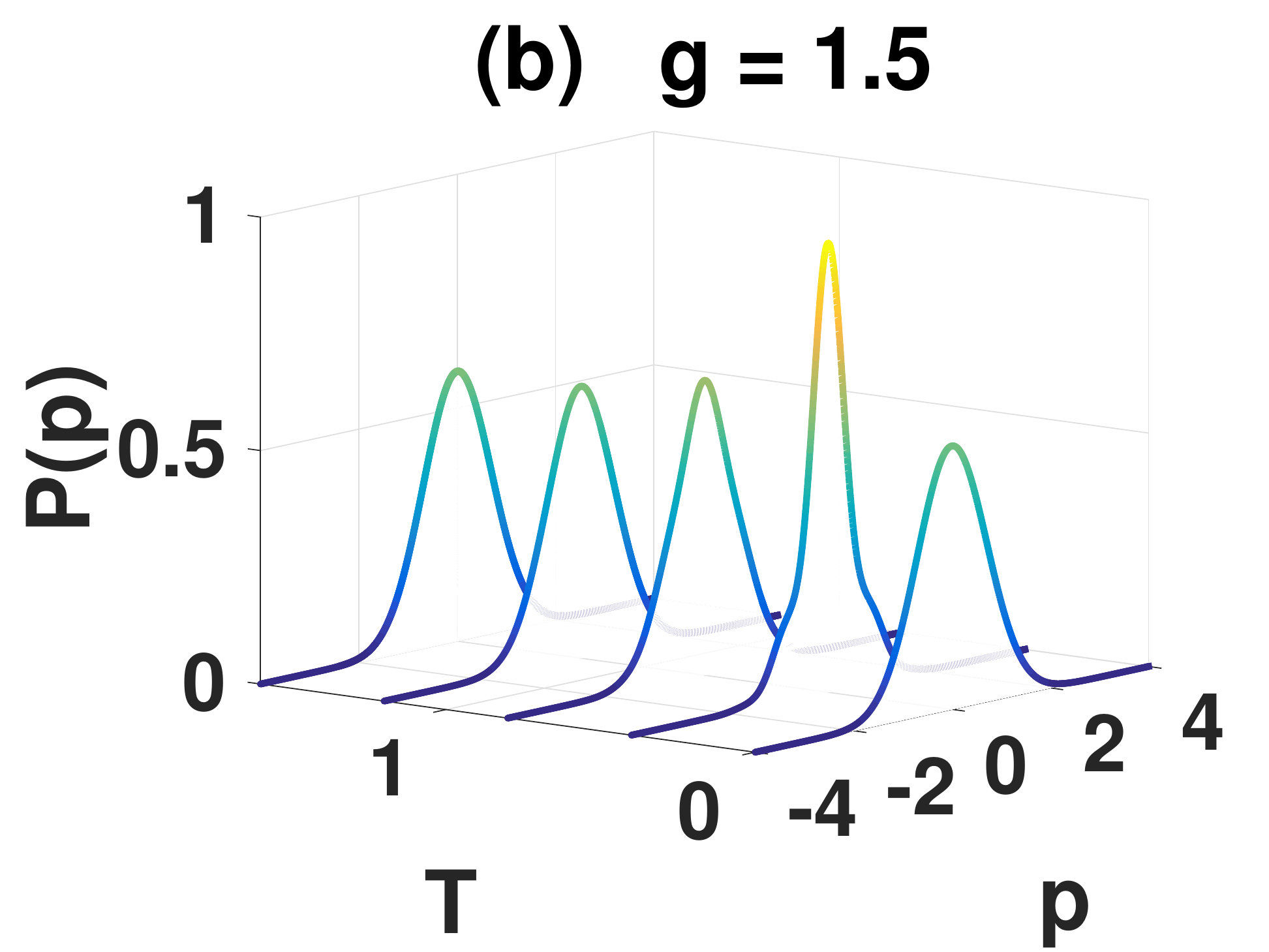}

\bigskip{}

\includegraphics[width=0.51\columnwidth]{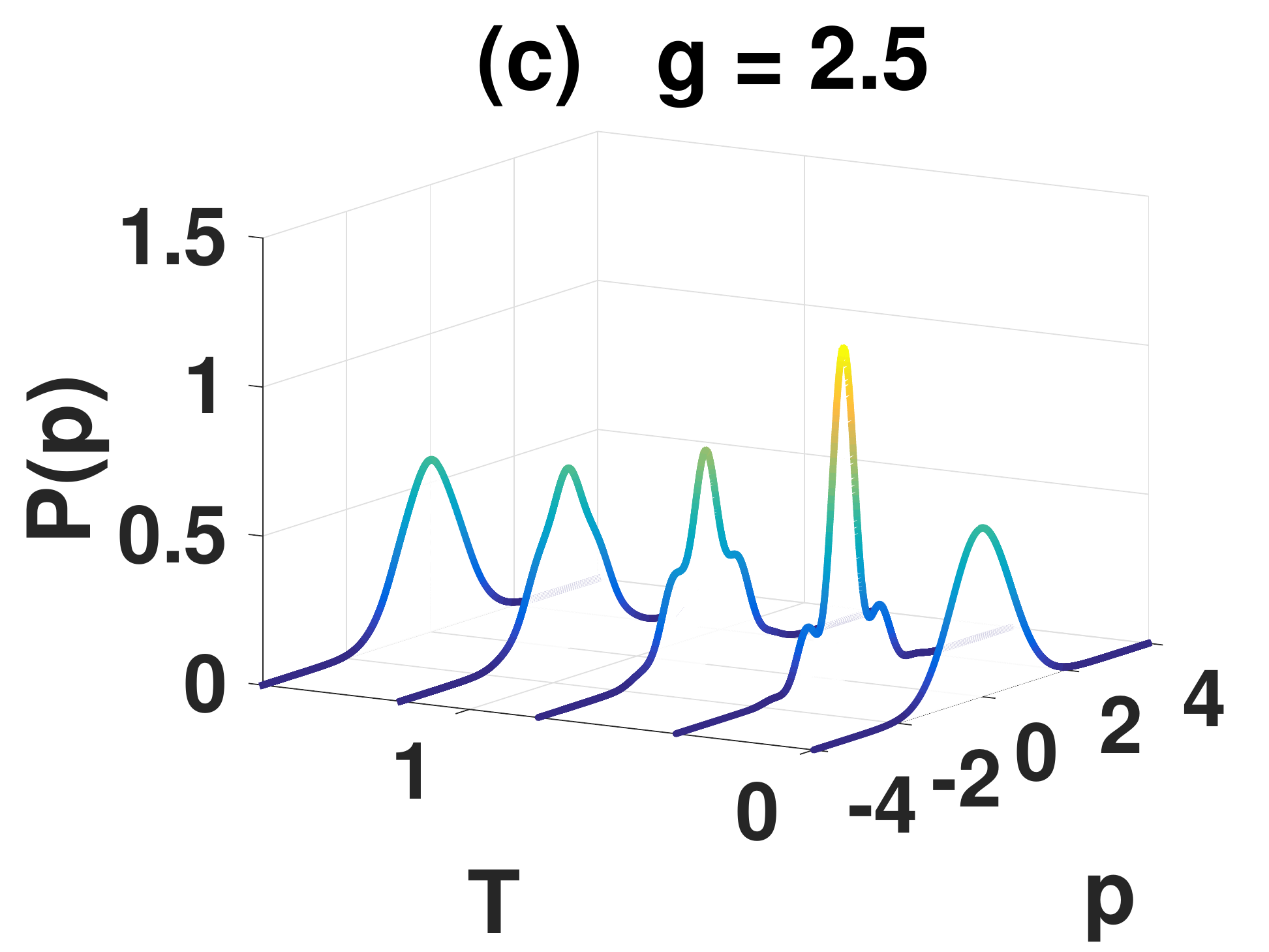}\includegraphics[width=0.51\columnwidth]{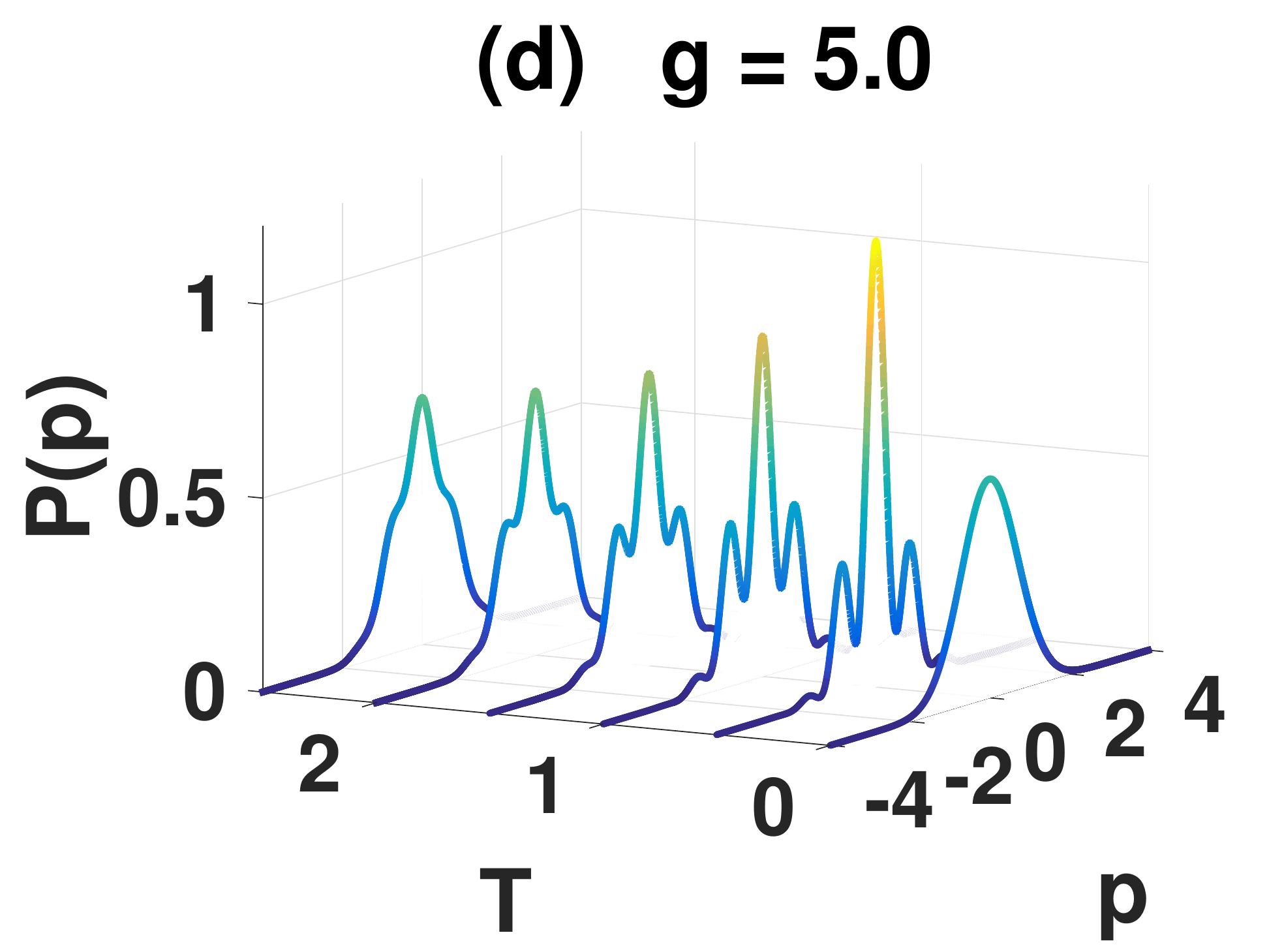}

\caption{The $p$-quadrature probability distribution as a function of scaled
time $T=G^{2}t$ for different values of $g=G/\sqrt{\gamma_{1}}$.
Here, $\alpha_{0}=2.5$. For (a), (b), and (c), the time range for
$T$ is $0-1.6$. For (d), the time range is $0-2.5$. \label{fig:lossy_Pp_ao2.5}
\textcolor{blue}{}}
\end{figure}
\begin{figure}[H]
\includegraphics[width=0.51\columnwidth]{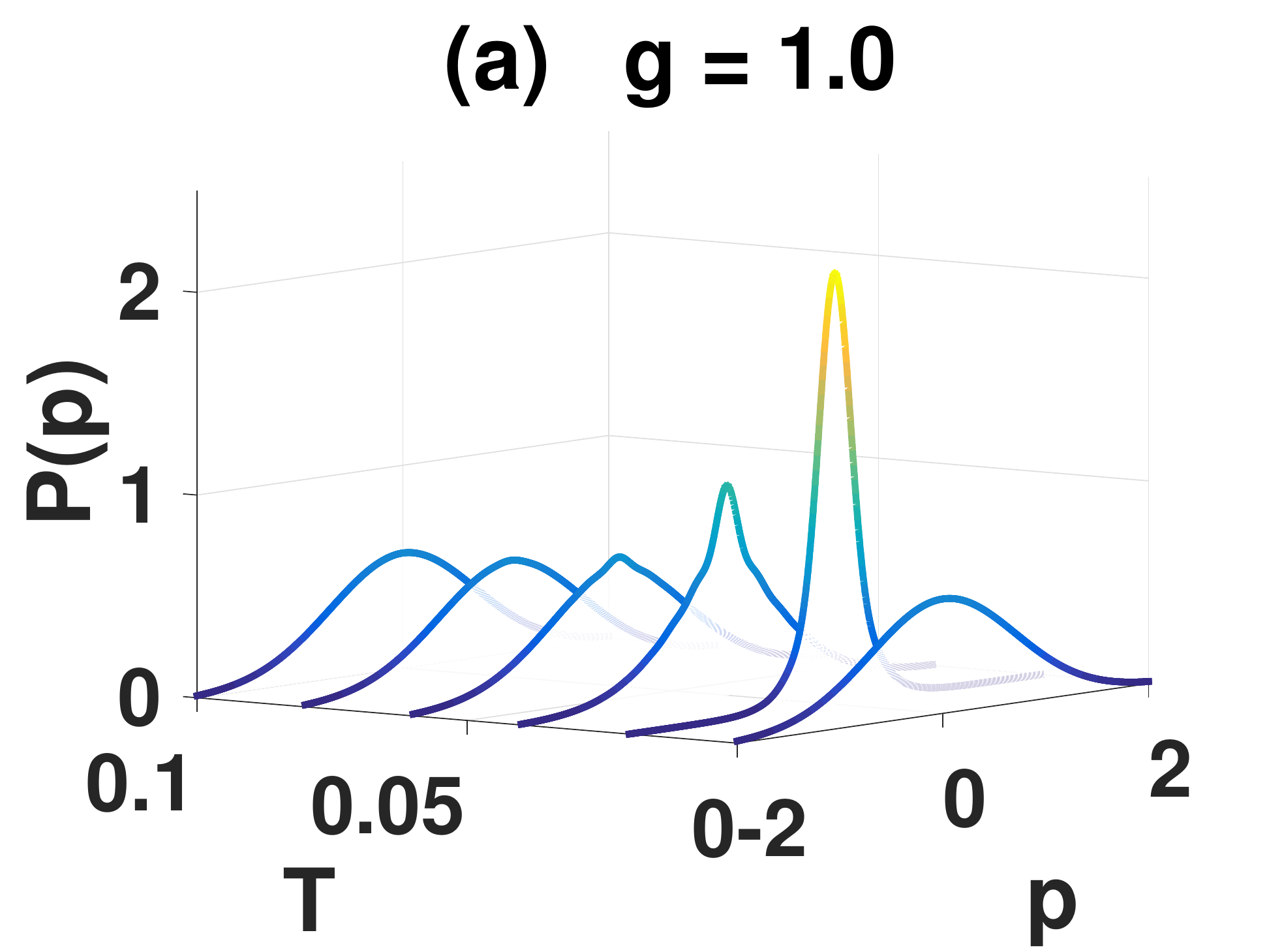}\includegraphics[width=0.51\columnwidth]{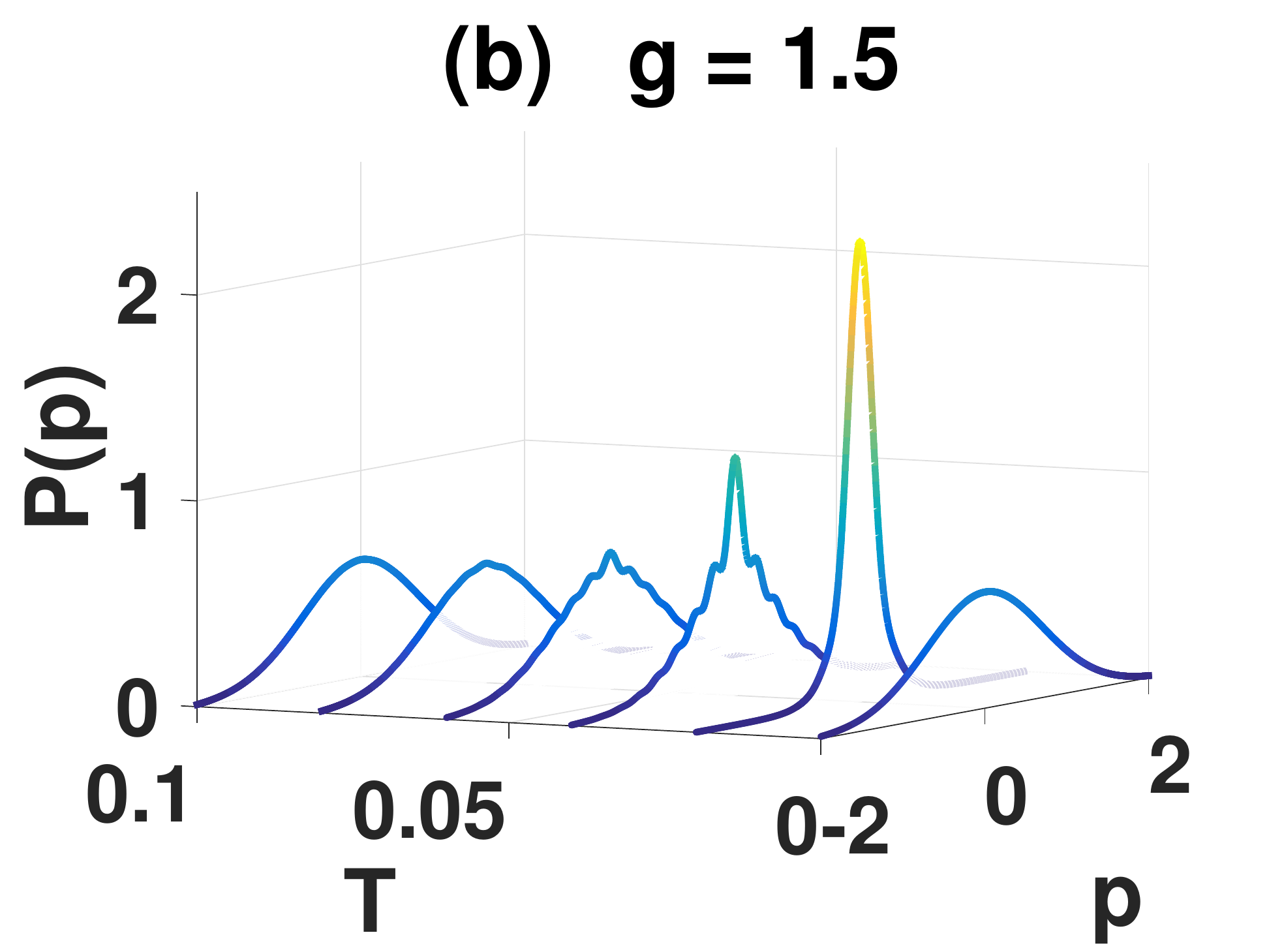}

\bigskip{}

\includegraphics[width=0.51\columnwidth]{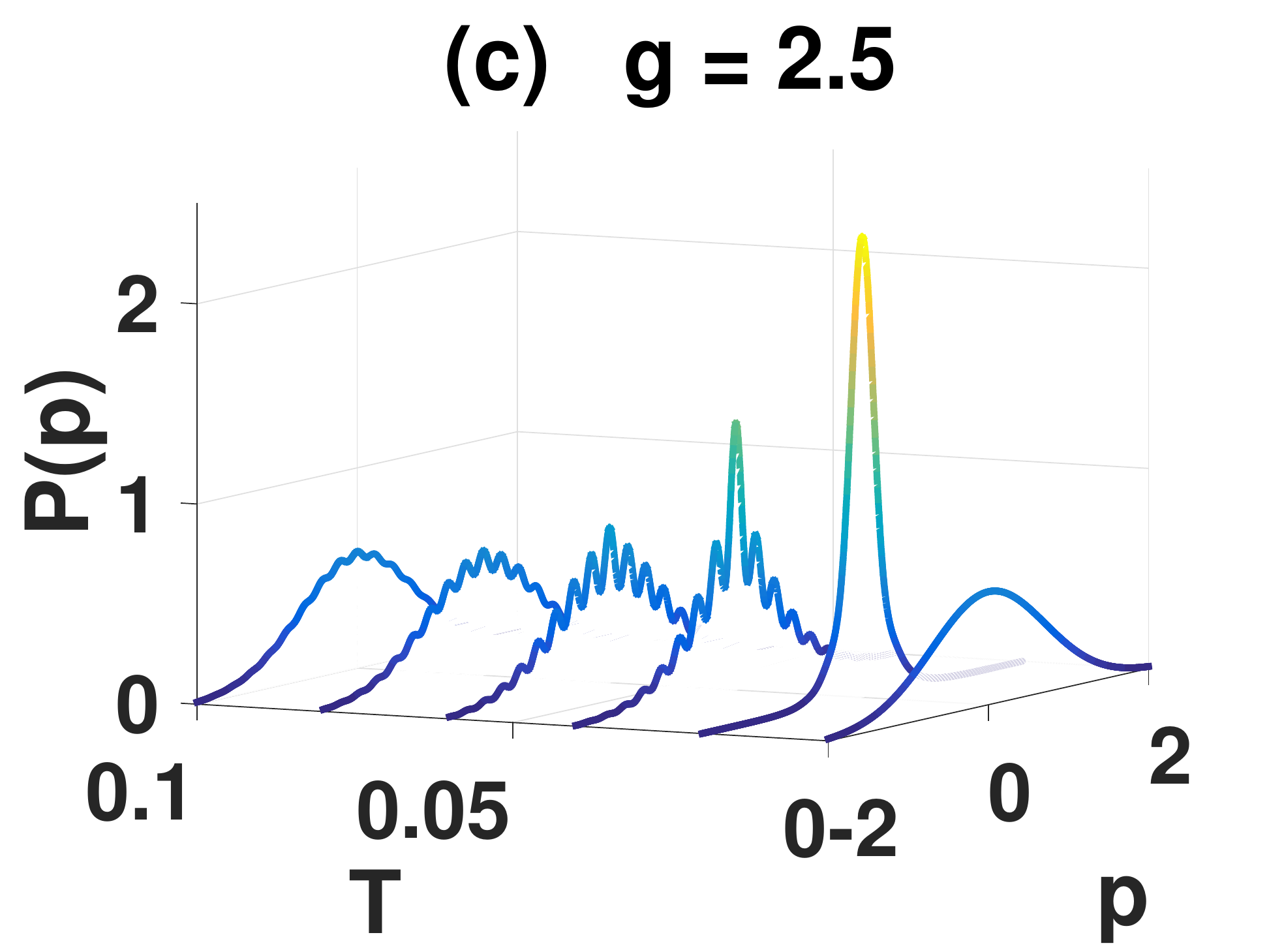}\includegraphics[width=0.51\columnwidth]{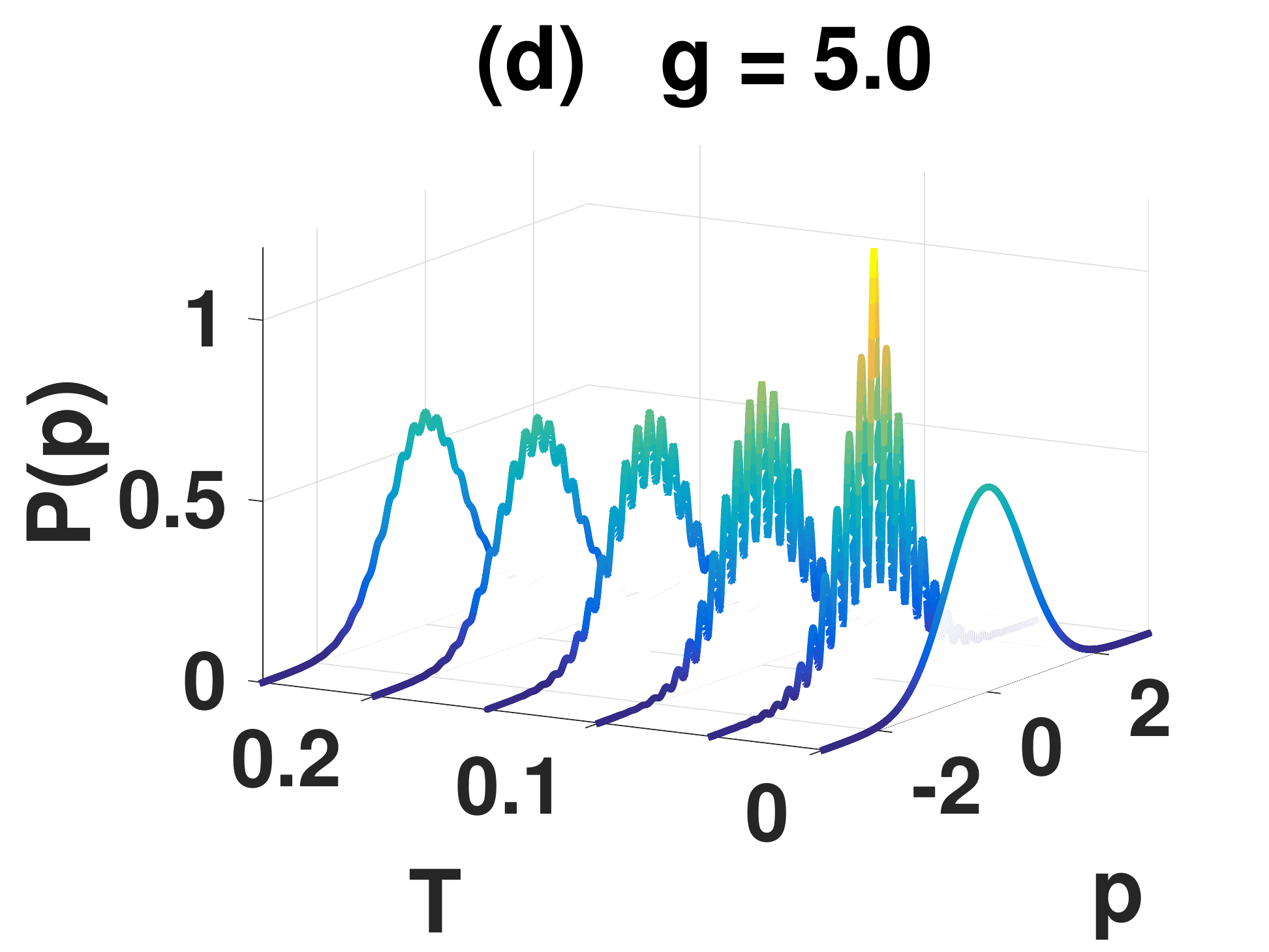}

\caption{The $p$-quadrature probability distribution as a function of scaled
time $T=G^{2}t$ for different values of $g=G/\sqrt{\gamma_{1}}$.
Here,\textcolor{red}{{} }$\alpha_{0}=10$.\label{fig:lossy_Pp_ao10}}
\end{figure}

\begin{figure}[H]
\centering{}\includegraphics[width=0.51\columnwidth]{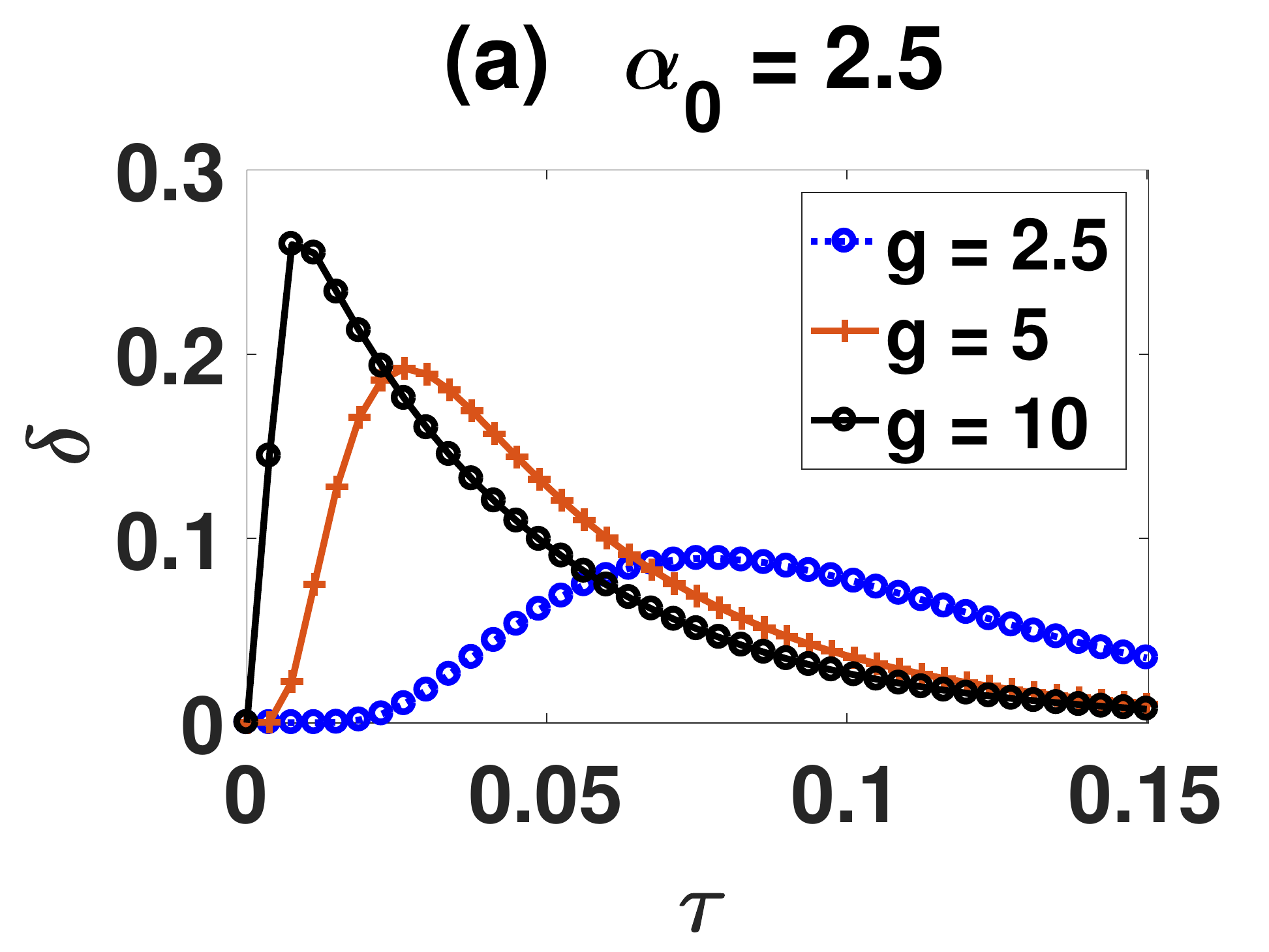}\includegraphics[width=0.51\columnwidth]{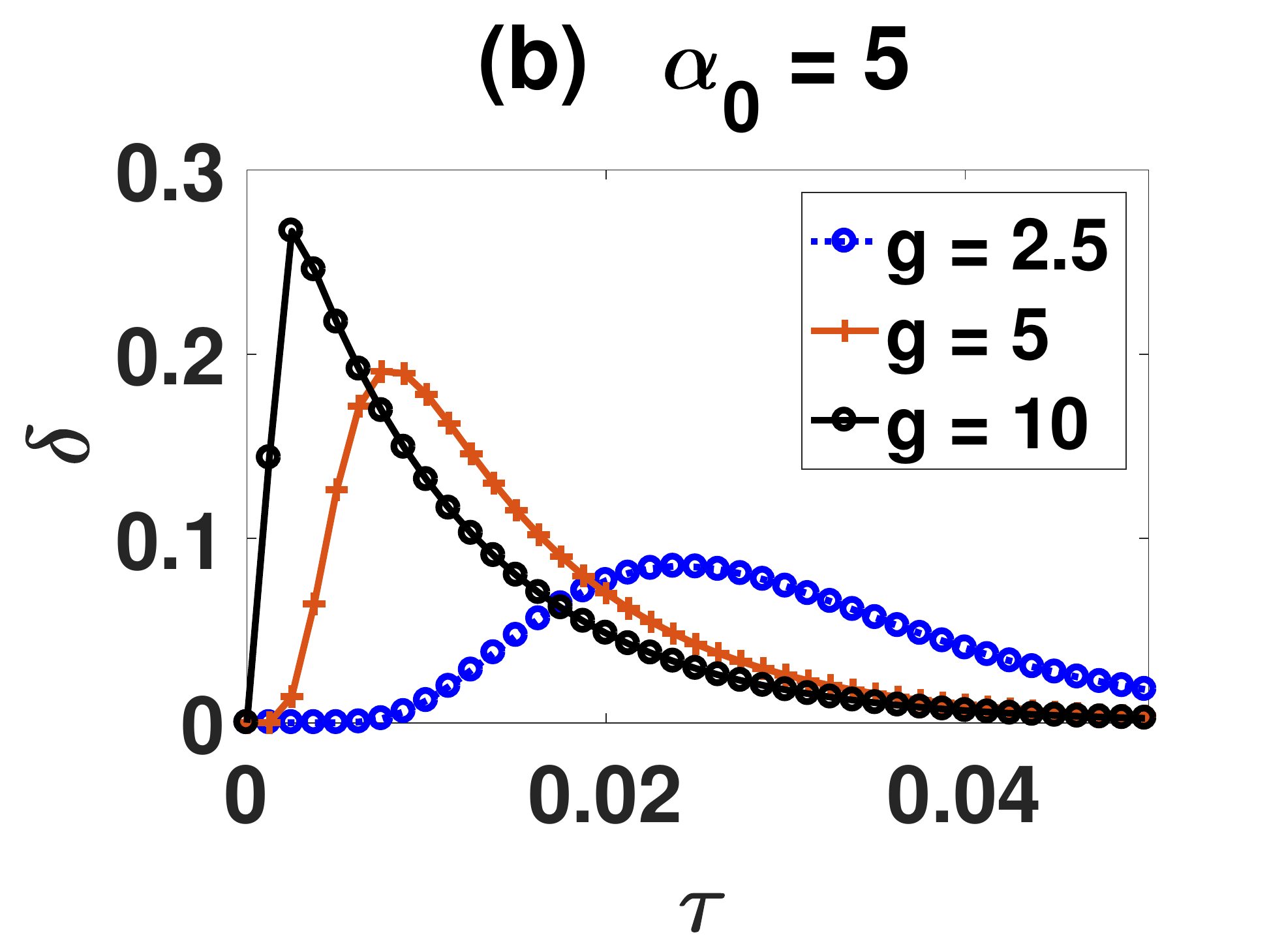}\bigskip{}
\includegraphics[width=0.51\columnwidth]{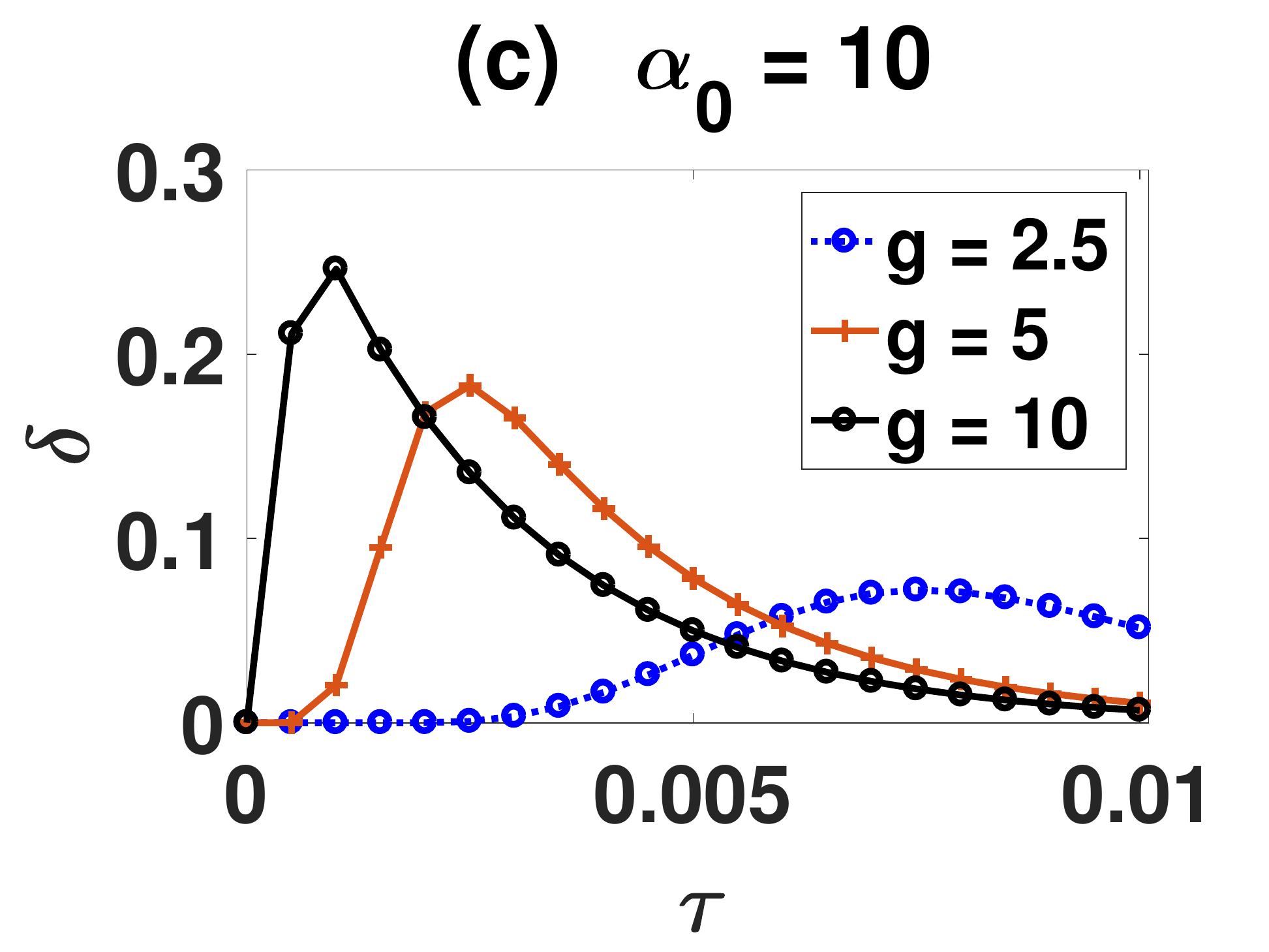}\caption{The evolution of Wigner negativity in time $\tau$ (in units of the
cavity lifetime $\gamma_{1}^{-1}$) for different values of $\alpha_{0}$.
In each plot, the different lines correspond to different $g$ values.
The blue, orange and black lines correspond to $g=2.5,\,5$ and $10$
respectively. \label{fig:scaling_gamma_delta_damping}\textcolor{red}{}}
\end{figure}

It is interesting to know the experimental run-time needed to
obtain a cat-like state with the maximal non-classicality. This is
quantified by the Wigner negativity. We computed the time evolution
of the Wigner negativity. This allows us to estimate the time of formation
of a transient state with the largest Wigner negativity, given $\alpha_{0}$
and $g$. In Fig. \ref{fig:scaling_gamma_delta_damping}, we present
the Wigner negativity results with different $g$'s, for $\alpha_{0}=2.5,\,5$
and $10$ respectively.   The results are presented with respect
to the time $\tau=\gamma_{1}t=T/g^{2}$ relative to the signal-cavity
lifetime. We see first the formation of the cat-state, followed by
its decay. Assuming the cavity lifetime is unchanged, for fixed $|\alpha_{0}|$
a larger $g$ implies a quicker formation, but also a quicker decay.
Larger cat-sizes $\alpha_{0}$ imply quicker timescales.

We define the cat-state lifetime as the time $\tau$ taken for the
Wigner negativity to reduce from the maximum value to $\delta\leq0.05$.
We note that this choice is rather arbitrary and is mainly motivated
by the practical consideration that a state with $\delta=0.05$ is
too small to be treated as a cat-state or any useful nonclassical
state, while at the same time not too small that the numerical simulations
remain tractable. A much longer simulation time is needed to reach
a state with $\delta=0$, which would be a more natural choice as
the cat-state lifetime. The cat-state lifetimes for different values
of $g$ and $\alpha_{0}$ are tabulated in Table \ref{tab:cat_lifetime_table}.
From the table and Fig. \ref{fig:scaling_gamma_delta_damping}, we
see that for a fixed $\alpha_{0}$, the cat-states with larger $g$
have a shorter lifetime, even though a larger Wigner negativity can
be reached. Also, for fixed $g$, the smaller cat-states have a longer
lifetime.
\begin{table}[H]
\begin{tabular}{|c|c|c|c|c|c|c|}
\hline 
\multirow{2}{*}{$g$} & \multicolumn{3}{c|}{$T$} & \multicolumn{3}{c|}{$\tau$}\tabularnewline
\cline{2-7} 
 & $\alpha_{0}=2.5$ & $\alpha_{0}=5$ & $\alpha_{0}=10$ & $\alpha_{0}=2.5$ & $\alpha_{0}=5$ & $\alpha_{0}=10$\tabularnewline
\hline 
\hline 
$2.5$ & $0.8206$ & $0.2344$ & $0.0625$ & $0.1313$ & $0.0375$ & $0.0100$\tabularnewline
\hline 
$5$ & $2.250$ & $0.5950$ & $0.1625$ & $0.0900$ & $0.0238$ & $0.0065$\tabularnewline
\hline 
10 & $7.880$ & $2.000$ & $0.50$ & $0.0788$ & $0.0200$ & $0.0050$\tabularnewline
\hline 
\end{tabular}

\caption{The cat-state lifetimes for different $g$ and $\alpha_{0}$ values
as given in units of the signal cavity decay time, $\tau=\gamma_{1}t$.
Here, $T=G^{2}t$. The cat-state lifetime is defined as the time taken
for the Wigner negativity to reach $\delta\protect\leq0.05$. \label{tab:cat_lifetime_table}}
\end{table}

\begin{figure}[H]
\includegraphics[width=0.51\columnwidth]{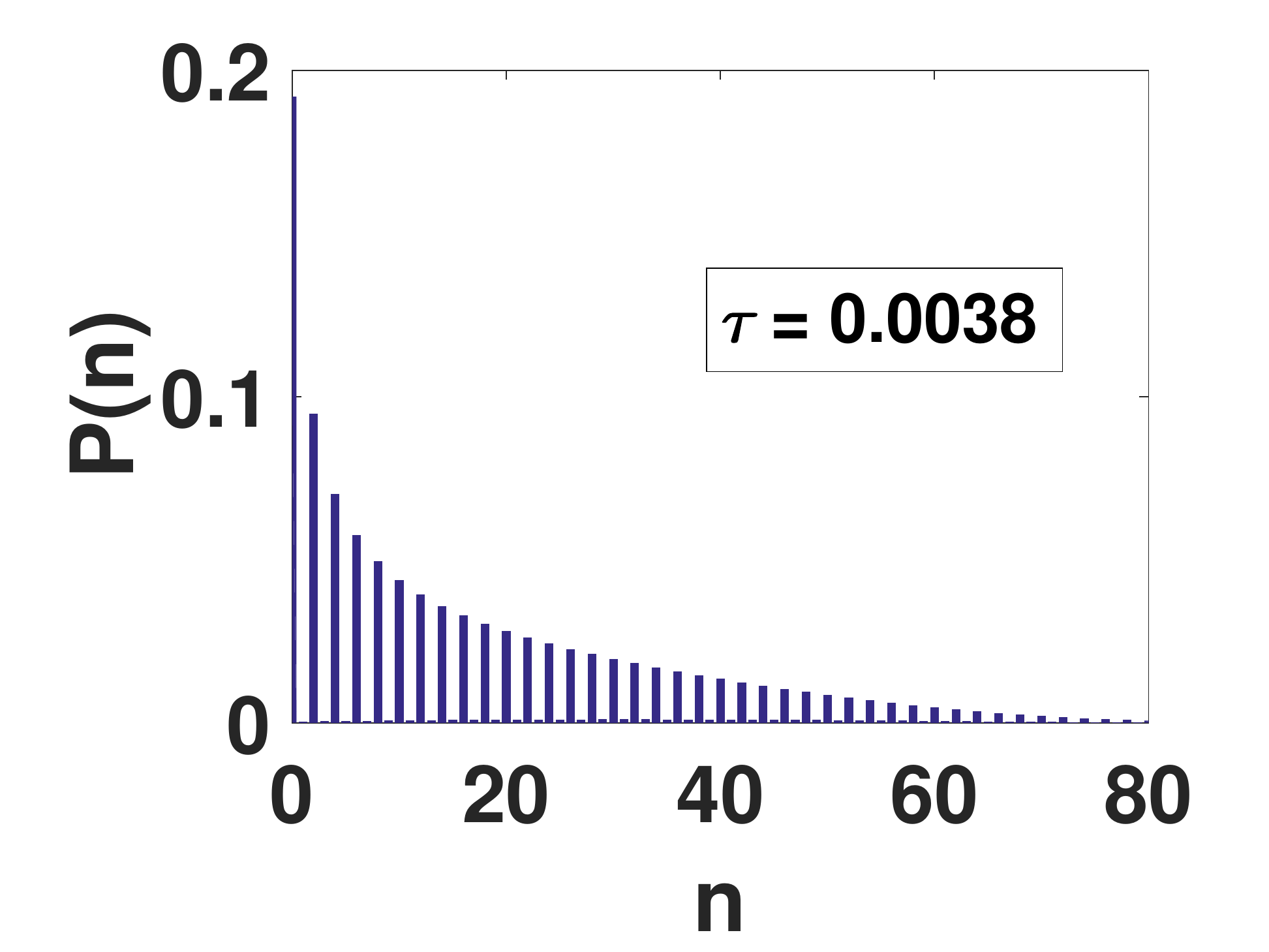}\includegraphics[width=0.51\columnwidth]{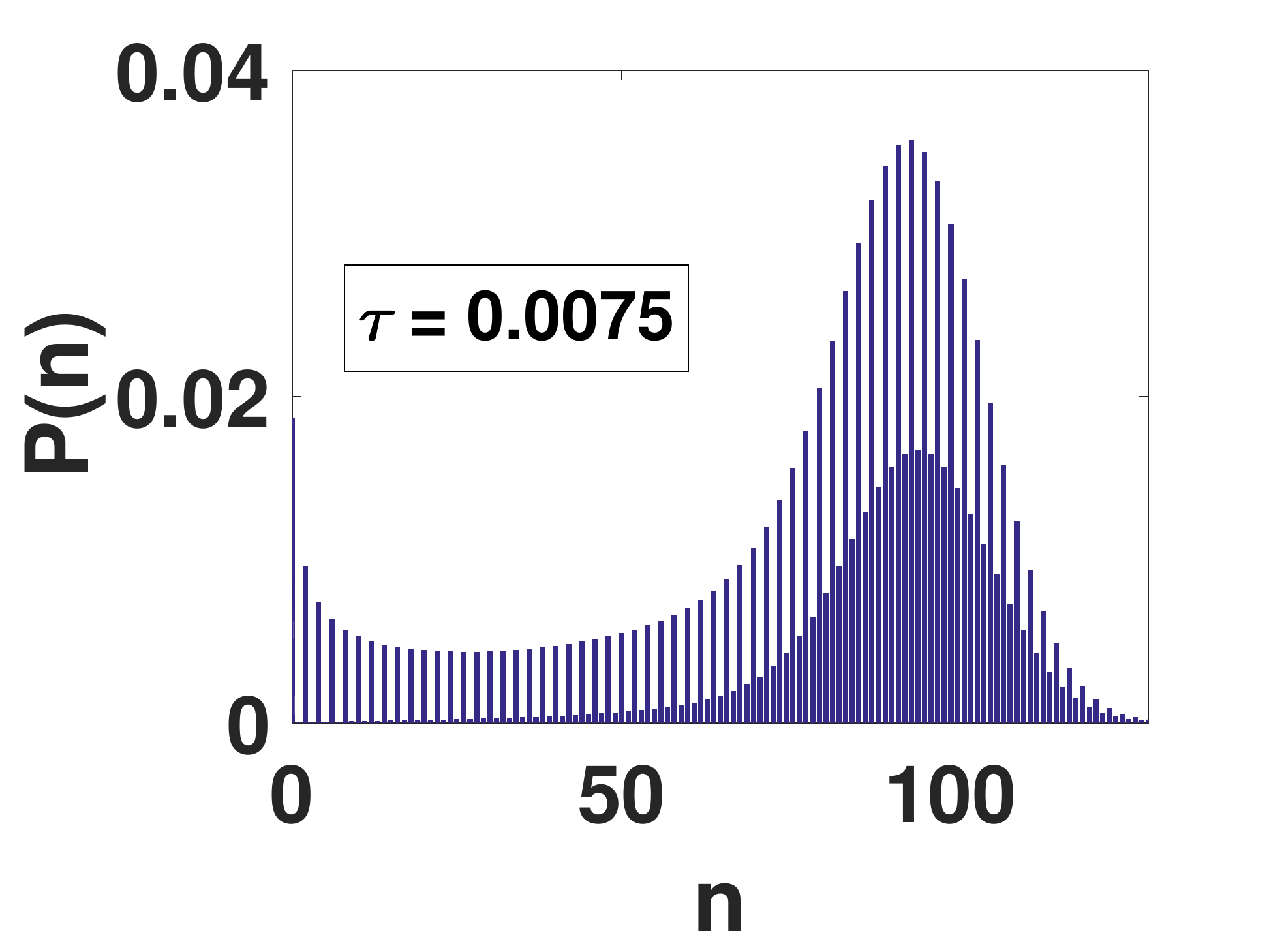}
\begin{centering}
\includegraphics[width=0.51\columnwidth]{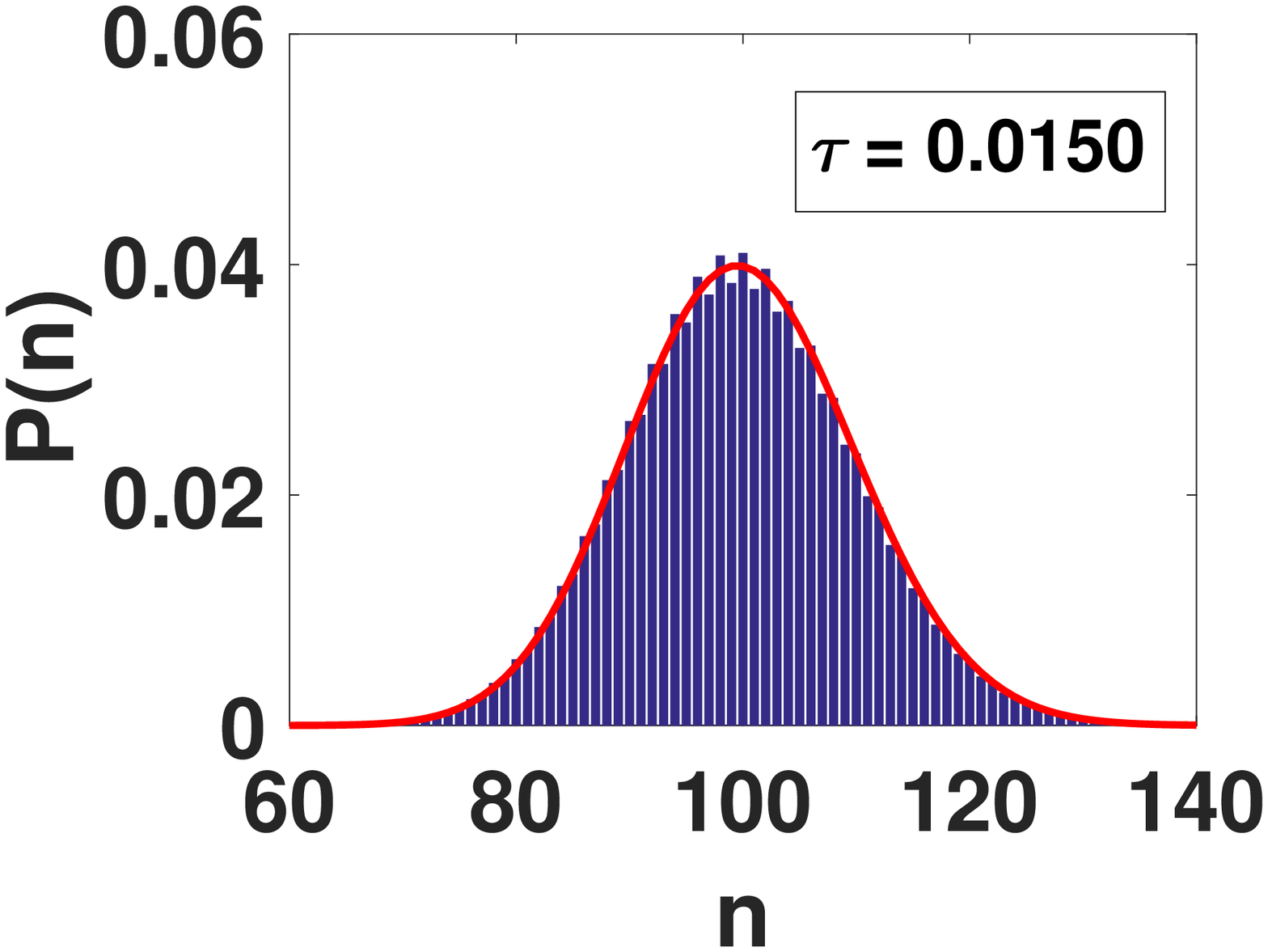}
\par\end{centering}
\caption{The photon number probability distribution at different times. Here
$g=2.5$ and $\alpha_{0}=10$. A state that only allows even photon
numbers eventually settles into a state that has a Poissonian distribution.
This can be seen in the plot at time $\tau=0.015$, where a Poissonian
distribution with a mean photon number of $100$ is fitted in red.
\label{fig:P_n_times}\textcolor{red}{}}
\end{figure}

Figure \ref{fig:P_n_times} shows the photon-number probability distribution
at different times, evolving from the vacuum state. The system evolves
from a vacuum state into an even cat-state (\ref{eq:even_odd_cat}).
The single-photon loss, however, will cause decoherence and the state
evolves into a classical mixture of even and odd cat-states. The time-step
errors for the results in Fig. \ref{fig:P_n_times} are negligible.
The photon number probability distribution at dimensionless time $\tau=0.0150$
centered around $n=100$, which agrees well with the steady state
prediction $\left|\alpha\right|^{2}=\lambda/g^{2}=100$. This distribution
resembles a Poissonian distribution, as expected for a coherent state.

\begin{figure}[H]
\begin{centering}
\includegraphics[width=0.7\columnwidth]{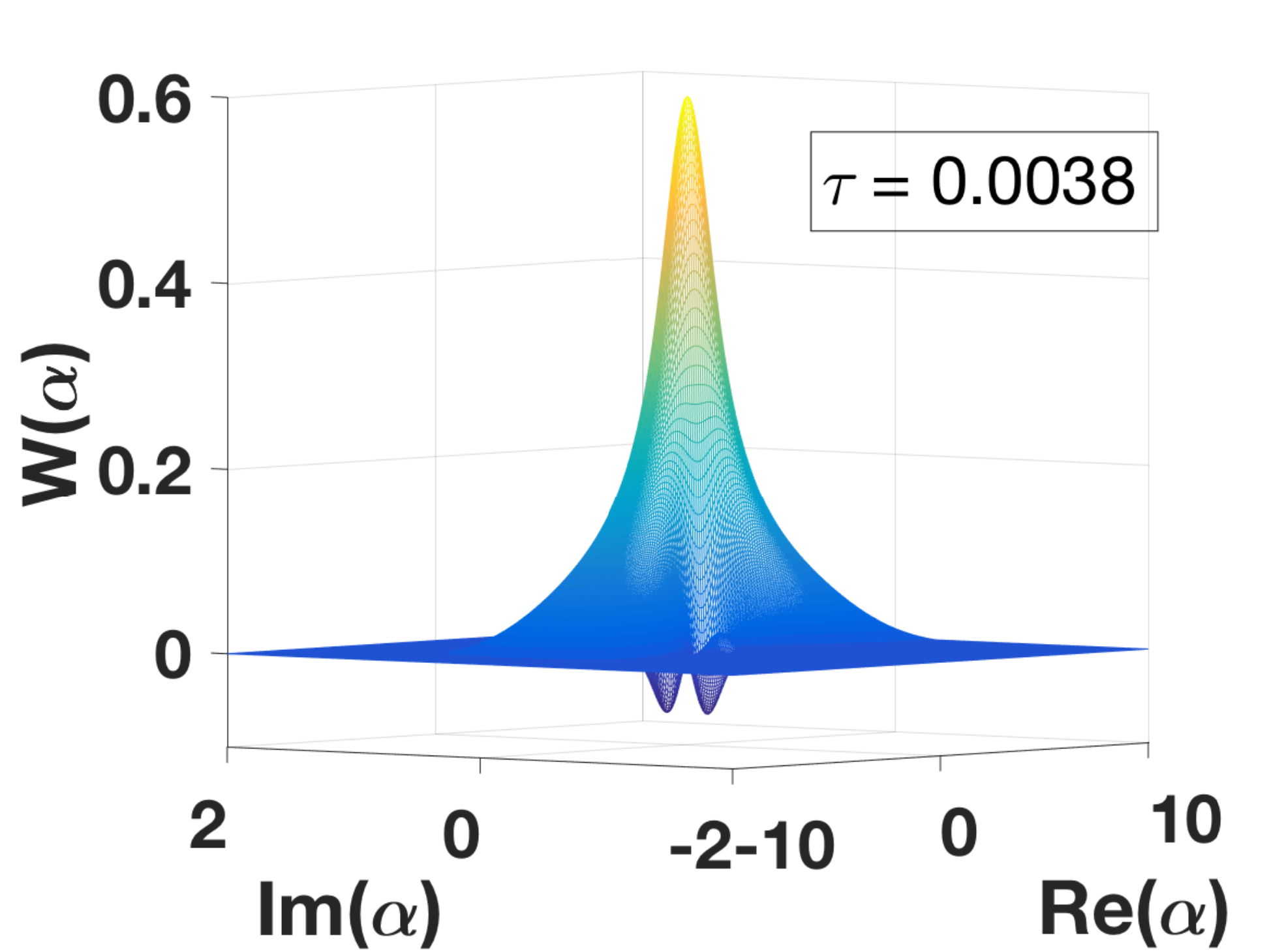}
\par\end{centering}
\begin{centering}
\bigskip{}
\par\end{centering}
\begin{centering}
\includegraphics[width=0.7\columnwidth]{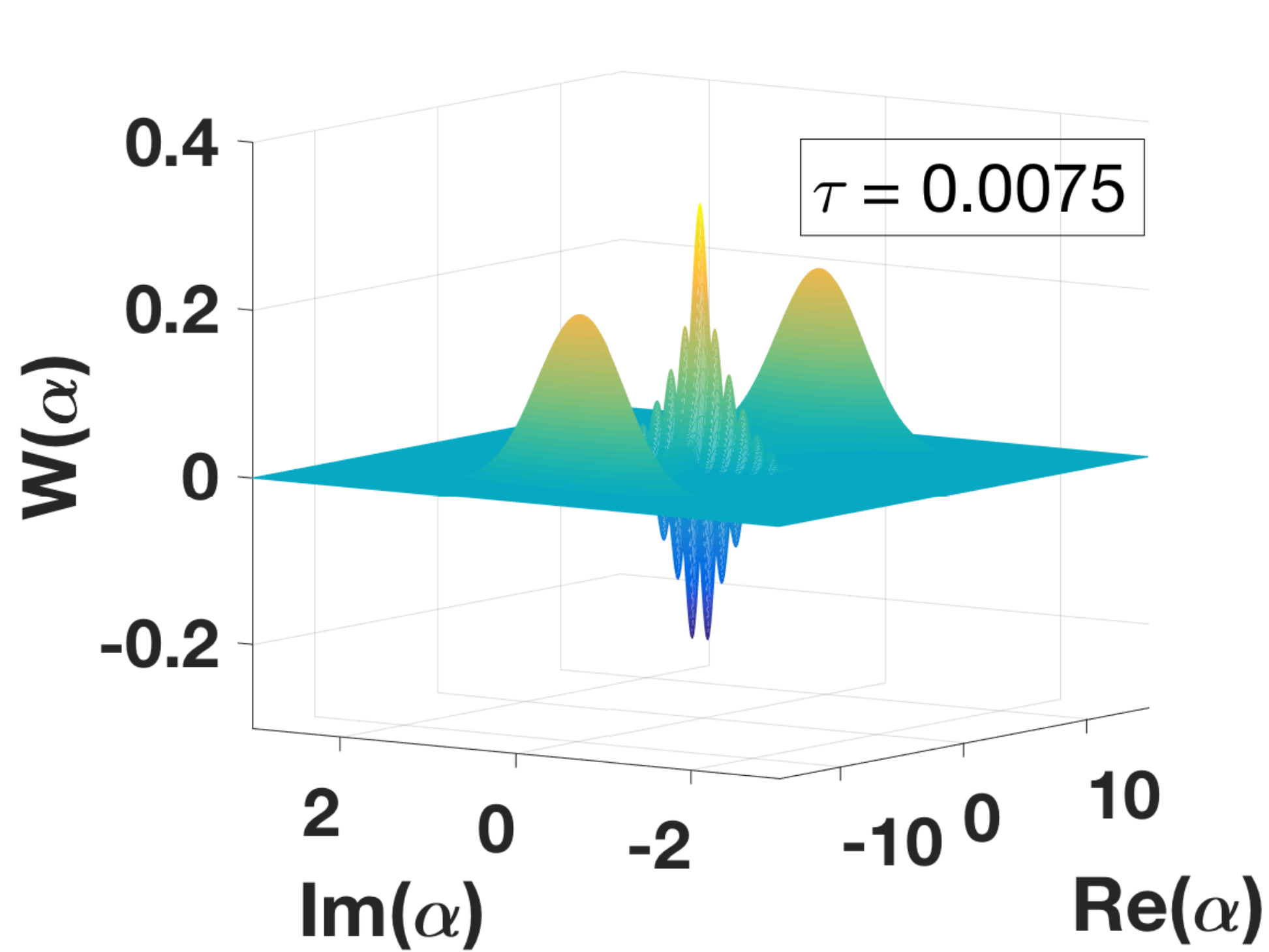}
\par\end{centering}
\caption{The Wigner function at different times. The parameters are $g=2.5$
and $\alpha_{0}=10$.\textcolor{red}{{} }\label{fig:wigner_times}}
\end{figure}

The Wigner functions at different times are computed according to
Eq. (\ref{eq:Wigner_new_expression}) and the results are presented
in Fig. \ref{fig:wigner_times}. The function around the origin admits
negative values which demonstrates the nonclassical nature of the
cat-state.
 
\section{Detuning\label{sec:Detuning}}

We now briefly consider the effect of a detuning ($\omega_{1}-\omega_{p}/2$)
between the signal mode and the external field frequency. Under conditions
of detuning, the system can display bistability in the intensity of
the signal mode as a function of the external driving intensity, which
is manifested as a hysteresis cycle \citep{Lugiato_1988detune,Fabre_1990detune}.
The system can also display self pulsing where the outputs give oscillations
in their intensities \citep{Lugiato_1988detune,Fabre_1990detune}.
These behaviors can, in turn, affect other quantum properties such
as the squeezing amplitudes. A full semiclassical analysis is given
in Sun et al. \citep{sun2019discrete}.

Here, we investigate the effect of detuning on the transient cat-state.
In this work, we consider only the detuning $\Delta=(\omega_{1}-\omega_{p}/2)/\gamma_{1}$
of the signal mode, and only the regime where $\Delta\leq\lambda$
in which case the steady-state semiclassical solution has two stable
values \citep{sun2019discrete}. We ignore thermal noise and select
$\chi=0$.

\begin{figure}[H]
\includegraphics[width=0.51\columnwidth]{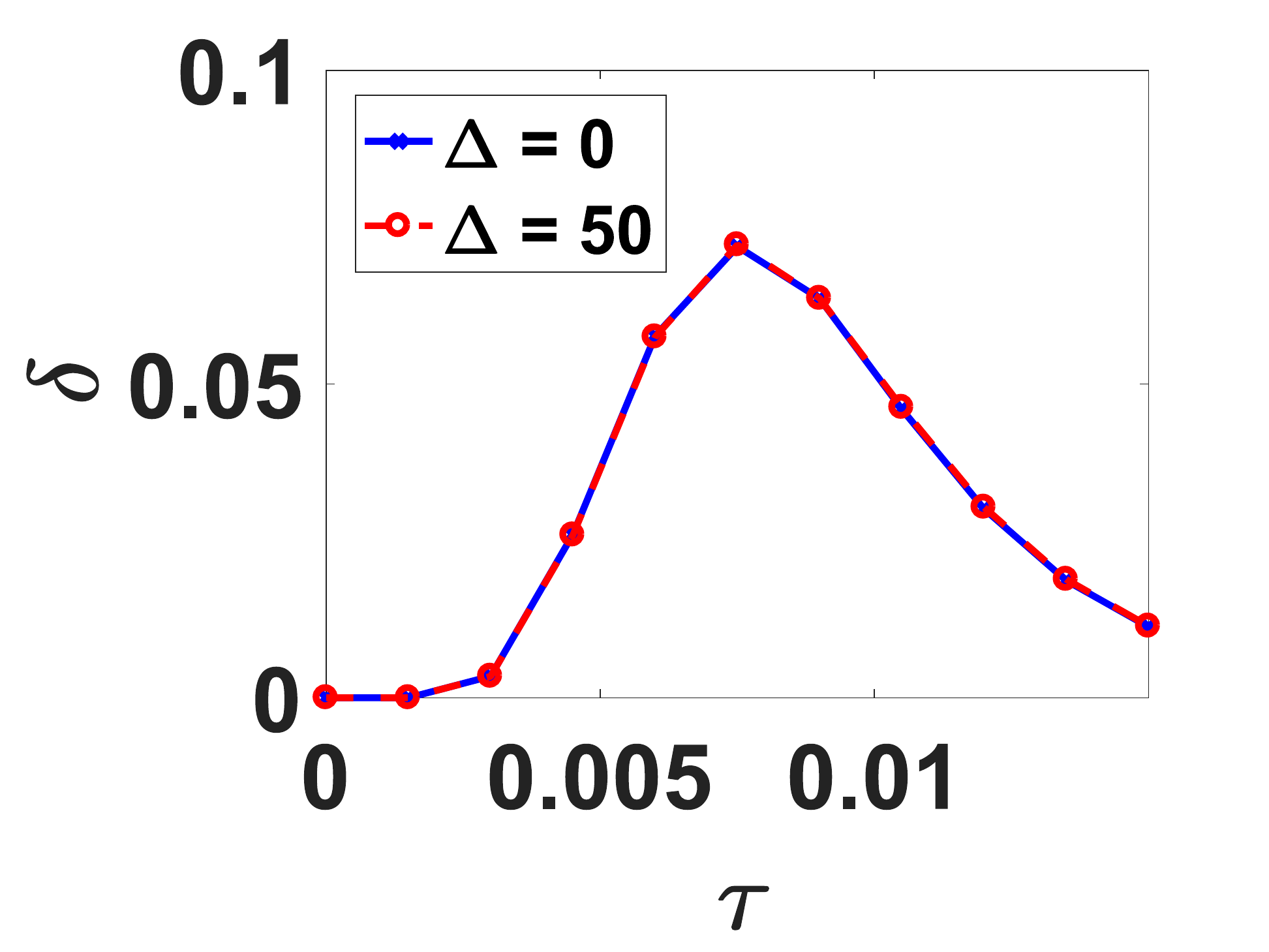}\includegraphics[width=0.51\columnwidth]{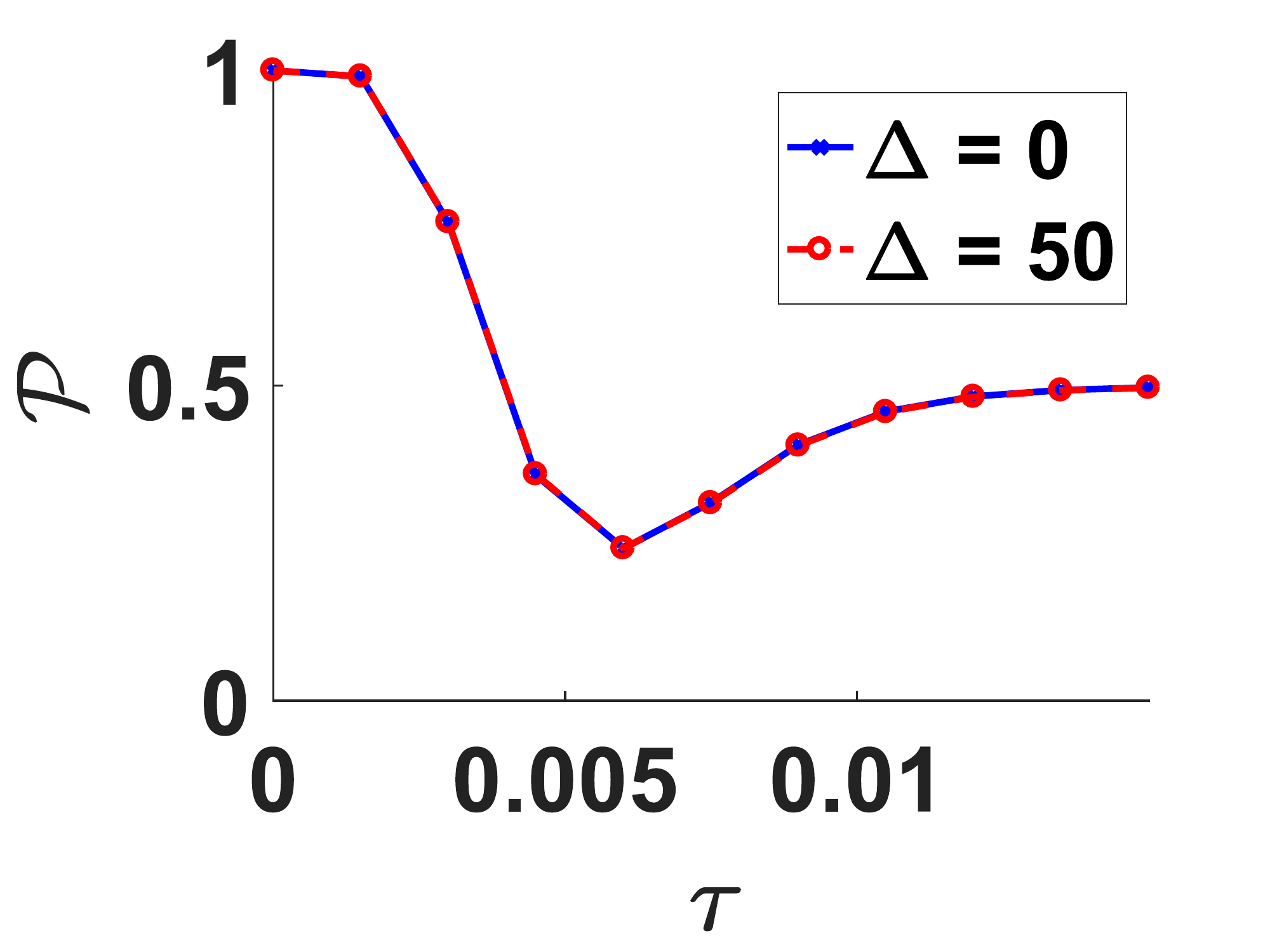}

\caption{The evolution of (left) the Wigner negativity and (right) the purity
for $\Delta=0$ and $\Delta=50$. The parameters are $g=2.5$ and
$\lambda/g^{2}=100$. Here $\chi=0$. The results show no difference
between the two cases with different detunings. \label{fig:delta_purity_detune}}
\end{figure}

\begin{figure}[H]
\begin{centering}
\includegraphics[width=0.6\columnwidth]{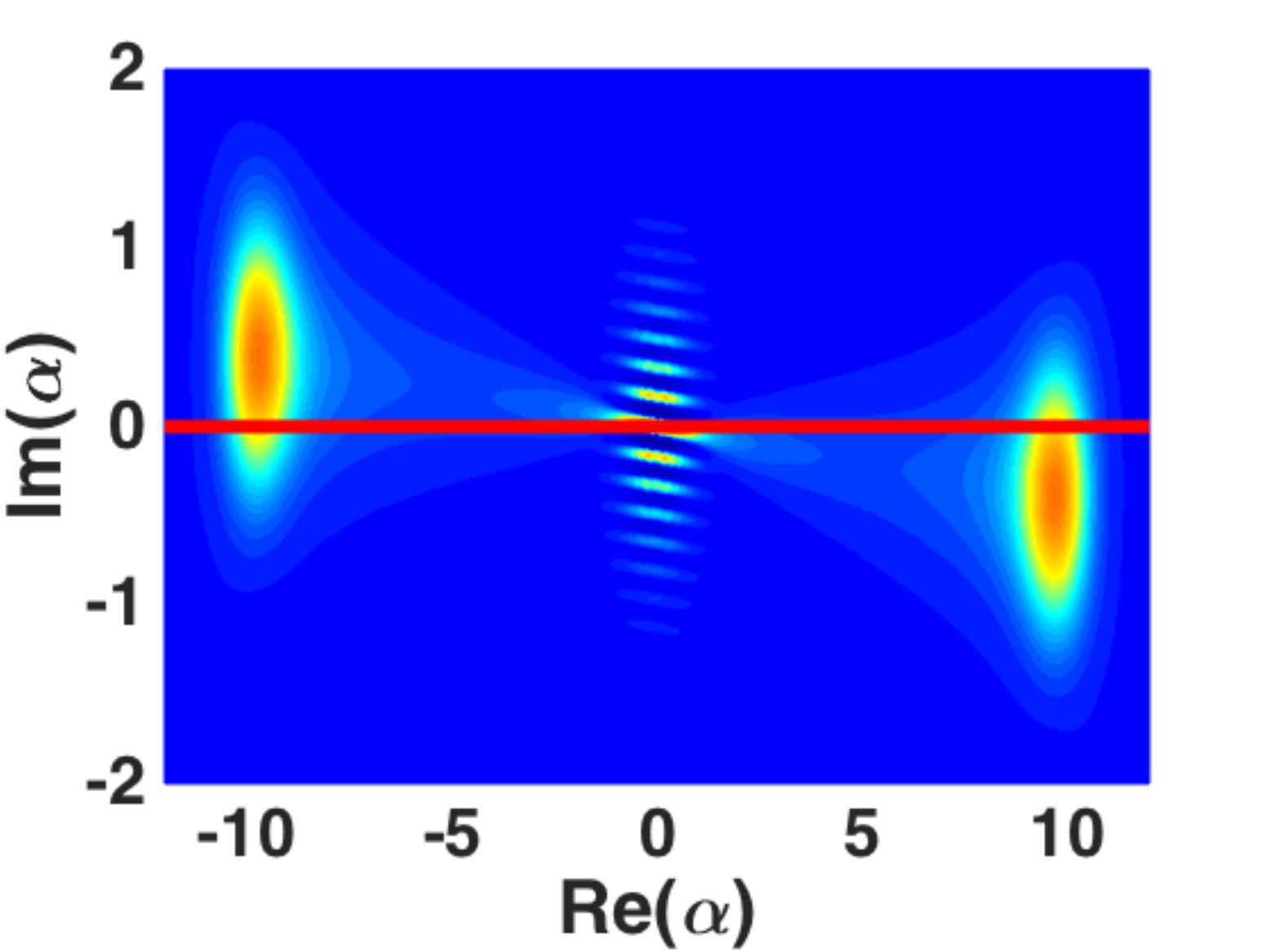}
\par\end{centering}
\caption{The Wigner function at dimensionless time $\tau=0.0075$. The parameters
are $g=2.5$ and $\alpha_{0}=10$. Here $\chi=0$. The detuning is
$\Delta=50$. In the presence of detuning $\Delta$, the physical
state is rotated in phase space. \label{Wigner_detune}}
\end{figure}

\begin{figure}[H]
\centering{}\includegraphics[width=0.55\columnwidth]{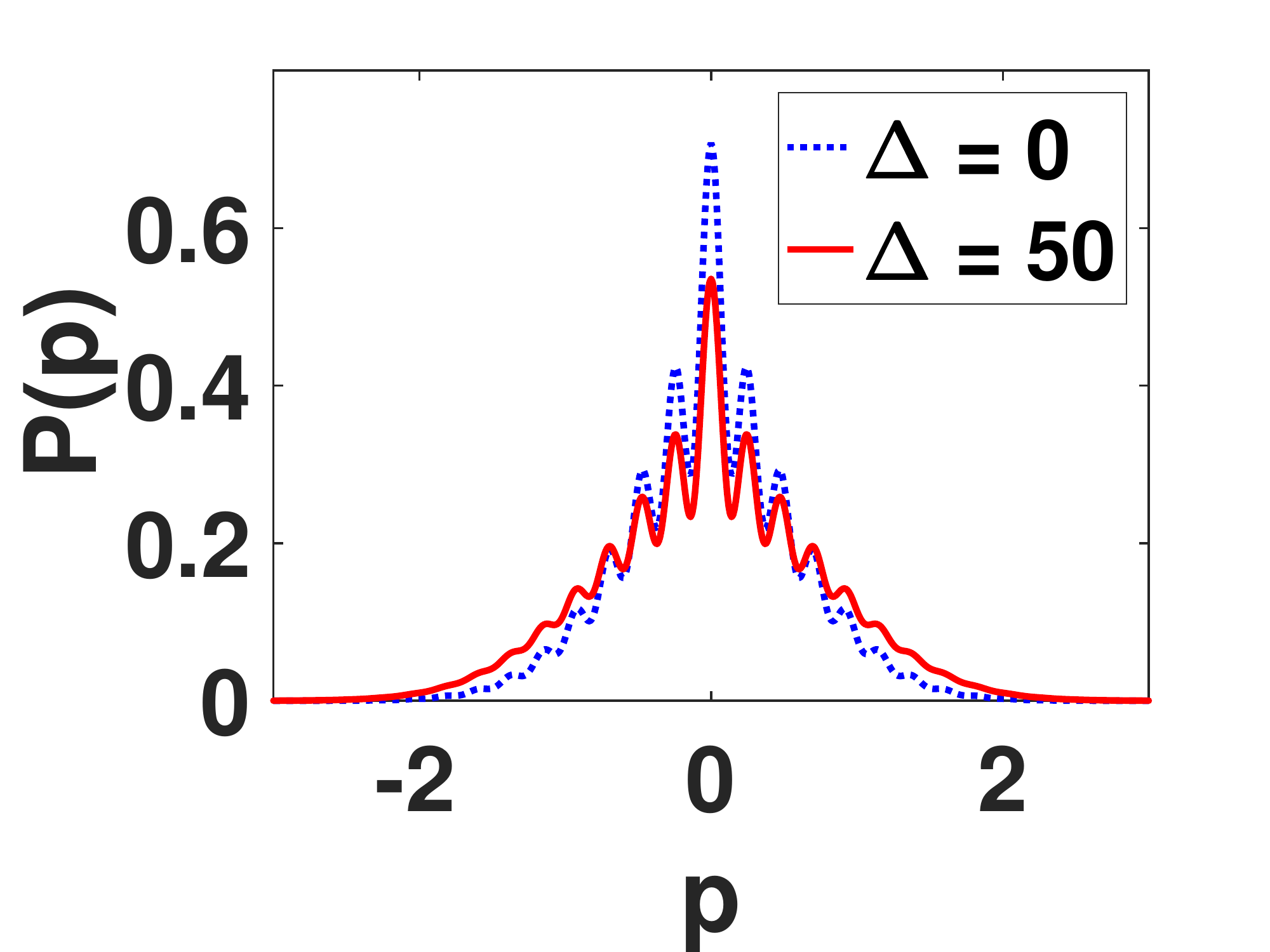}\caption{The $p$-quadrature probability distribution at dimensionless time
$\tau=0.0067$. The parameters are $g=2.5$ and $\alpha_{0}=10$.
Here $\chi=0$. The blue dashed line corresponds to zero detuning
and the orange solid line corresponds to $\Delta=50$. Since the detuning
rotates the physical state in phase space, the $p$-quadrature is
not at an optimal angle to observe the interference fringes. This
leads to a reduction in fringe visibility. The Wigner negativity and
purity are unaltered from the zero-detuning case. \label{fig:P_p_detune}}
\end{figure}

The Wigner negativity and purity calculations given in Fig. \ref{fig:delta_purity_detune}
reveal no observable differences in the physical states in the cases
with and without detuning. To this end, we plot a Wigner function
at an instant in time in Fig. \ref{Wigner_detune}. This shows that
the two mean values of the Gaussian peaks are no longer situated along
the real axis, but are rotated and have acquired complex values. The
effect of detuning is to rotate the physical state in phase space,
as consistent with the steady state analysis given by Sun et al. \citep{sun2019discrete}.
This explains the apparent reduction in the visibility of the interference
fringes as shown in Fig. \ref{fig:P_p_detune}; the $p$-quadrature
is not at an optimal angle to observe the interference fringes.

\section{Degenerate parametric oscillation with the anharmonic Kerr interaction\label{sec:Degenerate-parametric-oscillatio}}

A proposal to generate cat-states with a Kerr interaction is put forward
by Yurke and Stoler \citep{PhysRevLett.57.13,Yurke_physicalB1988}.
They showed that a coherent state can evolve into a multi-component
cat-state. Depending on the interaction time, a two-component cat-state
can also be created. The mechanism of cat-state creation in a Kerr
interaction originates from the fact that the phase acquired by the
state is photon-number dependent. This means that this method of creating
a cat-state is hard to achieve in the presence of single-photon losses.
However, the Yurke and Stoler proposal has been realized in a superconducting
circuit experiment \citep{Kirchmair_Nature2013Kerr}, where the Kerr
nonlinearity is larger than $30$ times the single-photon decay rate.
Drummond and Walls \citep{Drummond_JournalofPhysics1980} have provided
an exact steady-state solution to a driven, dissipative system with
a Kerr interaction at zero temperature, which gives quantum predictions
that are different from those of a semiclassical analysis.

The combined Kerr and parametric case was studied recently \citep{sun2019discrete,sunschrodinger}.
These authors gave a derivation of the adiabatic master equation and
both semiclassical and exact steady-state solutions.  The semiclassical
solutions have bistable regimes. There are also tristable regimes,
with detunings included. Here, we assume there are no detunings and
ignore thermal noise. In this case, the main effect of the additional
Kerr nonlinearities is to change the nature of the Schr\"odinger
cat solutions.

Below, we will give more detail by solving the master equation using
a particular choice of scaled variables. The master equation including
the Kerr nonlinearity is given by
\begin{align}
\frac{\partial}{\partial t}\rho & =\frac{\Lambda}{2}\left[a^{\dagger2}-a^{2},\rho\right]+\gamma_{1}\left(2a\rho a^{\dagger}-a^{\dagger}a\rho-\rho a^{\dagger}a\right)\nonumber \\
 & -i\frac{\bar{\chi}}{2}\left[a^{\dagger2}a^{2},\rho\right]+\frac{1}{2}G^{2}\left(2a^{2}\rho a^{\dagger2}-a^{\dagger2}a^{2}\rho-\rho a^{\dagger2}a^{2}\right)\,\nonumber \\
\end{align}
where $G=\sqrt{\bar{g}^{2}/\left(2\gamma_{2}\right)}$ and $\Lambda=\left|\bar{g}\epsilon\right|/\gamma_{2}$
as defined previously. We consider $\sqrt{G^{4}+\bar{\chi}^{2}}=G^{2}\sqrt{\left(1+\bar{\chi}^{2}/G^{4}\right)}$
which defines a dimensionless time $\mathcal{T}=\sqrt{G^{4}+\bar{\chi}^{2}}t$.
The master equation is then
\begin{align}
\frac{\partial}{\partial\mathcal{T}}\rho & =\!\frac{\Lambda}{2\sqrt{G^{4}+\bar{\chi}^{2}}}\left[a^{\dagger2}\!-\!a^{2},\rho\right]\!-\!\frac{i\bar{\chi}}{2G^{2}\sqrt{1+\frac{\bar{\chi}^{2}}{G^{4}}}}\left[a^{\dagger2}a^{2},\rho\right]\nonumber \\
 & +\frac{\gamma_{1}}{G^{2}\sqrt{1+\frac{\bar{\chi}^{2}}{G^{4}}}}\left(2a\rho a^{\dagger}-a^{\dagger}a\rho-\rho a^{\dagger}a\right)\nonumber \\
 & +\frac{1}{2}\frac{1}{\sqrt{1+\frac{\bar{\chi}^{2}}{G^{4}}}}\left(2a^{2}\rho a^{\dagger2}-a^{\dagger2}a^{2}\rho-\rho a^{\dagger2}a^{2}\right)\,.
\end{align}

The steady state in the presence of Kerr nonlinearity has a coherent
amplitude $\alpha_{0}$ given by (\ref{eq:defn-alpha-kerr}), with
an absolute value $\left|\alpha_{0}\right|=\sqrt{\lambda/\sqrt{g^{4}+\chi'^{2}}}\equiv\sqrt{\Lambda/G^{2}\sqrt{\left(1+\chi^{2}\right)}}$
where $\chi\equiv\bar{\chi}/G^{2}=\chi'/g^{2}$. With this choice
of scaling factor, the master equation above can be expressed in terms
of $\alpha_{0}$, $g$, and $\chi$ as follows:
\begin{align}
\frac{\partial}{\partial\mathcal{T}}\rho & =\frac{1}{2}\left|\alpha_{0}\right|^{2}\left[a^{\dagger2}-a^{2},\rho\right]-\frac{i}{2}\frac{\chi}{\sqrt{1+\chi^{2}}}\left[a^{\dagger2}a^{2},\rho\right]\nonumber \\
 & +\frac{1}{g^{2}\sqrt{1+\chi^{2}}}\left(2a\rho a^{\dagger}-a^{\dagger}a\rho-\rho a^{\dagger}a\right)\nonumber \\
 & +\frac{1}{2}\frac{1}{\sqrt{1+\chi^{2}}}\left(2a^{2}\rho a^{\dagger2}-a^{\dagger2}a^{2}\rho-\rho a^{\dagger2}a^{2}\right)\,.\label{eq:master_eqn_Kerr}
\end{align}
In the lossless case ($\gamma_{1}=0$, $g\rightarrow\infty$), the
third term does not contribute.

\subsection{No single-photon signal damping \label{subsec:lossless_Kerr}}

To study the behavior, we first examine the case with no signal damping,
corresponding to $\gamma_{1}=0$ (the third term in the master equation
is zero). From (\ref{eq:master_eqn_Kerr}), we see that the free parameters
in this case are the coherent amplitude $\left|\alpha_{0}\right|,\,g$
and $\chi\equiv\bar{\chi}/G^{2}=\chi'/g^{2}$. We fix $\left|\alpha_{0}\right|$
while changing $\chi$. To keep $\alpha_{0}$ constant for large $r$,
we assume a sufficiently large driving field, $\Lambda$ or $\lambda$.
Detunings are assumed zero. 

\begin{figure}[H]
\includegraphics[width=0.51\columnwidth]{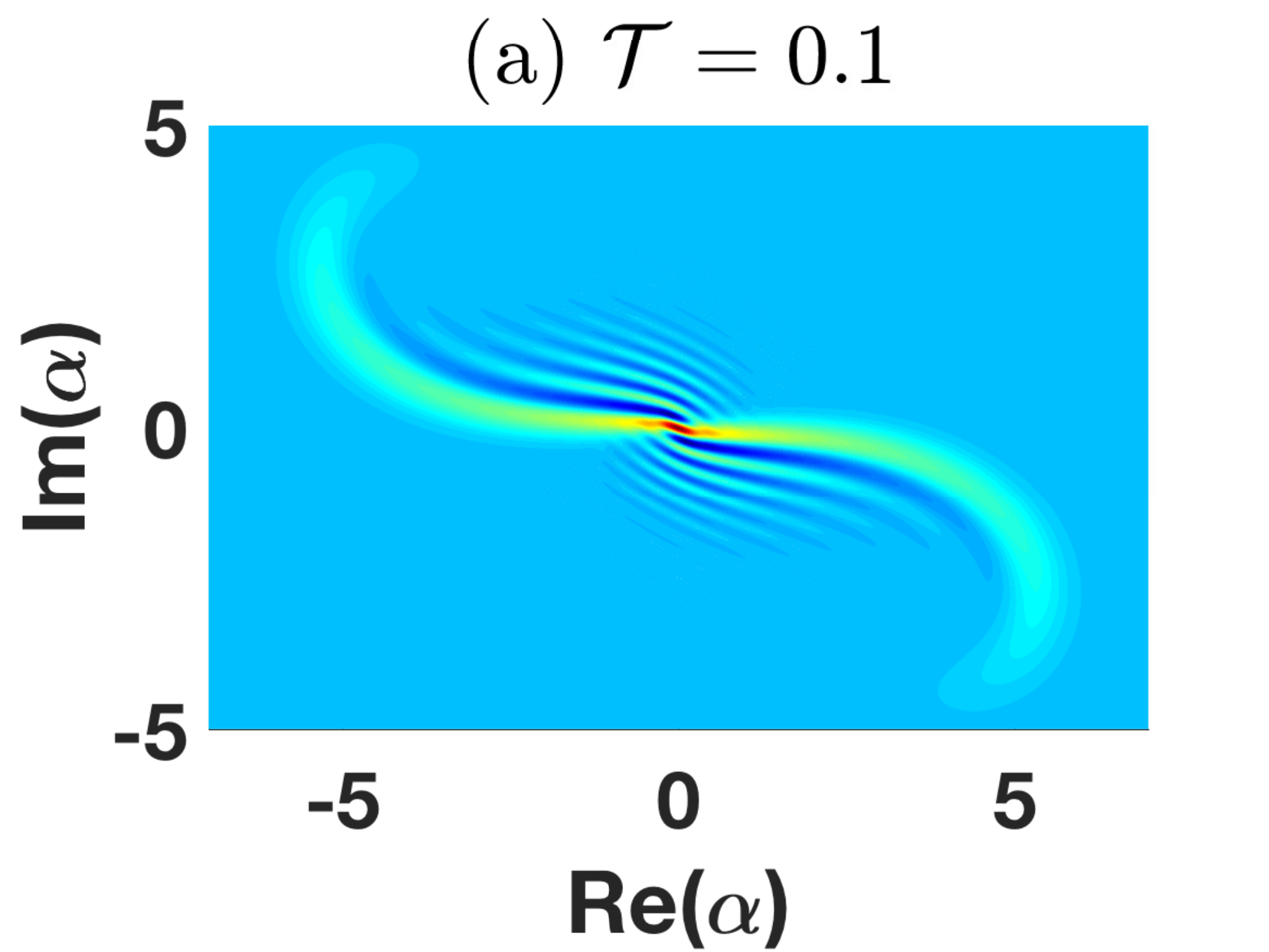}\includegraphics[width=0.51\columnwidth]{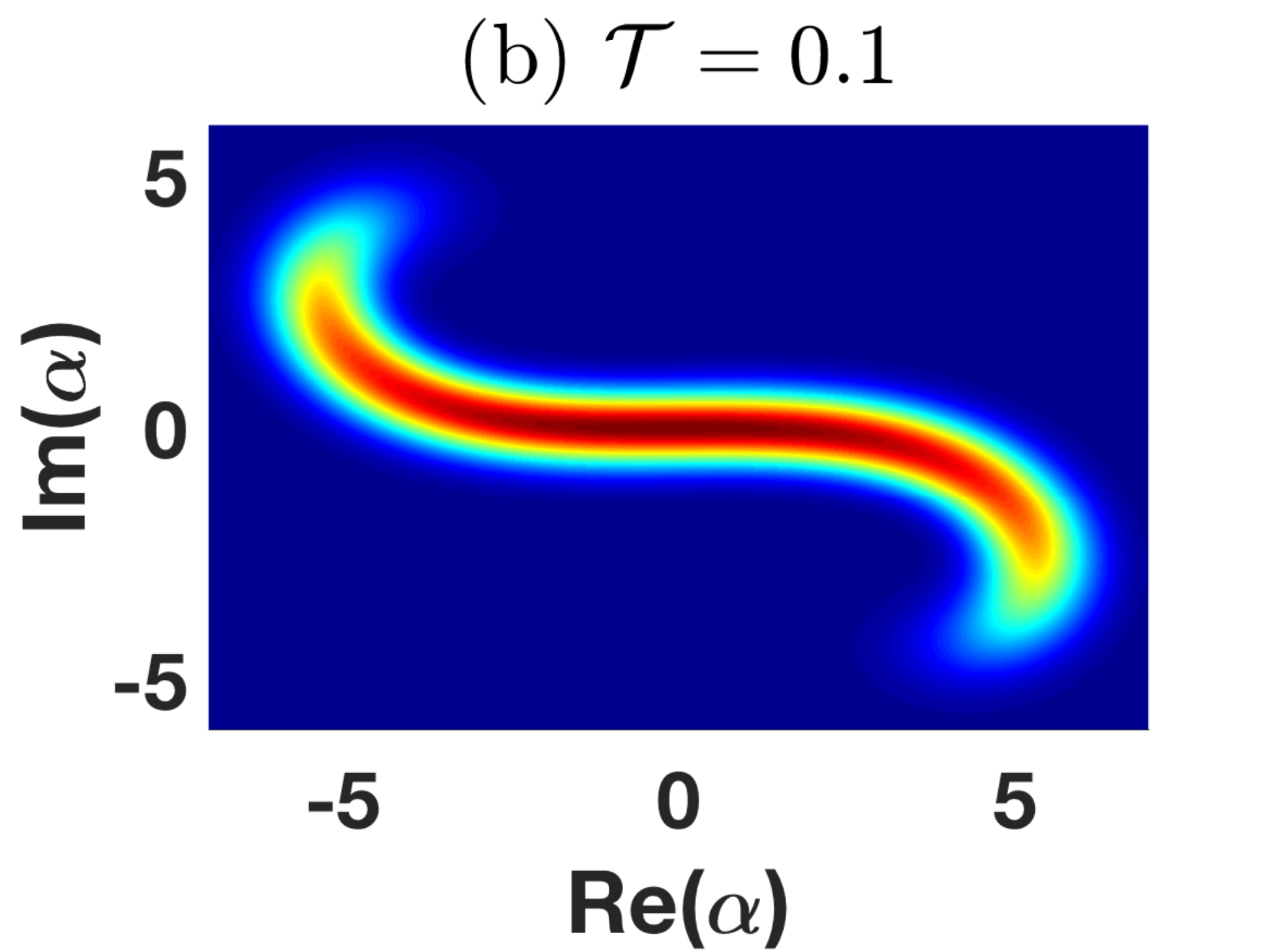}

\bigskip{}

\includegraphics[width=0.51\columnwidth]{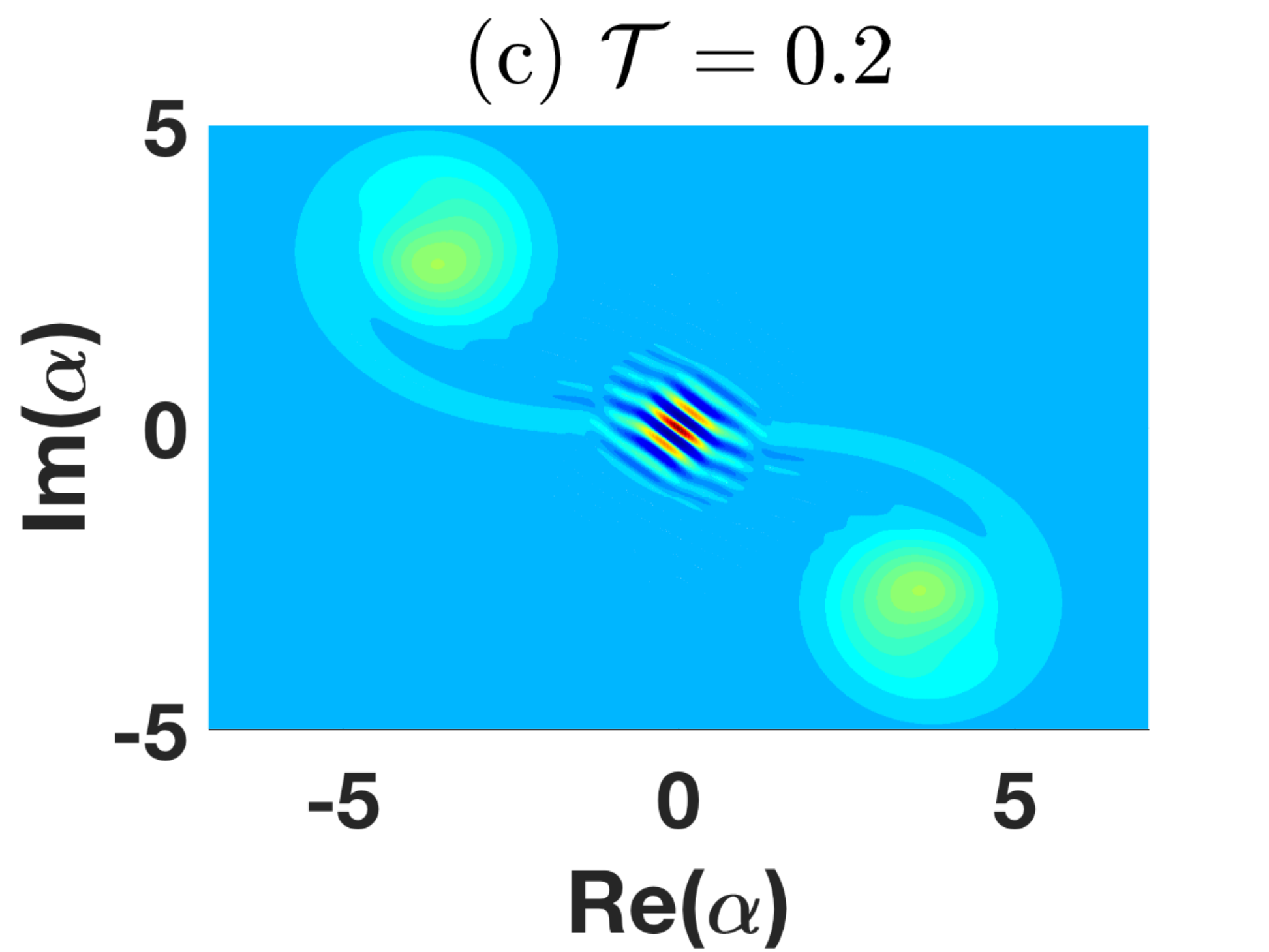}\includegraphics[width=0.51\columnwidth]{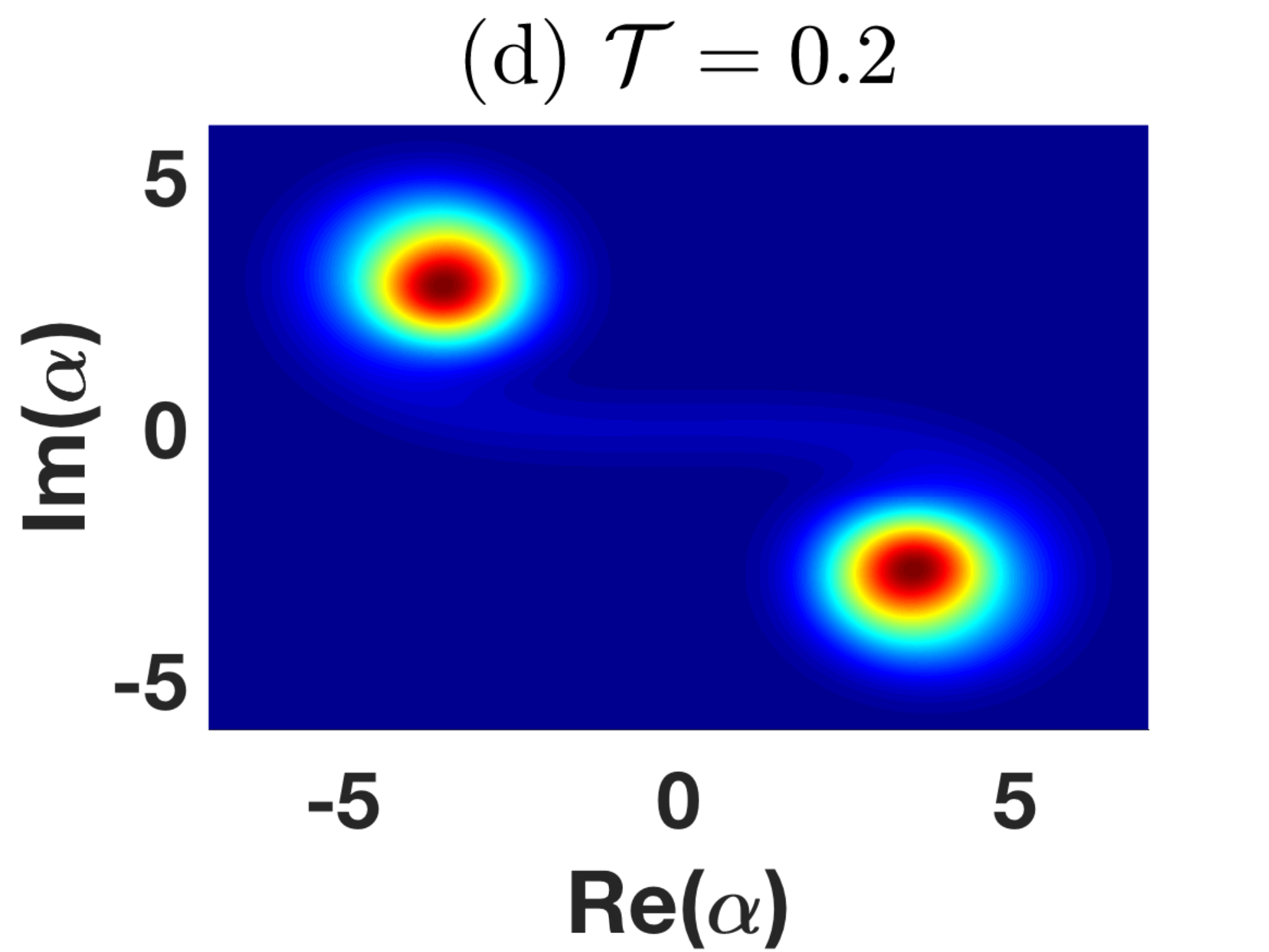}

\bigskip{}

\includegraphics[width=0.51\columnwidth]{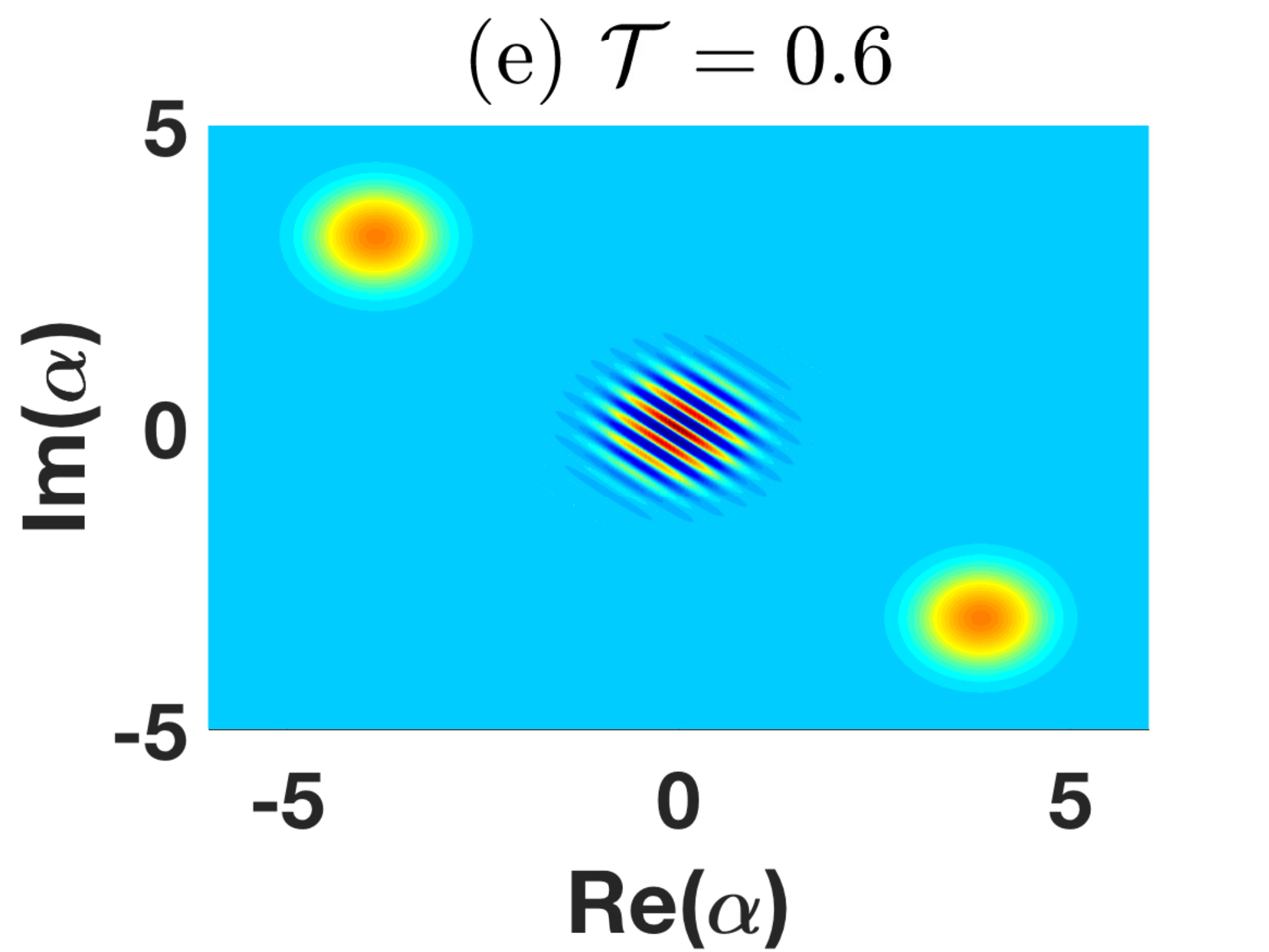}\includegraphics[width=0.51\columnwidth]{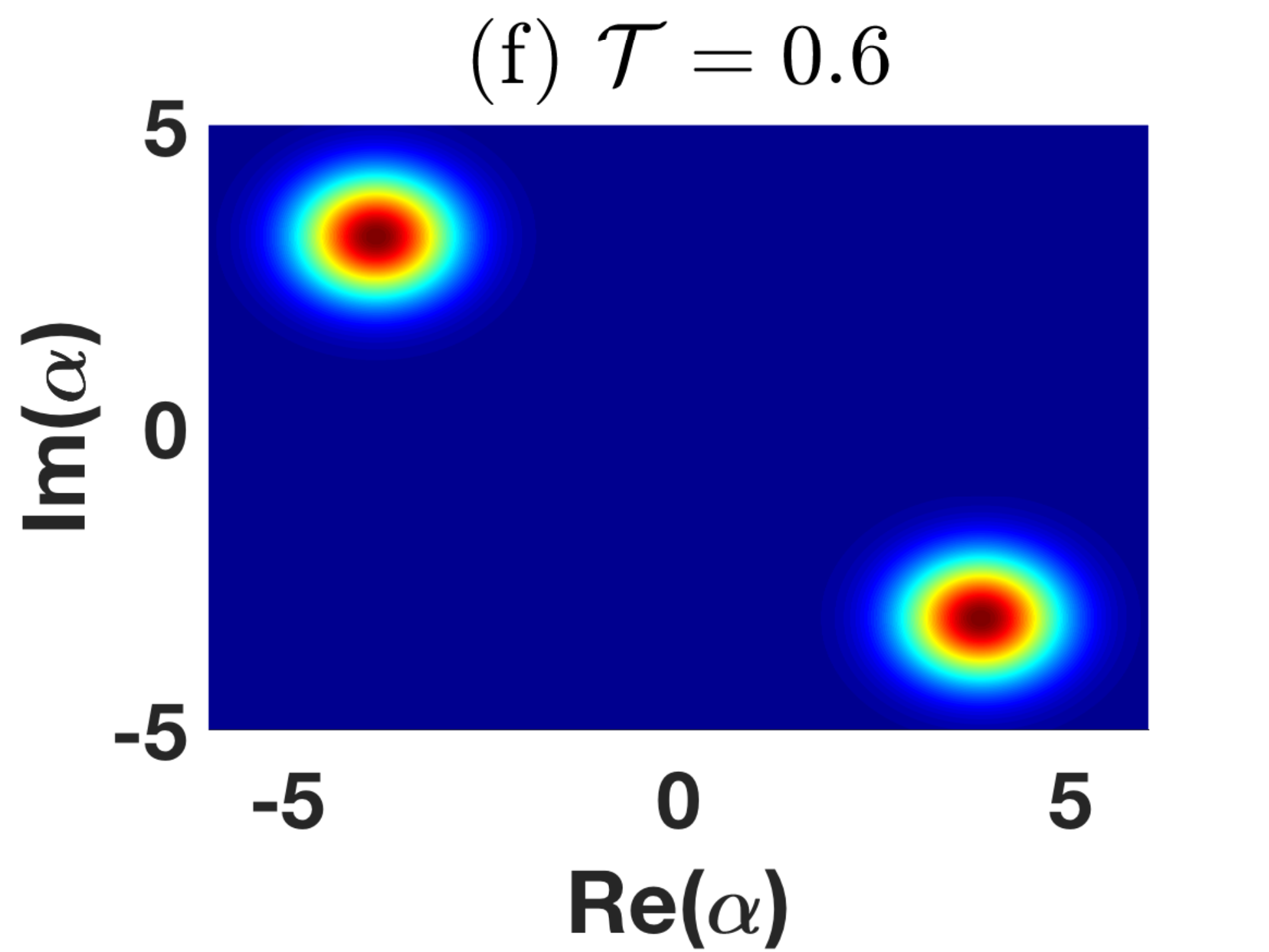}

\caption{The evolution of the Wigner function (a, c, e) and Q function (b,
d, f) with no single-photon damping. In this lossless case, the free
parameters determining the dynamics Eq. (\ref{eq:master_eqn_Kerr})
are $\chi$ and $\left|\alpha_{0}\right|$. Here, $\chi=5$ and $|\alpha_{0}|=5$.
\label{fig:Kerr_wigner_Q_a05}}
\end{figure}

In Fig. \ref{fig:Kerr_wigner_Q_a05}, we plot the evolution of the
Wigner and Q functions for $|\alpha_{0}|=5$ with $\chi=5$. These
phase space distributions show the dynamics of the system under the
presence of Kerr interaction. Starting with an initial vacuum state,
the state quickly turns into a squeezed state with a curved distribution
in the phase space distributions due to the large Kerr effect, as
shown in Fig. \ref{fig:Kerr_wigner_Q_a05} (a) and (b). Some time
later, we observe the build up of two Gaussian peaks that correspond
to the complex amplitudes with opposite phases as predicted in Eq.
(\ref{eq:defn-alpha-kerr}). Finally, the system reaches a steady
state, as shown in Fig. \ref{fig:Kerr_wigner_Q_a05} (e) and (f),
where the two Gaussian peaks are fully separated. In particular, in
the Wigner distribution of Fig. \ref{fig:Kerr_wigner_Q_a05} (e),
negative values around the origin suggests the presence of cat-states,
which is confirmed by computing the corresponding Wigner negativities
and compared with the analytical Wigner negativity value of a cat-state
as given in Eq. (\ref{eq:Wigner_cat_state}).

\begin{figure}[H]
\begin{centering}
\includegraphics[width=0.6\columnwidth]{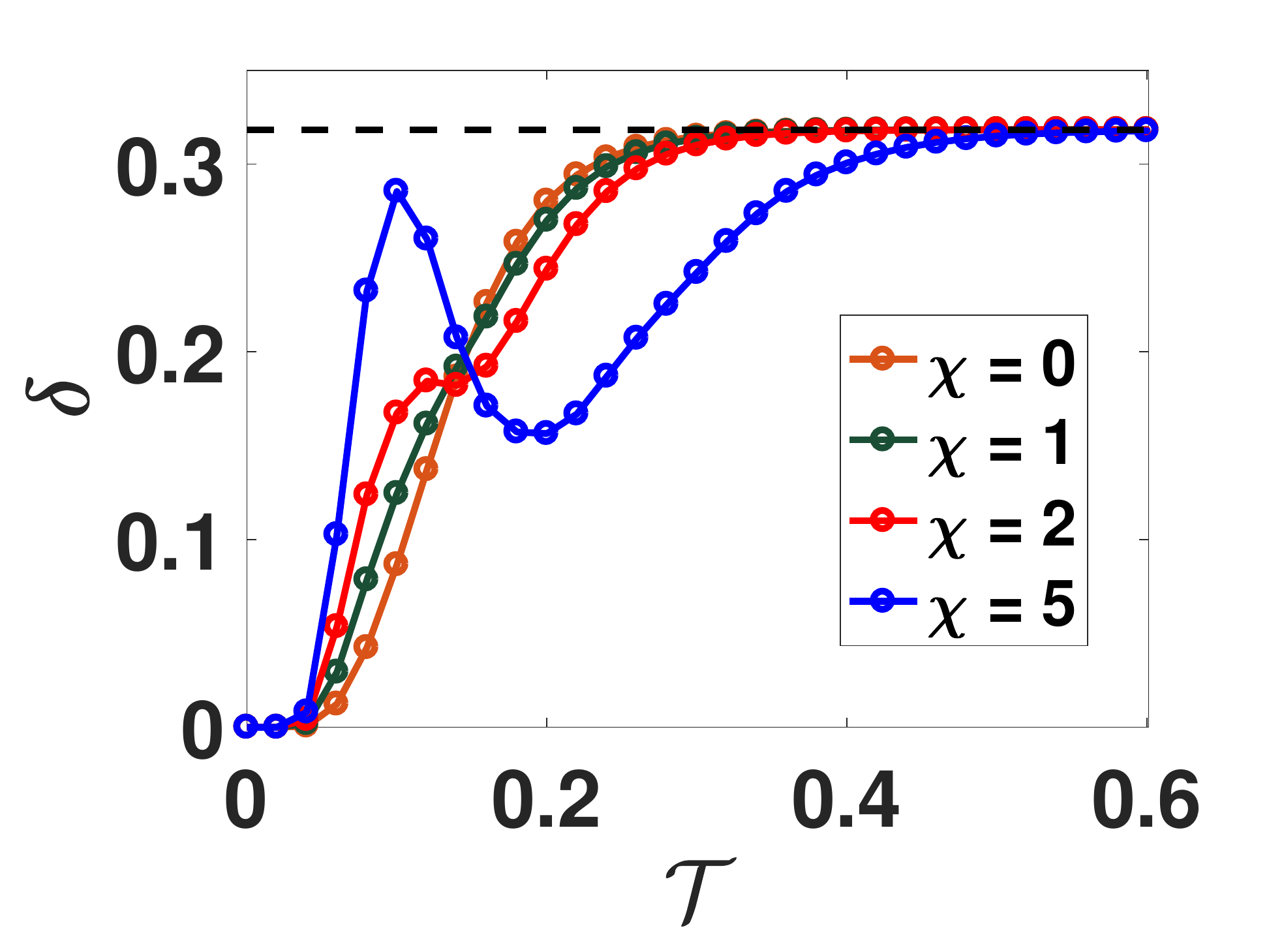}
\par\end{centering}
\caption{The evolution of the Wigner negativity with different $\chi$ ratios
for $\left|\alpha_{0}\right|=5$, in the lossless case $\gamma_{1}=0$.
By comparing the numerical Wigner negativity with the analytical
Wigner negativity (black dashed horizontal line) for a Wigner function
in Eq. (\ref{eq:Wigner_cat_state}), the cat formation time is determined
when the numerical value agrees with the analytical value to four
significant figures. \label{fig:kerr_delta_lossless_a05}}
\end{figure}

The evolution of the Wigner negativity for $|\alpha_{0}|=5$, for
different values of $\chi$ is presented in Fig. \ref{fig:kerr_delta_lossless_a05}.
For $\chi=1$, the Wigner negativity time evolution is similar to
that of the case without Kerr interaction. The Wigner negativity increases
until reaching a value corresponding to a cat-state. For larger $\chi$,
however, the dynamics is markedly different; the negativity rises
steadily initially, reaching a peak before decreasing and increasing
again until the value reaches the negativity corresponding to that
of a cat-state. 

An understanding of this dynamics for large $\chi$ can be obtained
from the corresponding Wigner function time evolution in Fig. \ref{fig:Kerr_wigner_Q_a05}.
In the earlier stage of the dynamics, the Kerr term dominates the
parametric gain term for large $\chi$. The large contribution from
the Kerr effect produces a nonclassical state; the larger the Kerr
strength, the larger the peak Wigner negativity. As the two Gaussian
peaks with the same amplitude but opposite phases are building, the
Wigner negativity value decreases, before increasing again due to
the formation of a cat-state as the system approaches the steady state.
We note that a cat-state corresponds to the case where the Wigner
function has two fully separated Gaussian peaks with the presence
of interference fringes around the origin.

\begin{figure}[H]
\includegraphics[width=0.51\columnwidth]{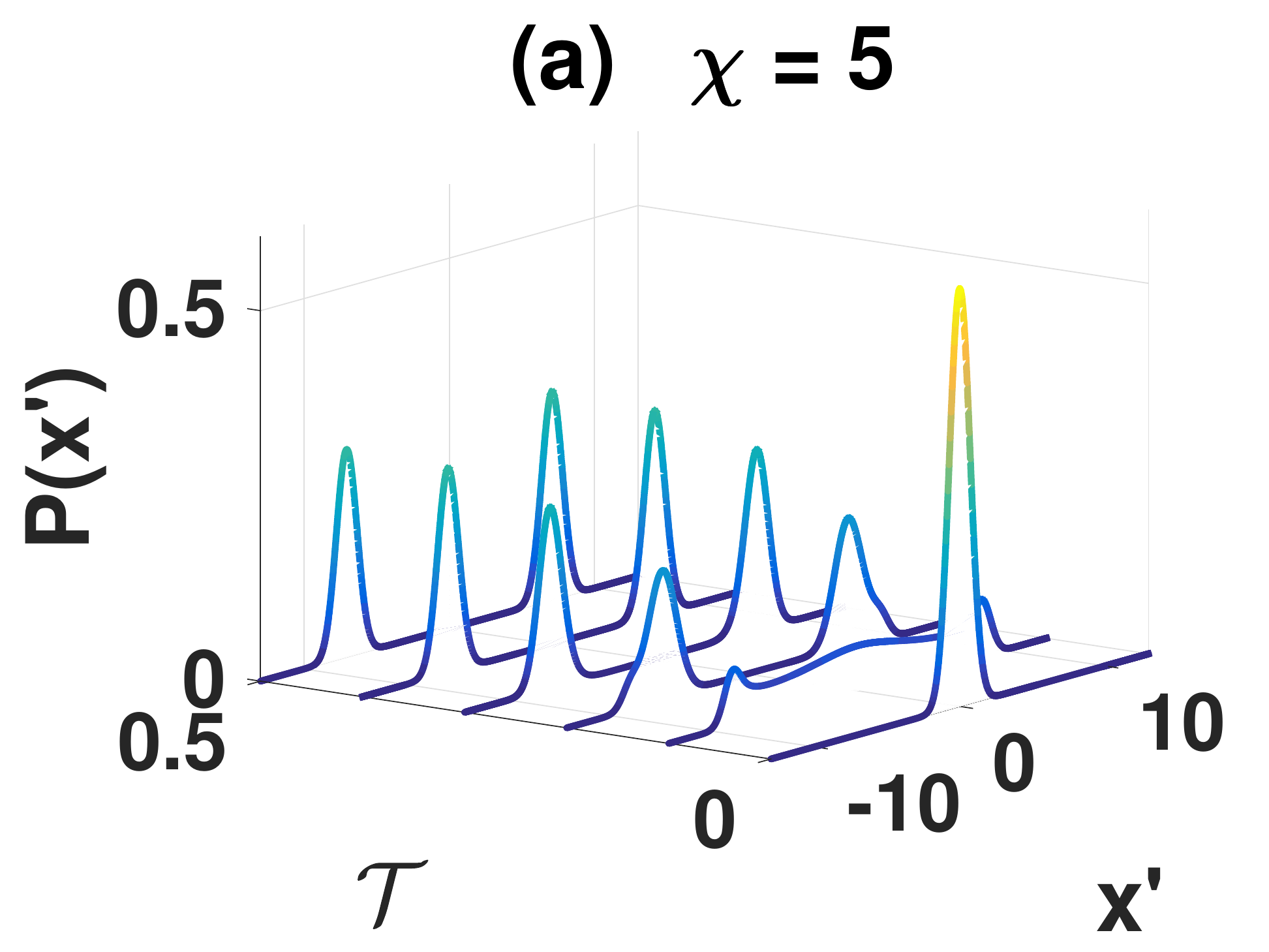}\includegraphics[width=0.51\columnwidth]{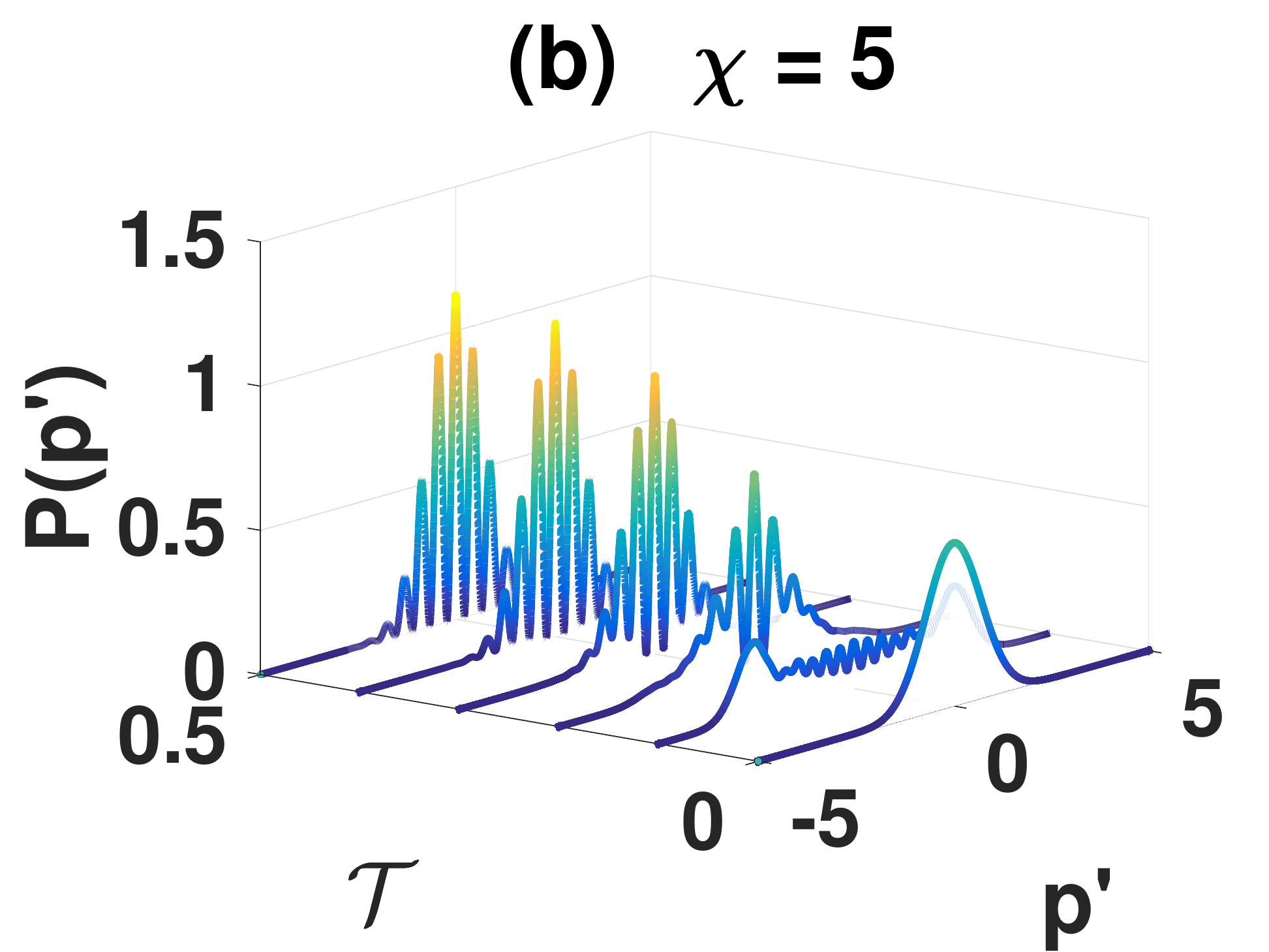}

\bigskip{}

\includegraphics[width=0.51\columnwidth]{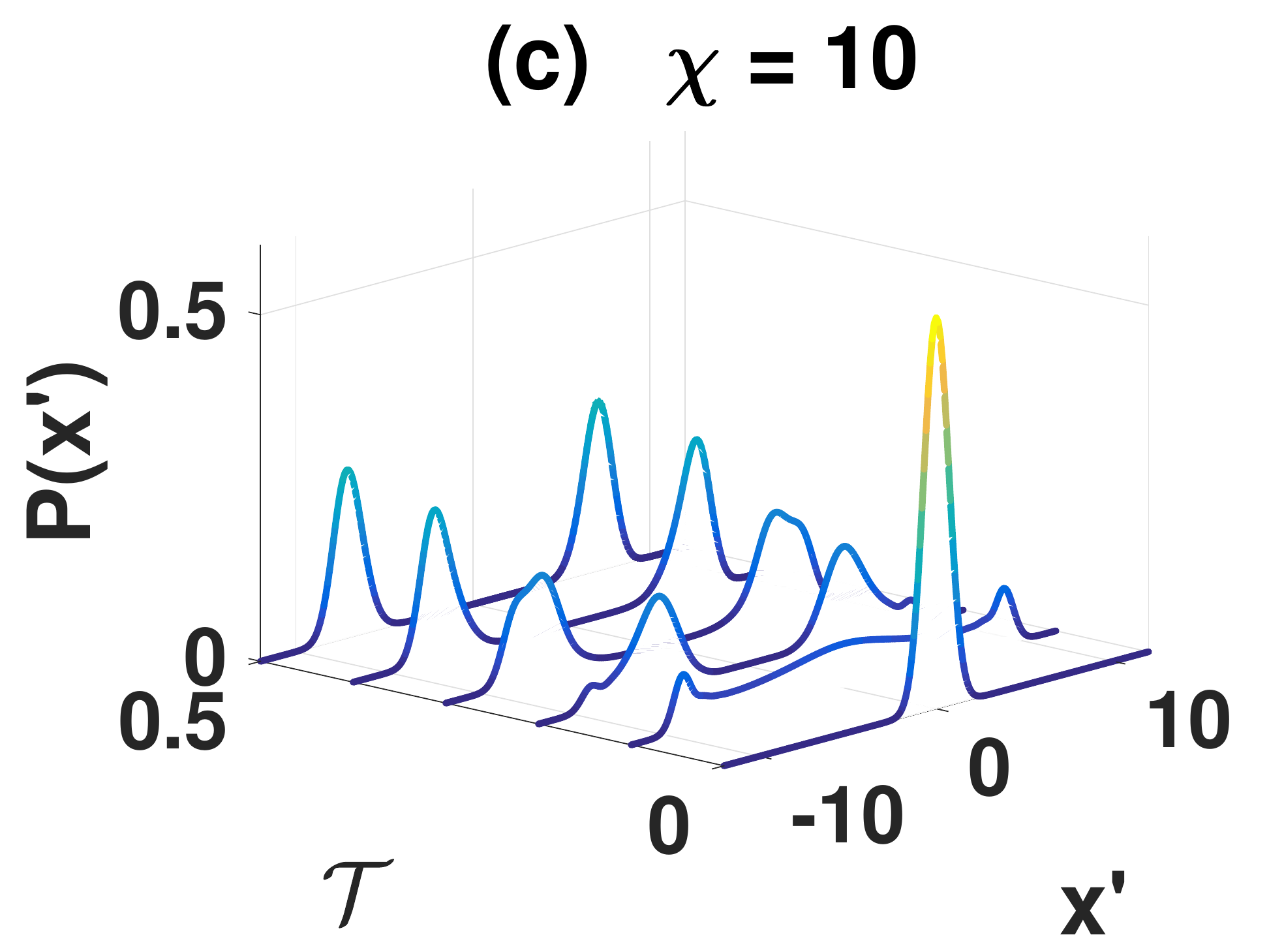}\includegraphics[width=0.51\columnwidth]{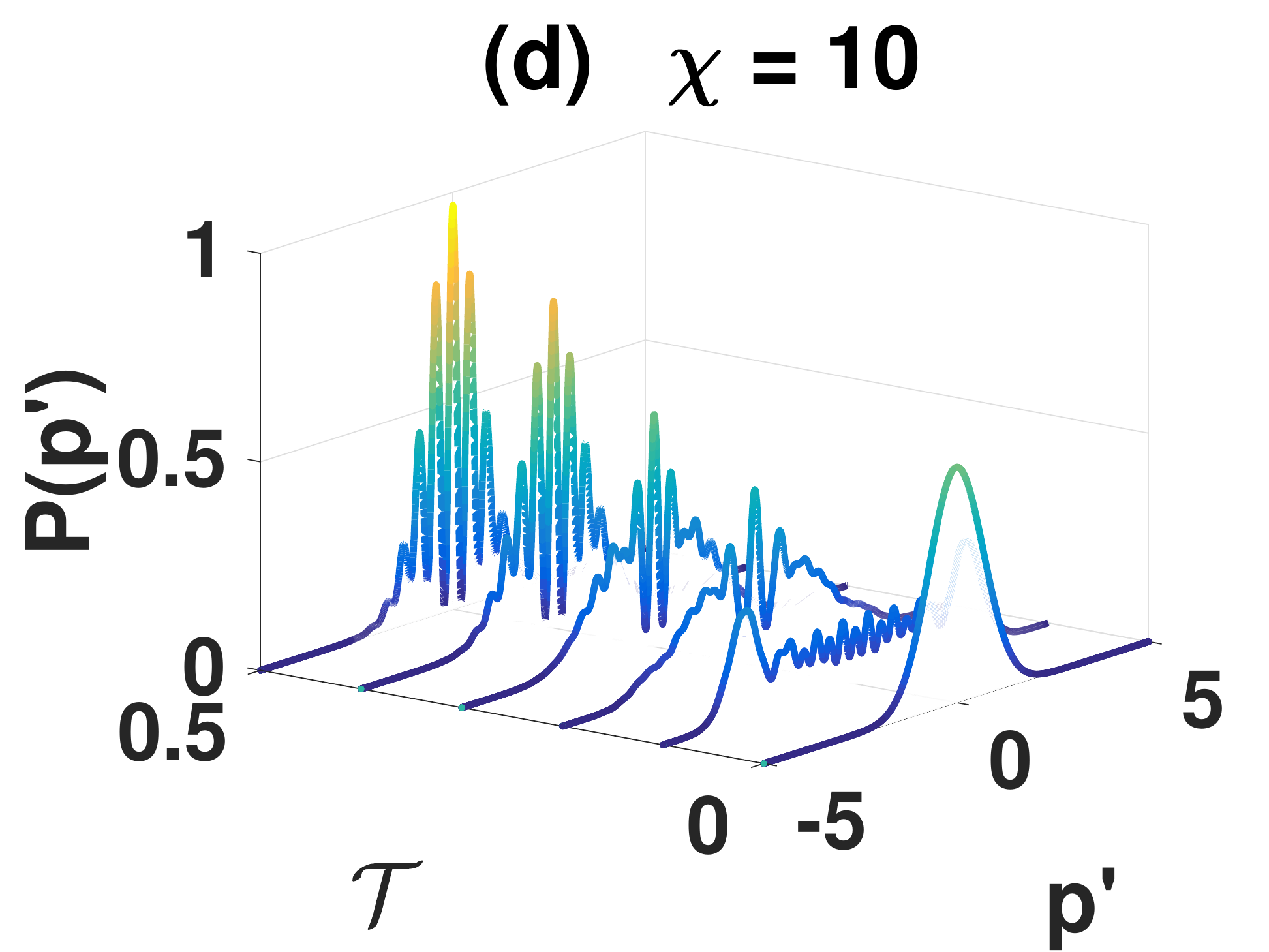}

\caption{The evolution of quadrature probability distributions $x'=x_{\phi}$
and $p'=x_{\phi+\pi/2}$ respectively, for (a,b) $\chi=5$ and (c,d)
$\chi=10$. Here, $|\alpha_{0}|=5$ and the angle $\phi$ is determined
from the predicted complex amplitude $\alpha_{0}=|\alpha_{0}|e^{i\phi}$
as given in Eq. (\ref{eq:defn-alpha-kerr}). \label{fig:quadrature_a05}\textbf{\textcolor{red}{}}}

\end{figure}

\begin{figure}[H]
\includegraphics[width=0.51\columnwidth]{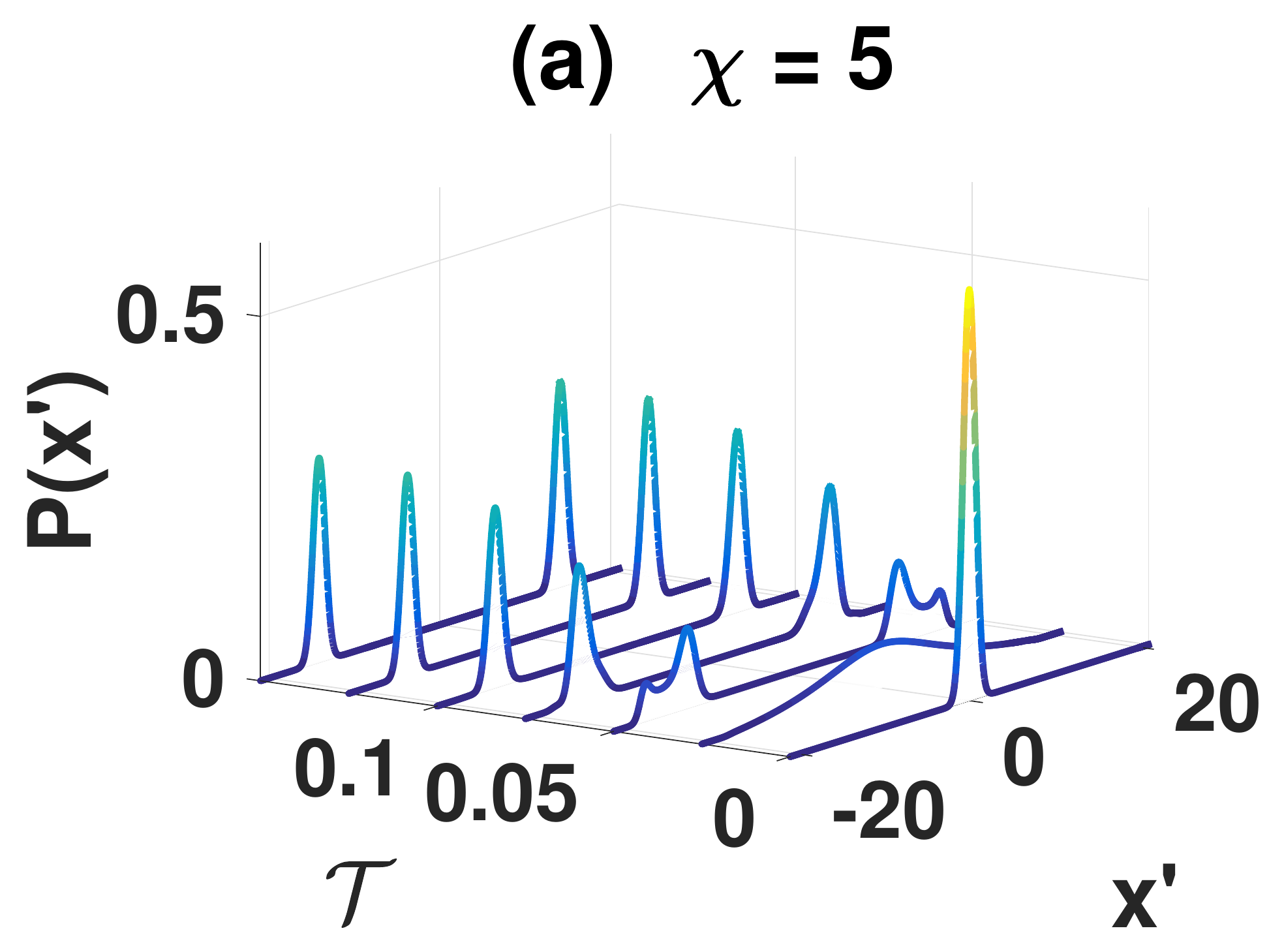}\includegraphics[width=0.51\columnwidth]{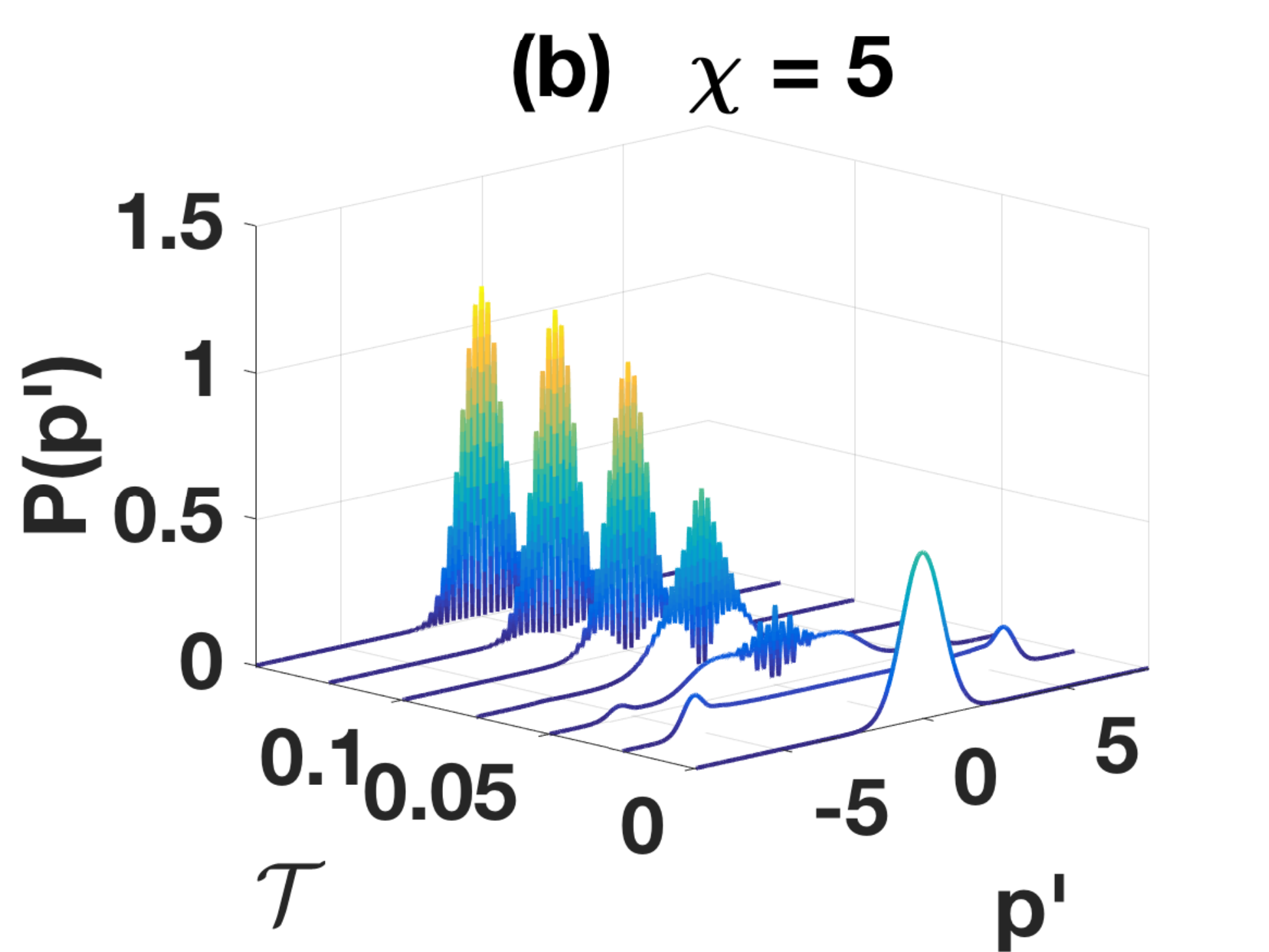}

\bigskip{}

\includegraphics[width=0.51\columnwidth]{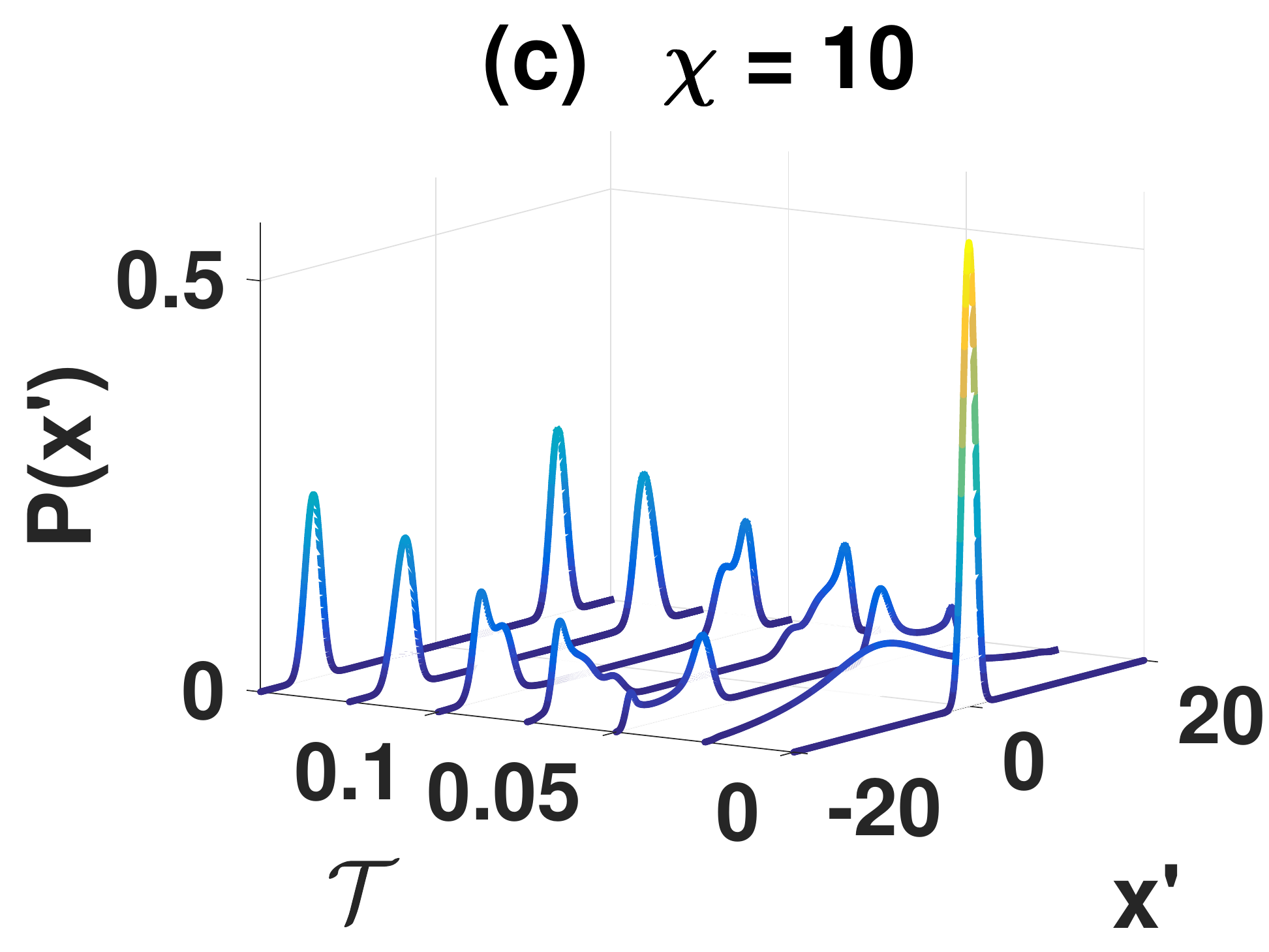}\includegraphics[width=0.51\columnwidth]{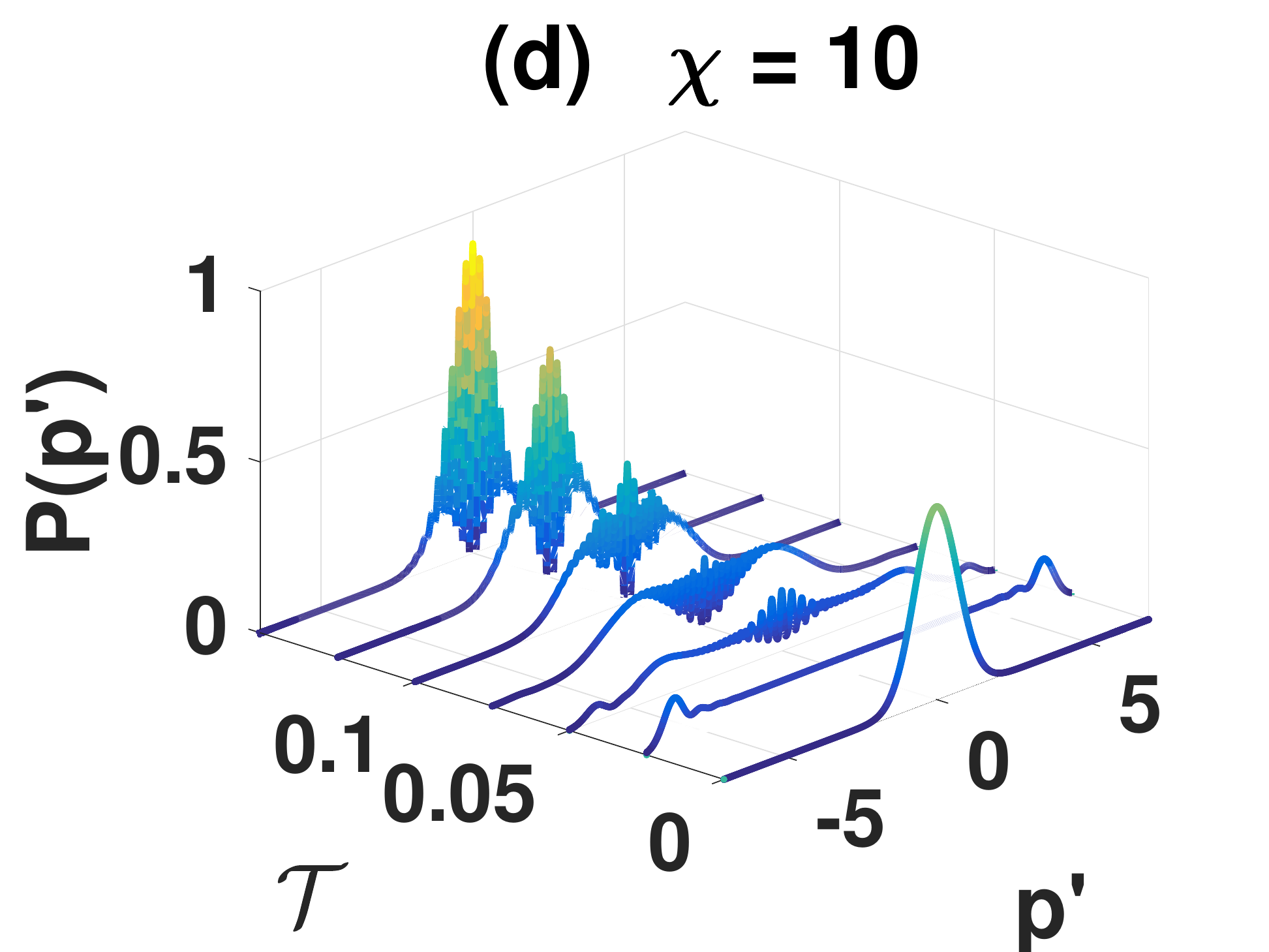}

\caption{The evolution of quadrature probability distributions $x'=x_{\phi}$
and $p'=x_{\phi+\pi/2}$ respectively, for (a,b) $\chi=5$ and (c,d)
$\chi=10$. Here, $|\alpha_{0}|=10$ and the angle $\phi$ is determined
from the predicted complex amplitude $\alpha_{0}=|\alpha_{0}|e^{i\phi}$
as given in Eq. (\ref{eq:defn-alpha-kerr}). The time range for all
plots is $0-0.15$. \label{fig:quadrature_a010} \textcolor{red}{}}
\end{figure}

We also plotted the evolution of the rotated quadrature probability
distributions $P\left(x_{\phi}\right)$ and $P\left(x_{\phi+\pi/2}\right)$,
where the angle $\phi$ is determined from the predicted complex amplitude
$\alpha_{0}=|\alpha_{0}|e^{i\phi}$ as given in Eq. (\ref{eq:defn-alpha-kerr}).
The results are plotted in Figs. \ref{fig:quadrature_a05} and \ref{fig:quadrature_a010}
for $|\alpha_{0}|=5$ and $|\alpha_{0}|=10$ respectively. In each
figure, the rotated quadrature probability distributions for different
$\chi$ values are also presented. For larger $\chi$, it takes a
similar dimensionless time $\mathcal{T}$ for the quadrature probability
distribution to reach the one that corresponds to a cat-state, which
implies a shorter real time. 

The cat-formation times for different $\chi$ and $|\alpha_{0}|$
values, in both the dimensionless time $\mathcal{T}_{\text{cat}}$
and \textit{real} time $t_{\text{cat}}=\mathcal{T}_{\text{cat}}/\left(G^{2}\sqrt{1+\chi^{2}}\right)$,
are presented in Table \ref{tab:cat_formation_time_Kerr} using the
value of $G=2.24\times10^{2}\sqrt{\text{Hz}}$  as taken from the
parameters of the experiment of \citet{Leghtas_Science2015} (refer
Section \ref{subsec:G^2_scaling_lossless}).  The cat-formation time
is determined by comparing the numerical Wigner negativity with that
of a pure cat-state Wigner function in Eq. (\ref{eq:Wigner_cat_state}).
From the table, we see that for a cat-state of fixed amplitude, a
similar nonlinearity $\chi$ has a larger $\mathcal{T}_{\text{cat}}$,
in agreement with the observations in Figs. \ref{fig:quadrature_a05}
and \ref{fig:quadrature_a010}. Also from the table, a longer $\mathcal{T}_{\text{cat}}$
corresponds to a shorter $t_{\text{cat}}$. Thus, a larger Kerr interaction
speeds up the cat-formation time.

\begin{table}[H]
\begin{centering}
\begin{tabular}{|c|c|c|c|c|}
\hline 
\multirow{2}{*}{$\chi$} & \multicolumn{2}{c|}{$\mathcal{T}_{\text{cat}}$} & \multicolumn{2}{c|}{$t_{\text{cat}}=\mathcal{T}_{\text{cat}}/\left(G^{2}\sqrt{1+\chi^{2}}\right)$
($\mu s$)}\tabularnewline
\cline{2-5} 
 & $\left|\alpha_{0}\right|=5$ & $\left|\alpha_{0}\right|=10$ & $\left|\alpha_{0}\right|=5$ & $\left|\alpha_{0}\right|=10$\tabularnewline
\hline 
$0$ & $0.40\pm0.02$ & $0.125\pm0.005$ & $8.00\pm0.40$ & $2.50\pm0.10$\tabularnewline
\hline 
$1$ & $0.44\pm0.02$ & $0.130\pm0.005$ & $6.22\pm0.28$ & $1.84\pm0.07$\tabularnewline
\hline 
$2$ & $0.52\pm0.02$ & $0.135\pm0.005$ & $4.65\pm0.18$ & $1.21\pm0.04$\tabularnewline
\hline 
$5$ & $0.74\pm0.02$ & $0.20\pm0.005$ & $2.90\pm0.08$ & $0.78\pm0.02$\tabularnewline
\hline 
\end{tabular}
\par\end{centering}
\caption{The cat-formation times for different values of the nonlinear parameter
$\chi$ and $\left|\alpha_{0}\right|$. The parameter \textcolor{black}{$G=2.24\times10^{2}\sqrt{\text{Hz}}$}
\textcolor{blue}{} is used to convert the dimensionless time $\mathcal{T}_{\text{cat}}$
to the real time $t_{\text{cat}}$. \label{tab:cat_formation_time_Kerr}\textcolor{red}{{} }}
\end{table}

\subsection{Single-photon signal damping}

Now we focus on the case where $\gamma_{1}\neq0$ i.e. $g$ is finite.
We examine the transient behavior of the signal field assuming the
initial state is the vacuum state. The free parameters in this case
are the coherent amplitude $\left|\alpha_{0}\right|,\,g$ and $\chi$,
as well as the scaled time $\mathcal{T}=\left(g^{2}\sqrt{1+\chi^{2}}\right)t$.

In the presence of single-photon damping, an ideal pure cat-state
cannot be formed even as a transient state. This is true without the
Kerr interaction, but becomes more noticeable in the solutions we
give for nonzero $\chi$. Rather, in an optimal situation, a cat-like
state is formed where two peaks are fully separated and interference
fringes are present around the origin. Here, we define the cat lifetime
as the time taken for the Wigner negativity to reach $\delta\leq0.05$,
provided the quadrature distributions are initially consistent with
a cat-state, being two-peaked for $x'$ and with fringes for $p'$.

\begin{figure}[H]
\includegraphics[width=0.51\columnwidth]{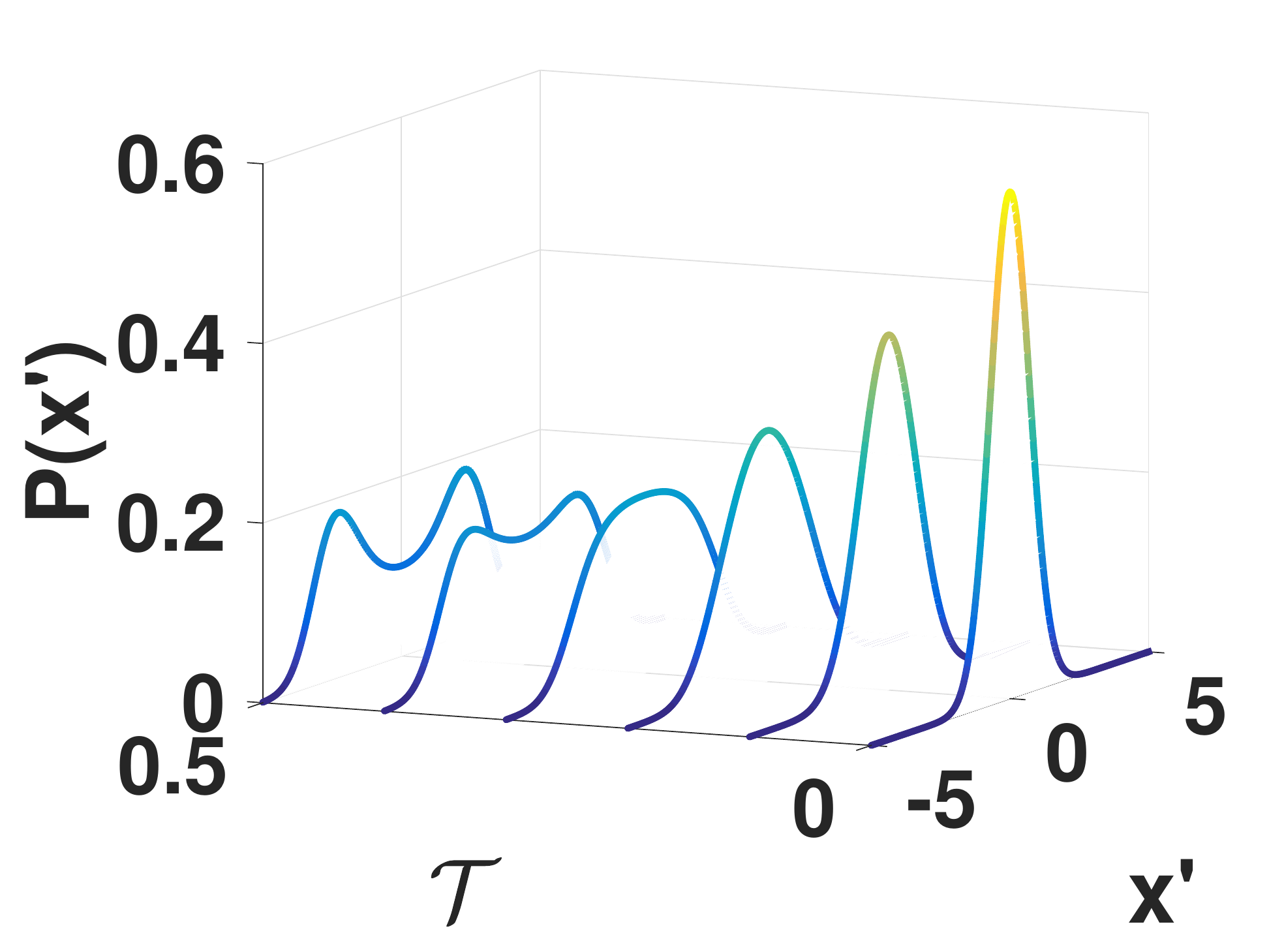}\includegraphics[width=0.51\columnwidth]{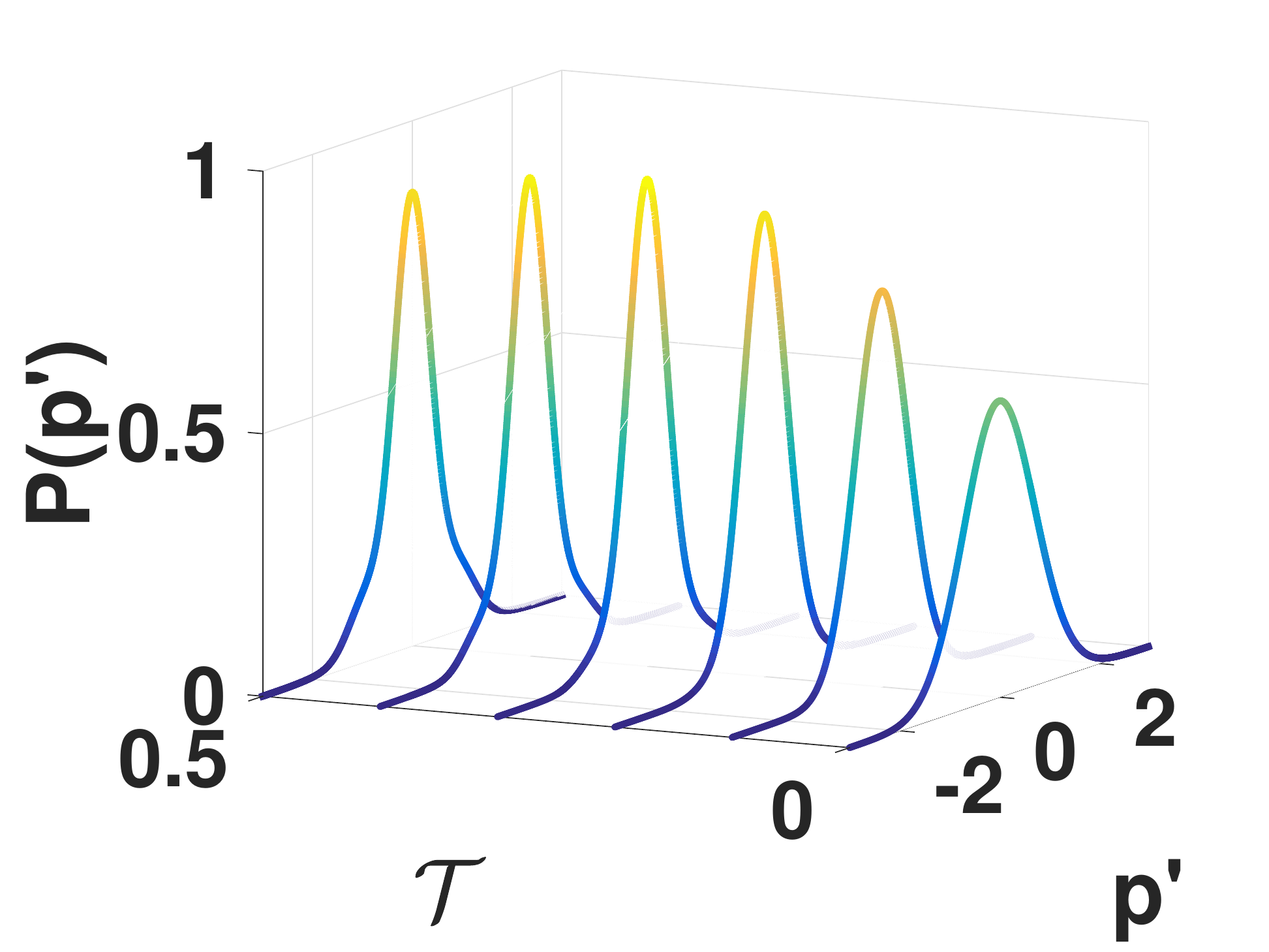}

\bigskip{}

\includegraphics[width=0.51\columnwidth]{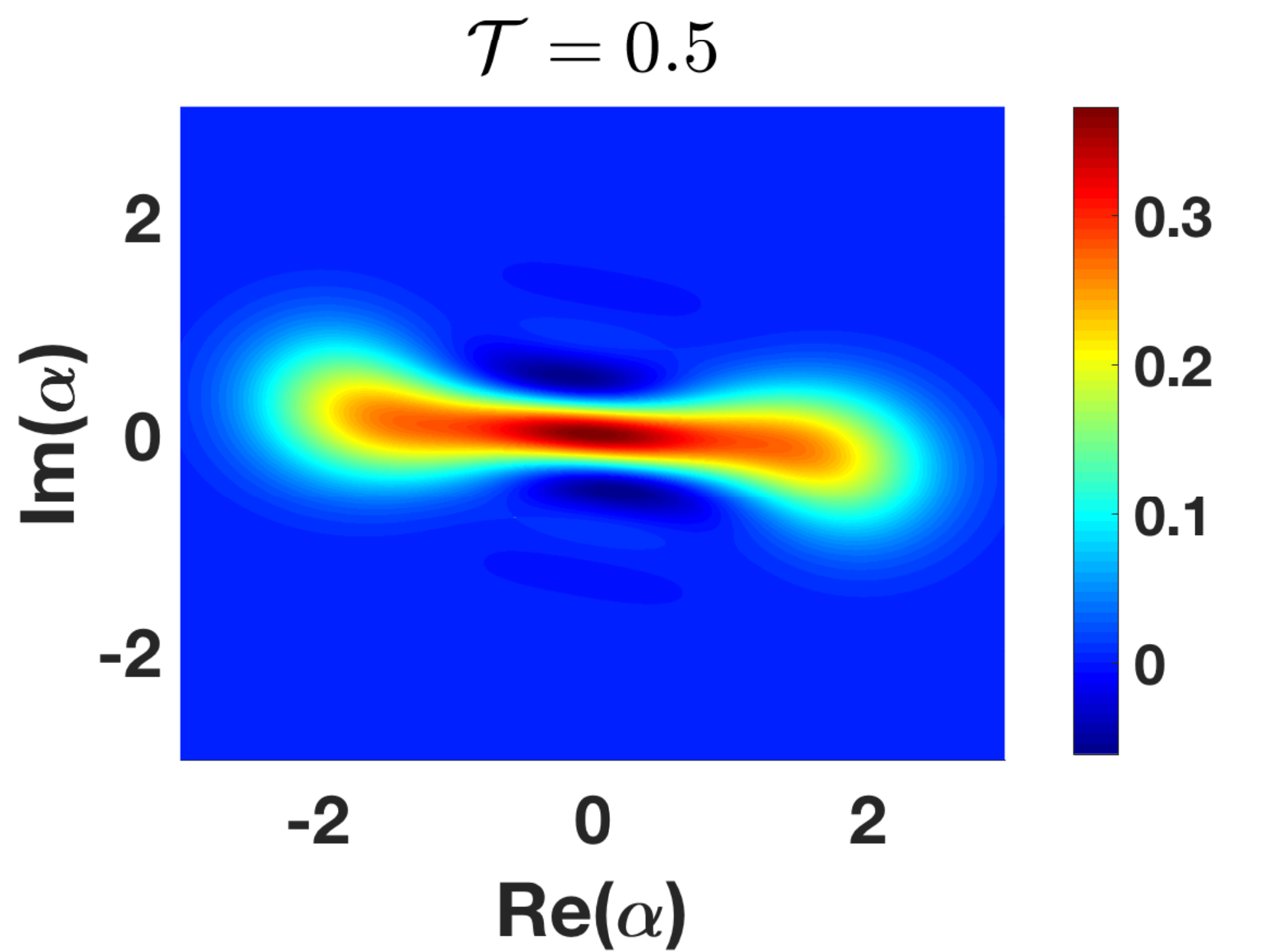}\includegraphics[width=0.51\columnwidth]{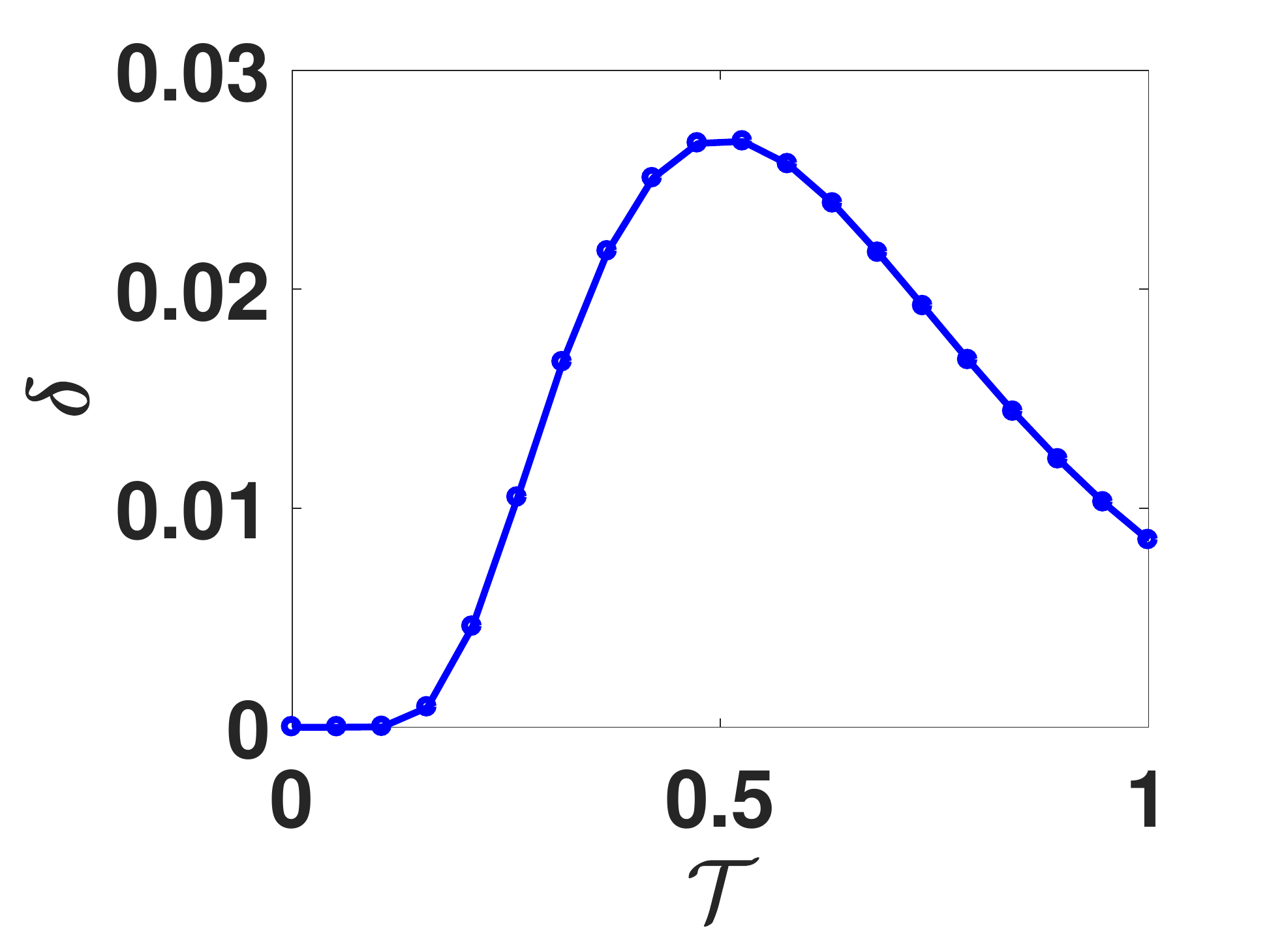}

\caption{The quadrature probability distributions as a function of time (top
left and right). The bottom left figure shows the Wigner function
at $\mathcal{T}=0.5$, which has the largest Wigner negativity $\delta$
as given in the bottom right plot. Here, the experimental parameters
\citep{Leghtas_Science2015} are $g=1.41$, $\chi'=1.01$, giving
$\chi=0.5$ and an estimated coherent amplitude $\left|\alpha_{0}\right|=2$.
\label{fig:experimental_sim}\textcolor{red}{}}
\end{figure}

It is reported that a cat-like state has been observed in the experiment
of Leghtas et al. \citep{Leghtas_Science2015}. In the following,
we carry out the numerical simulation of the experiment using the
published experimental parameters where $g=1.41$, $\chi'=1.01$,
giving $\chi=0.5$ and an estimated coherent amplitude $\left|\alpha_{0}\right|=2$.
The numerical results are shown in Fig. \ref{fig:experimental_sim}
where the time evolution of the quadrature probability distributions
and the Wigner negativity are plotted. We see from Fig. \ref{fig:experimental_sim}
(top left) that the coherent peaks in $x'$ with opposite phases are
never fully separated for $|\alpha_{0}|=2$. The largest Wigner negativity
value in the simulation, located around dimensionless time $\mathcal{T}=0.5$
is small ($\sim0.025$) and this is reflected by the absence of observable
interference fringes in the quadrature probability distribution in
Fig. \ref{fig:experimental_sim} (top right). This supports that,
while a nonclassical state is produced in the experiment, the state
is not a mesoscopic cat-state: The coherent peaks are not fully separated
and the non-classicality of the state as quantified by the Wigner
negativity is weak.

In Table \ref{tab:experiment_Kerr}, we evaluate the cat-lifetime
as defined in the previous section, by evaluating the time taken for
a cat-state to decay to a Wigner negativity smaller than $0.05$.
For the parameters of the experiment, we note again that for $|\alpha_{0}|=2$,
the steady state corresponds to two peaks in $x'$ that are not fully
separated. From the table, for $g\leq1.5$, the Wigner negativity
does not exceed $0.05$ and is too small (when compared to a pure
cat-state with amplitude $|\alpha_{0}|=2$, which has a Wigner negativity
of $0.2937$ as predicted by Eq. \ref{eq:Wigner_cat_state}) to be
considered a cat-state at any point of the simulation. True cat-states
are generated for higher $g$, however. Next, we investigate the
non-classicality of transient cat-states with larger coherent amplitudes
and Kerr strengths.

\begin{table}[H]
\begin{centering}
\begin{tabular}{|c|c|c|c|c|}
\hline 
\multirow{2}{*}{$\chi$} & \multicolumn{3}{c|}{$\mathcal{T}_{\text{life}}$} & $t_{\text{life}}=\mathcal{T}_{\text{life}}/\left(\text{\ensuremath{\gamma}}g^{2}\sqrt{1+\chi^{2}}\right)$
$(\mu s)$\tabularnewline
\cline{2-5} 
 & $g=1$ & $g=1.5$ & $g=2.5$ & $g=2.5$\tabularnewline
\hline 
$0.5$ & $0$ & $0$ & $1.225$ & $7.01$\tabularnewline
\hline 
$1.0$ & $0$ & $0$ & $1.375$ & $6.22$\tabularnewline
\hline 
\end{tabular}
\par\end{centering}
\caption{The cat-like state lifetime for different $\chi$ and $g$ values,
for $|\alpha_{0}|=2$. For comparison, the experimental parameters
of Leghtas et al. \citep{Leghtas_Science2015} are $g=1.41$, $\chi=0.5$
and $\gamma=2\pi\times3.98\text{kHz}$. \textcolor{red}{ \label{tab:experiment_Kerr}}}

\end{table}

To study the effect of single-photon damping, we compute the time
evolution of the quadrature phase amplitude distributions and Wigner
negativity, varying $g$ for different values of $\chi$ and $|\alpha_{0}|$.
Recall in Section \ref{subsec:lossless_Kerr} with no signal single-photon
loss that a large Kerr interaction speeds up the cat-formation time.
As a cat-state is highly nonclassical, the system parameters that
lead to earlier cat formation might also lead to a quicker decay/decrease
in the Wigner negativity and the corresponding cat-state lifetime.
This is confirmed by Table \ref{tab:experiment_Kerr} for the experimental
parameters of \citep{Leghtas_Science2015}. 

\begin{figure}[H]
\includegraphics[width=0.51\columnwidth]{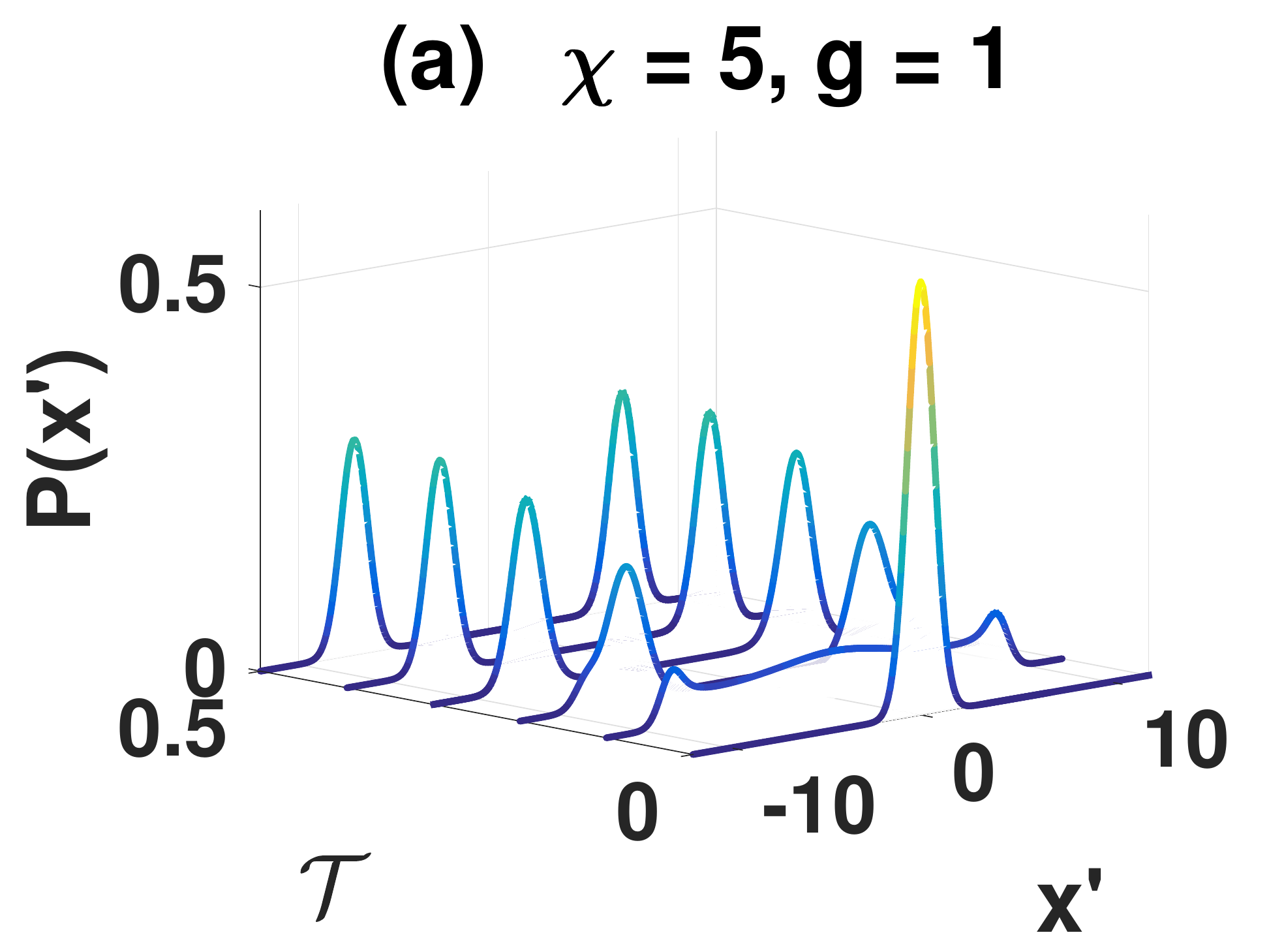}\includegraphics[width=0.51\columnwidth]{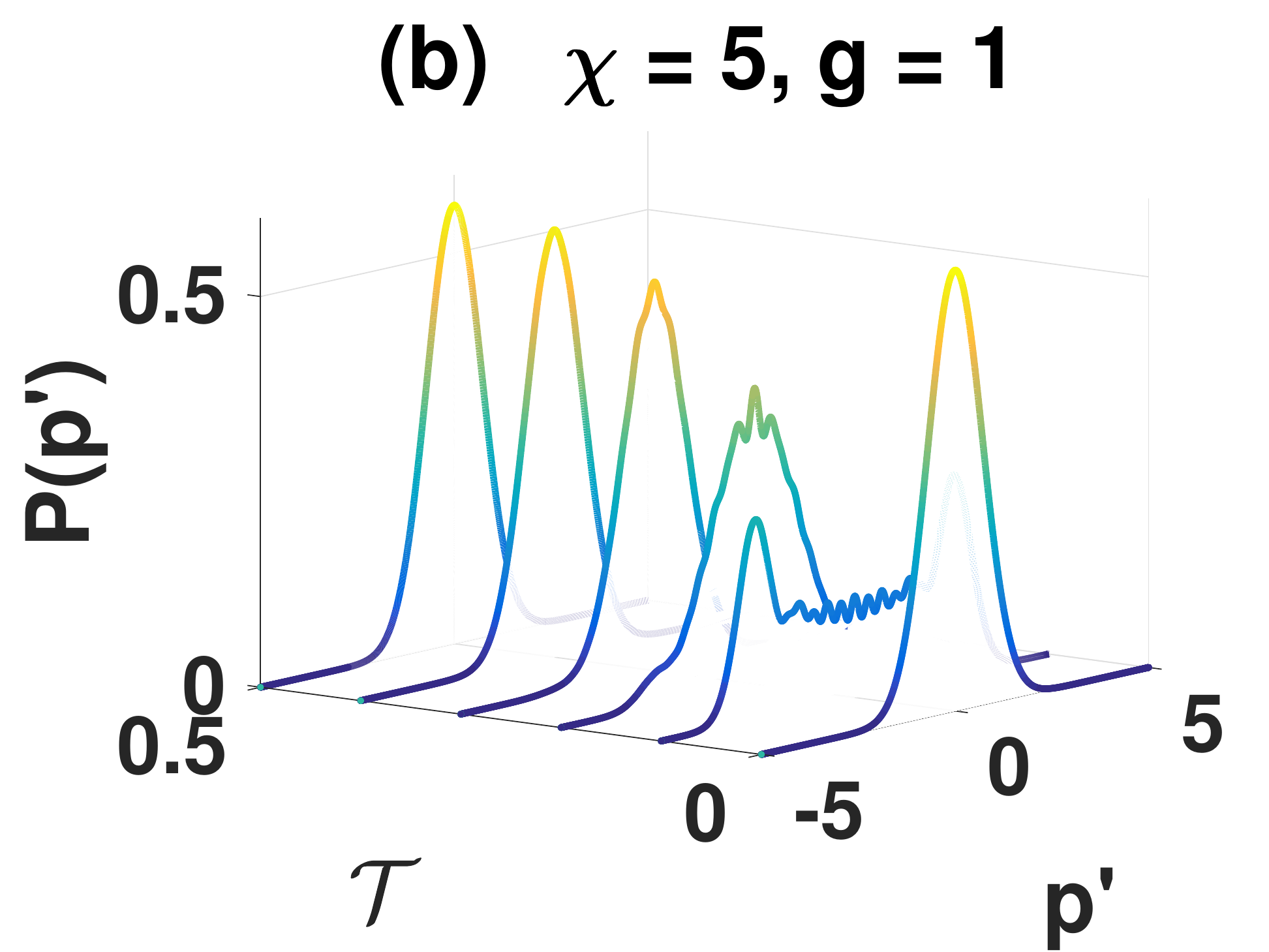}

\bigskip{}

\includegraphics[width=0.51\columnwidth]{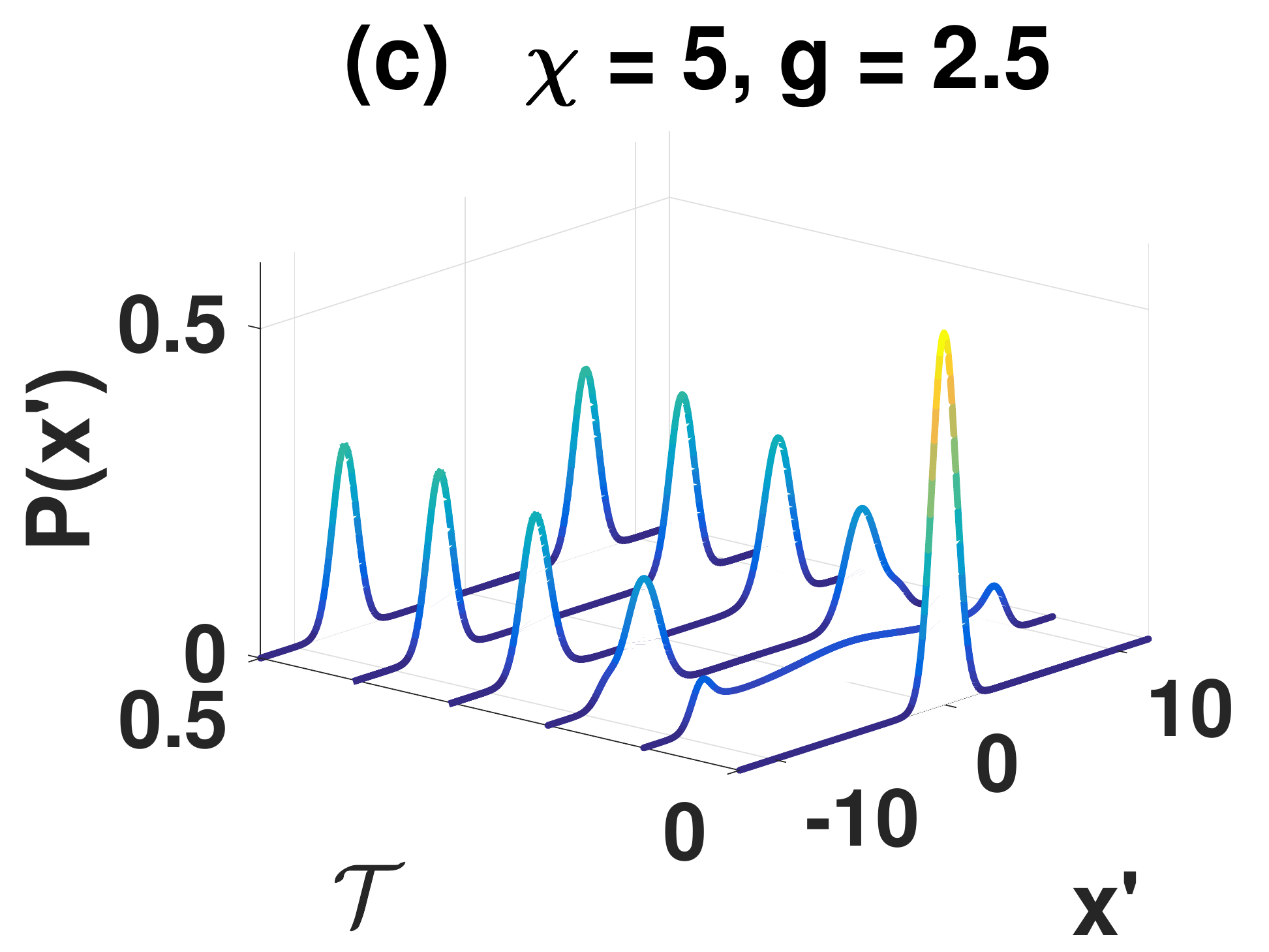}\includegraphics[width=0.51\columnwidth]{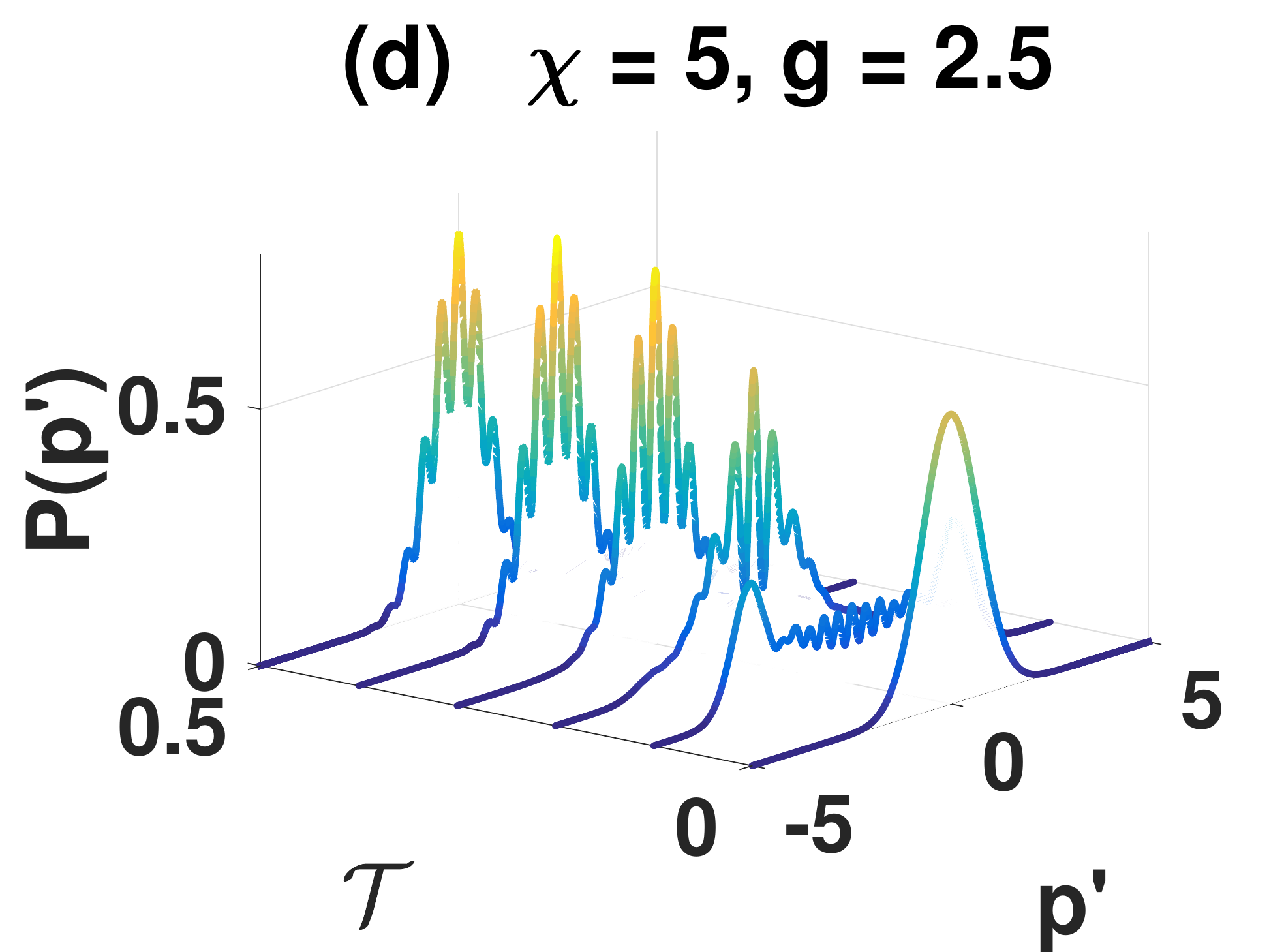}

\caption{The evolution of quadrature probability distributions $x'=x_{\phi}$
and $p'=x_{\phi+\pi/2}$ respectively, in the presence of single-photon
damping. The angle $\phi$ is determined from the predicted complex
amplitude $\alpha_{0}=|\alpha_{0}|e^{i\phi}$ as given in Eq. (\ref{eq:defn-alpha-kerr}).
Here, $\chi=5$, $|\alpha_{0}|=5$ with $g=1$ and $2.5$. \label{fig:lossy_kerr_quadrature_a05-r5}}
\end{figure}

\begin{figure}[H]

\includegraphics[width=0.51\columnwidth]{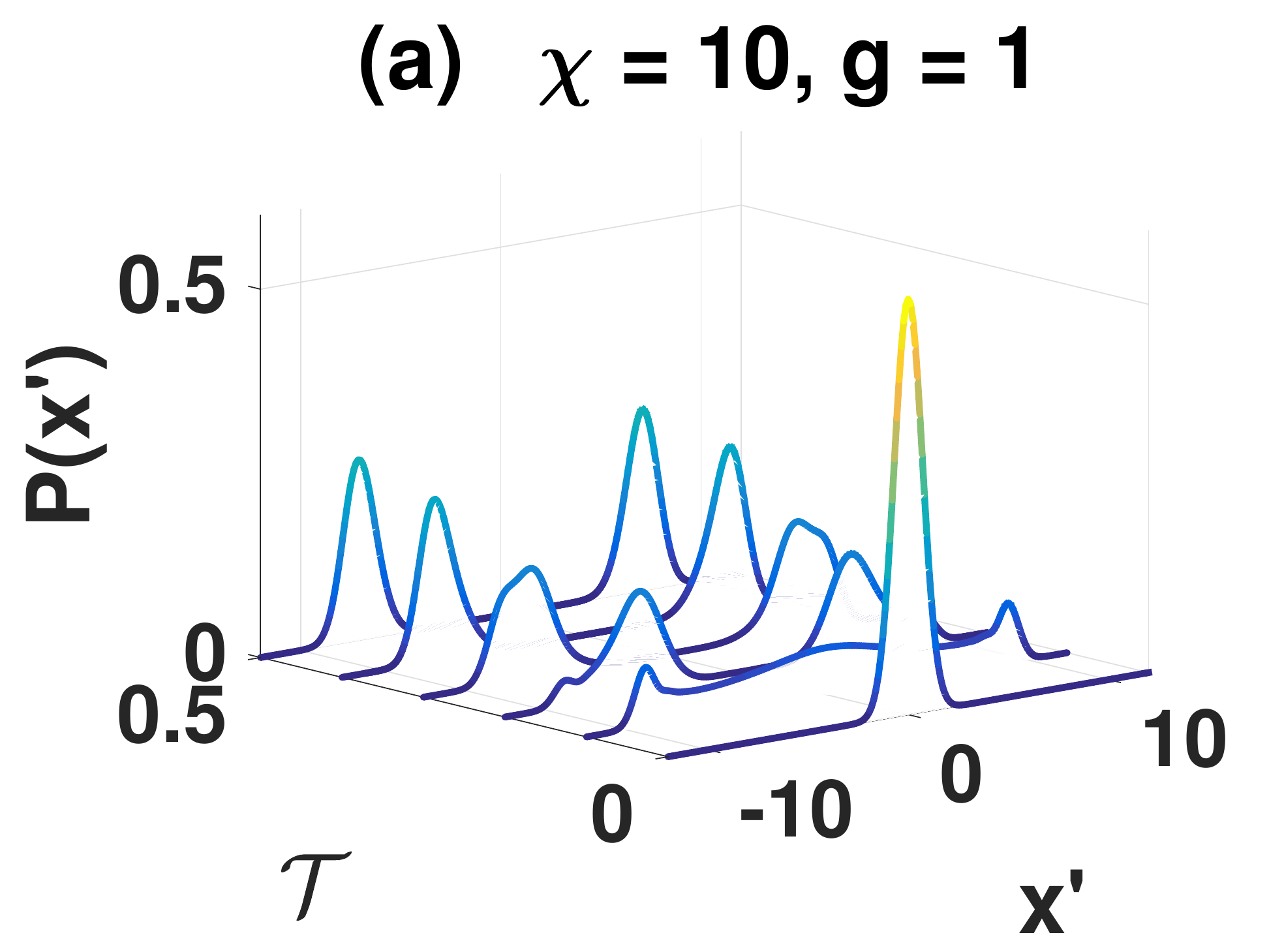}\includegraphics[width=0.51\columnwidth]{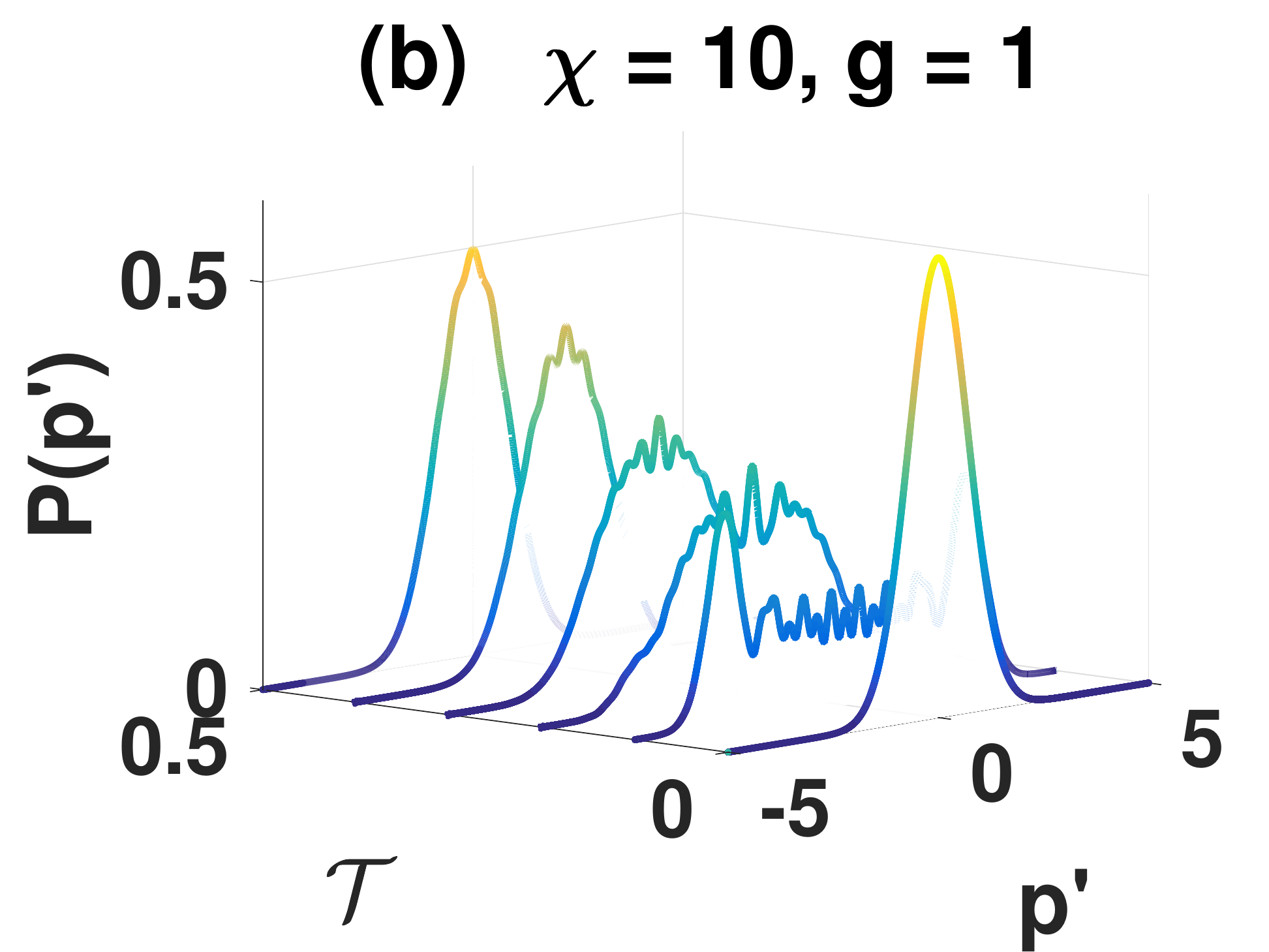}

\bigskip{}

\includegraphics[width=0.51\columnwidth]{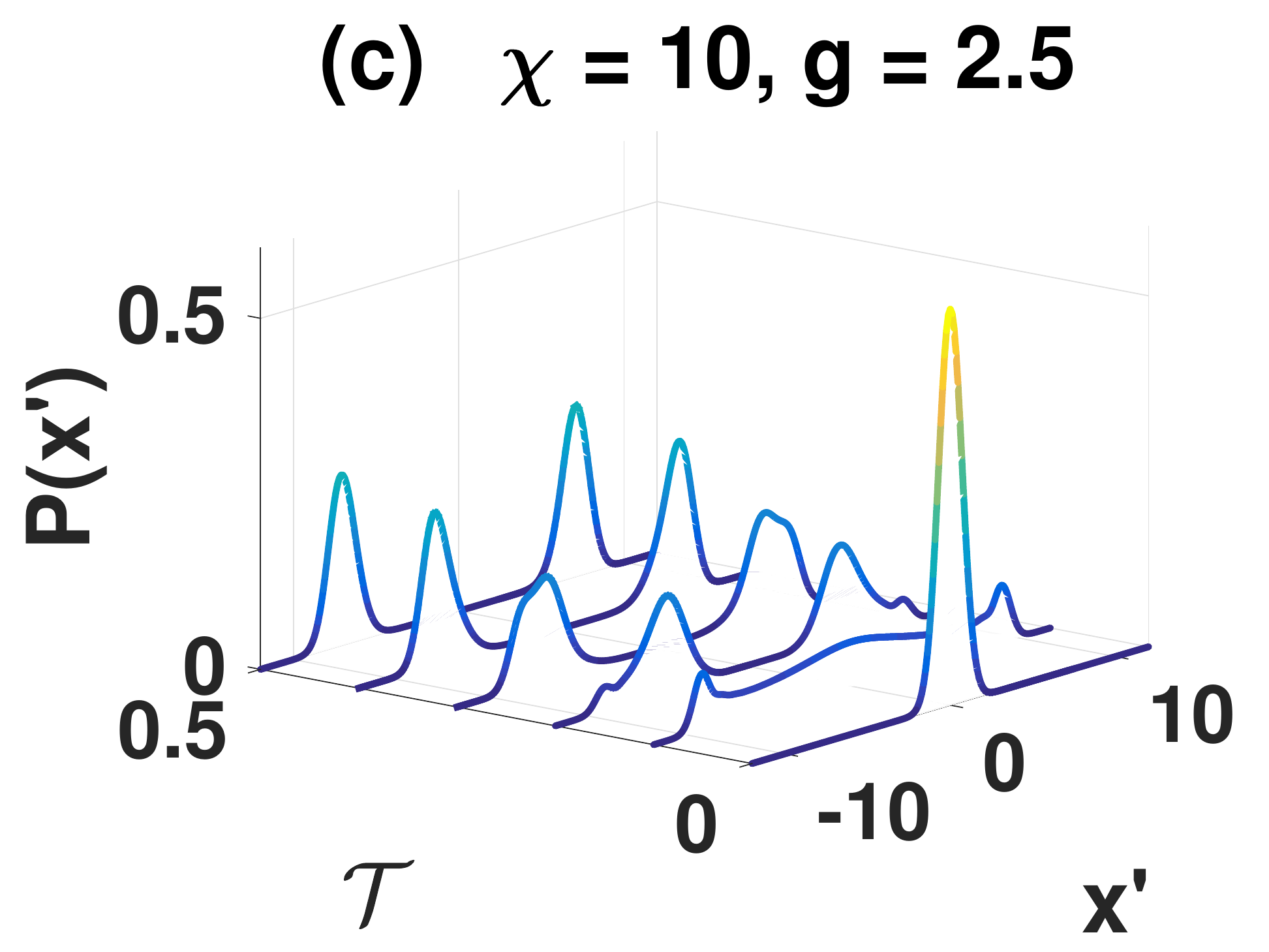}\includegraphics[width=0.51\columnwidth]{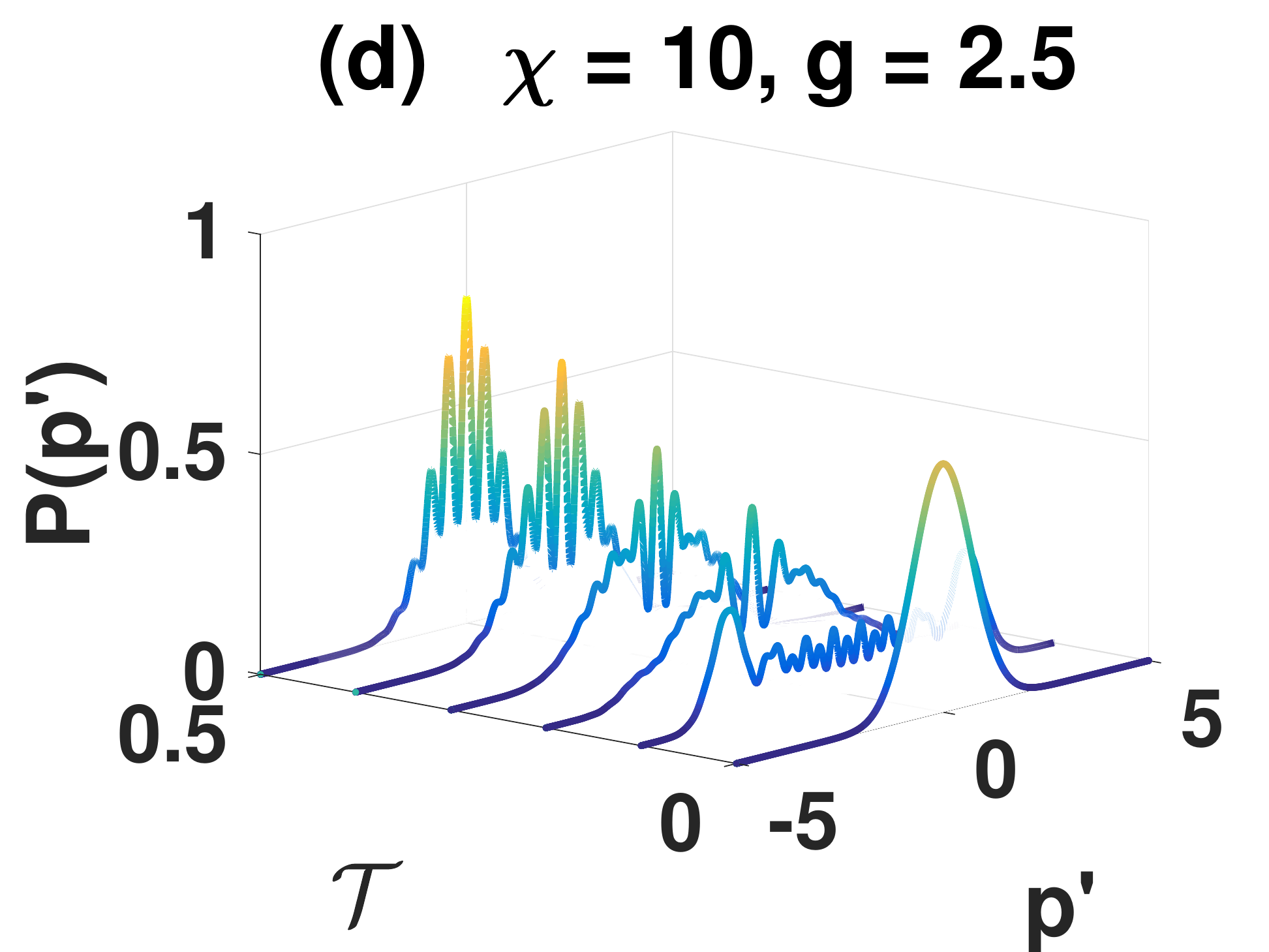}

\caption{Description as for Figure \ref{fig:lossy_kerr_quadrature_a05-r5}.
Here, $\chi=10$, $|\alpha_{0}|=5$ with $g=1$ and $2.5$. \label{fig:lossy_kerr_quadrature_a05_r10}}
\end{figure}

\begin{figure}[H]
\includegraphics[width=0.51\columnwidth]{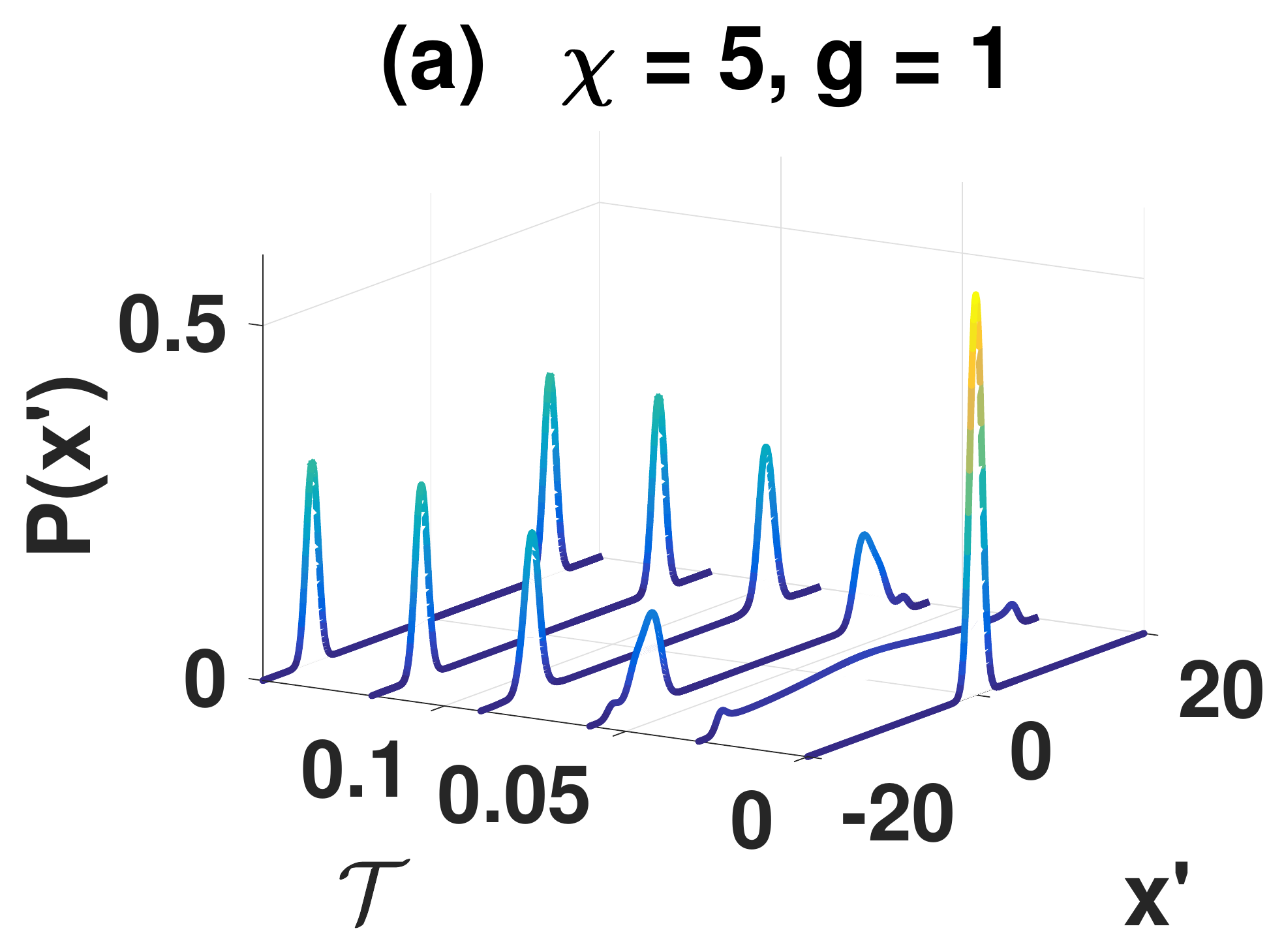}\includegraphics[width=0.51\columnwidth]{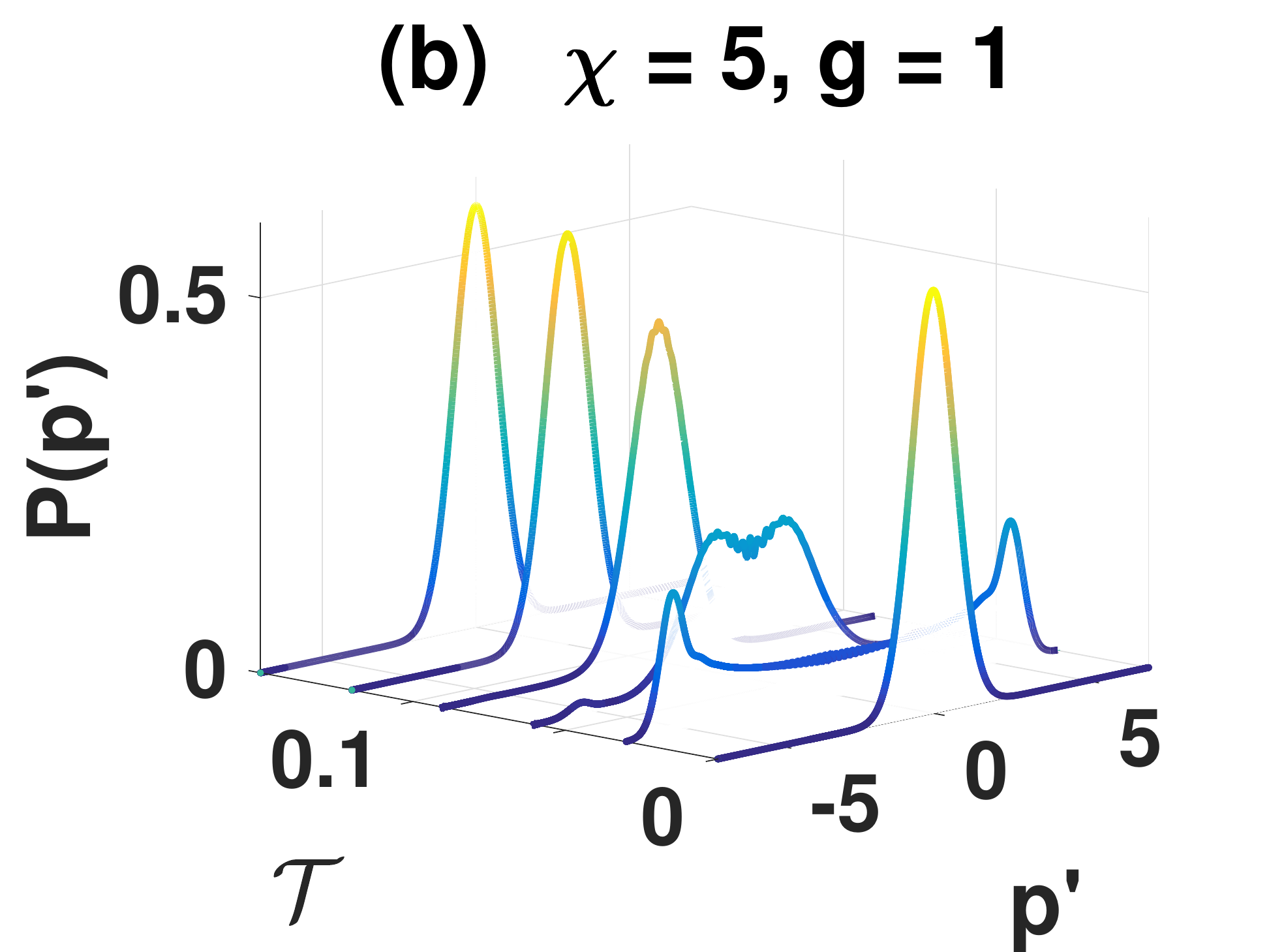}

\bigskip{}

\includegraphics[width=0.51\columnwidth]{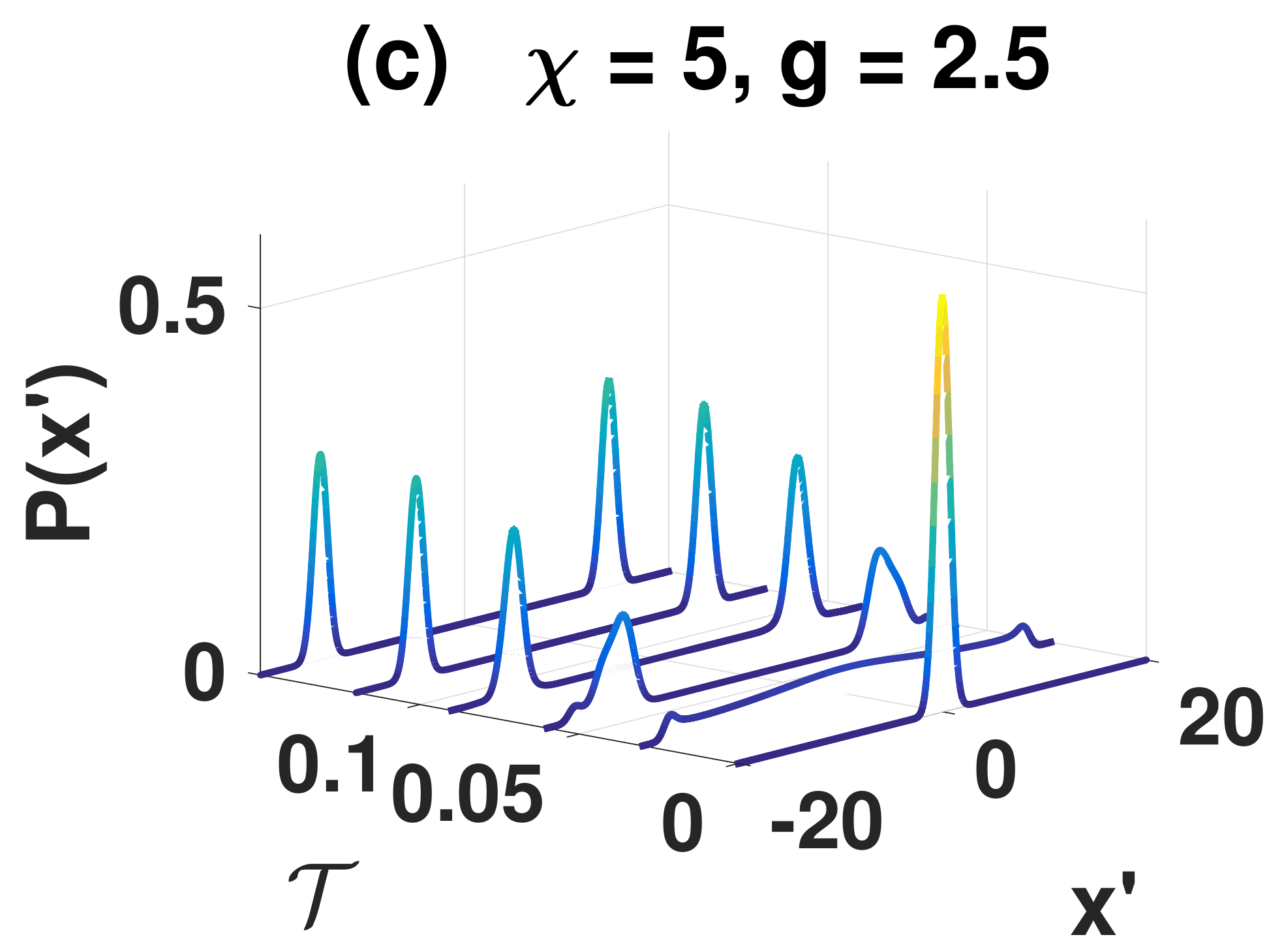}\includegraphics[width=0.51\columnwidth]{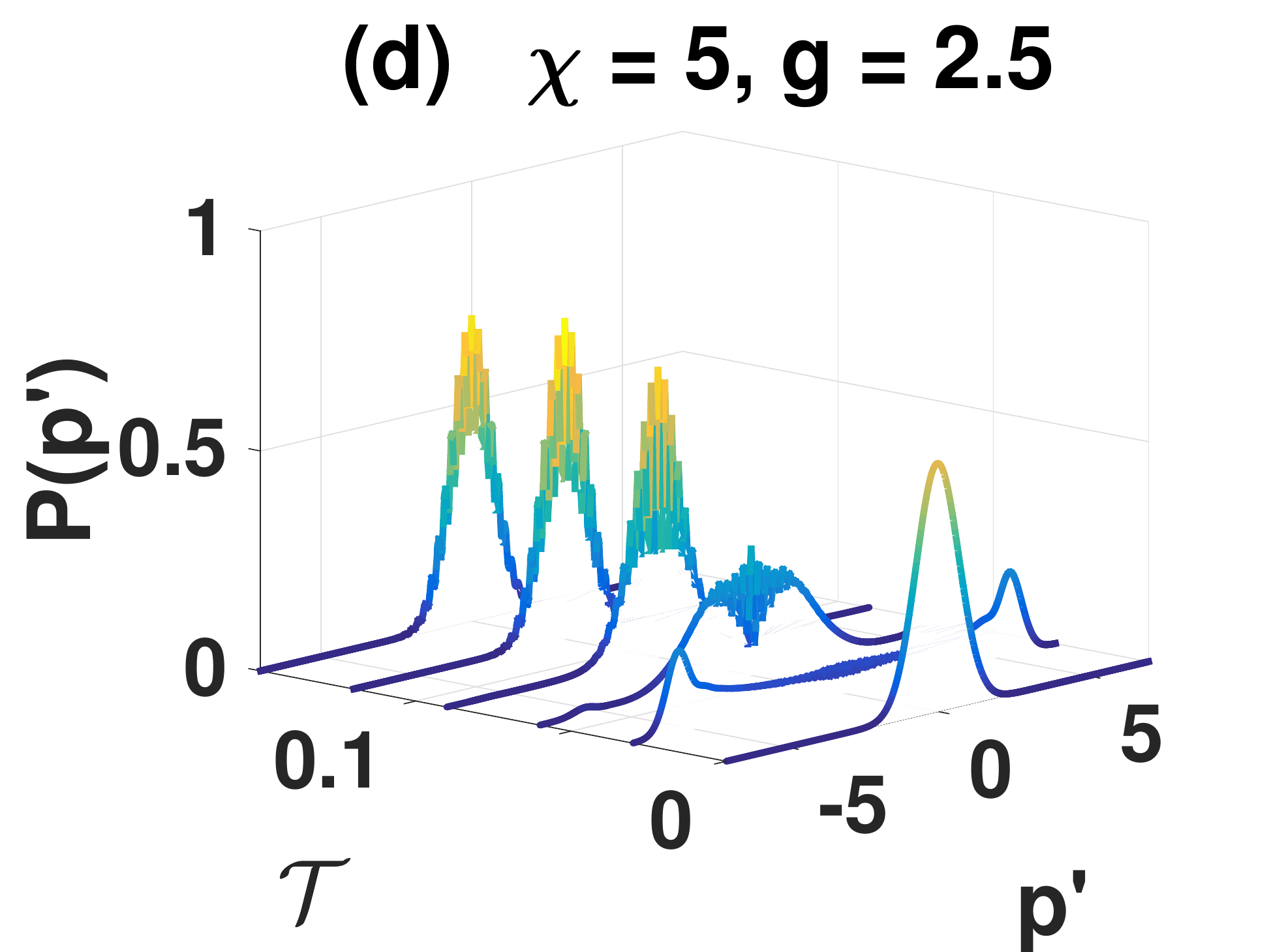}

\caption{Description as for Figure \ref{fig:lossy_kerr_quadrature_a05-r5}.
Here $|\alpha_{0}|=10$. The time range for all plots is $0-0.15$.
\label{fig:lossy_kerr_quadrature_a010-r5}}
\end{figure}

\begin{figure}[H]

\bigskip{}

\includegraphics[width=0.51\columnwidth]{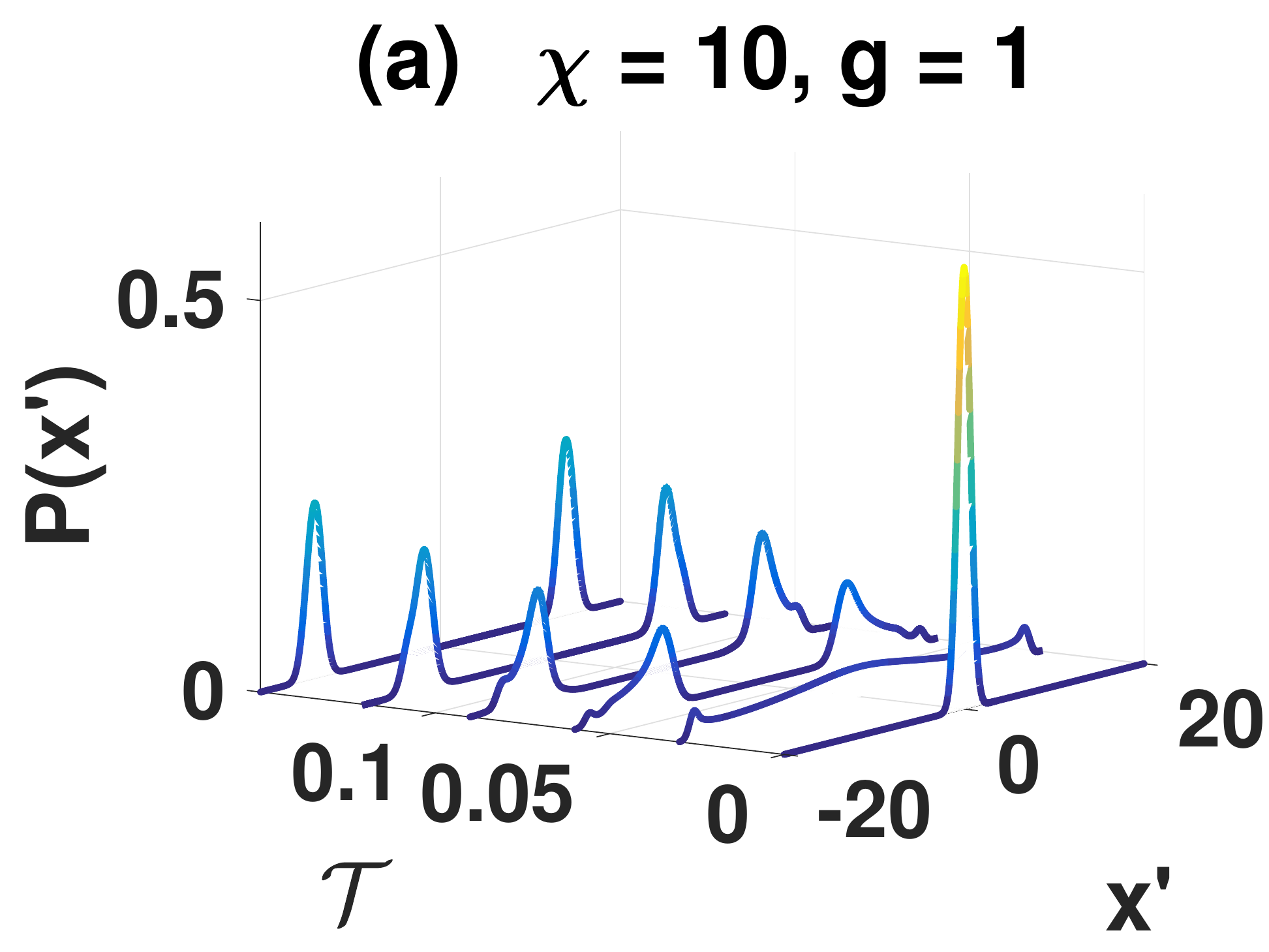}\includegraphics[width=0.51\columnwidth]{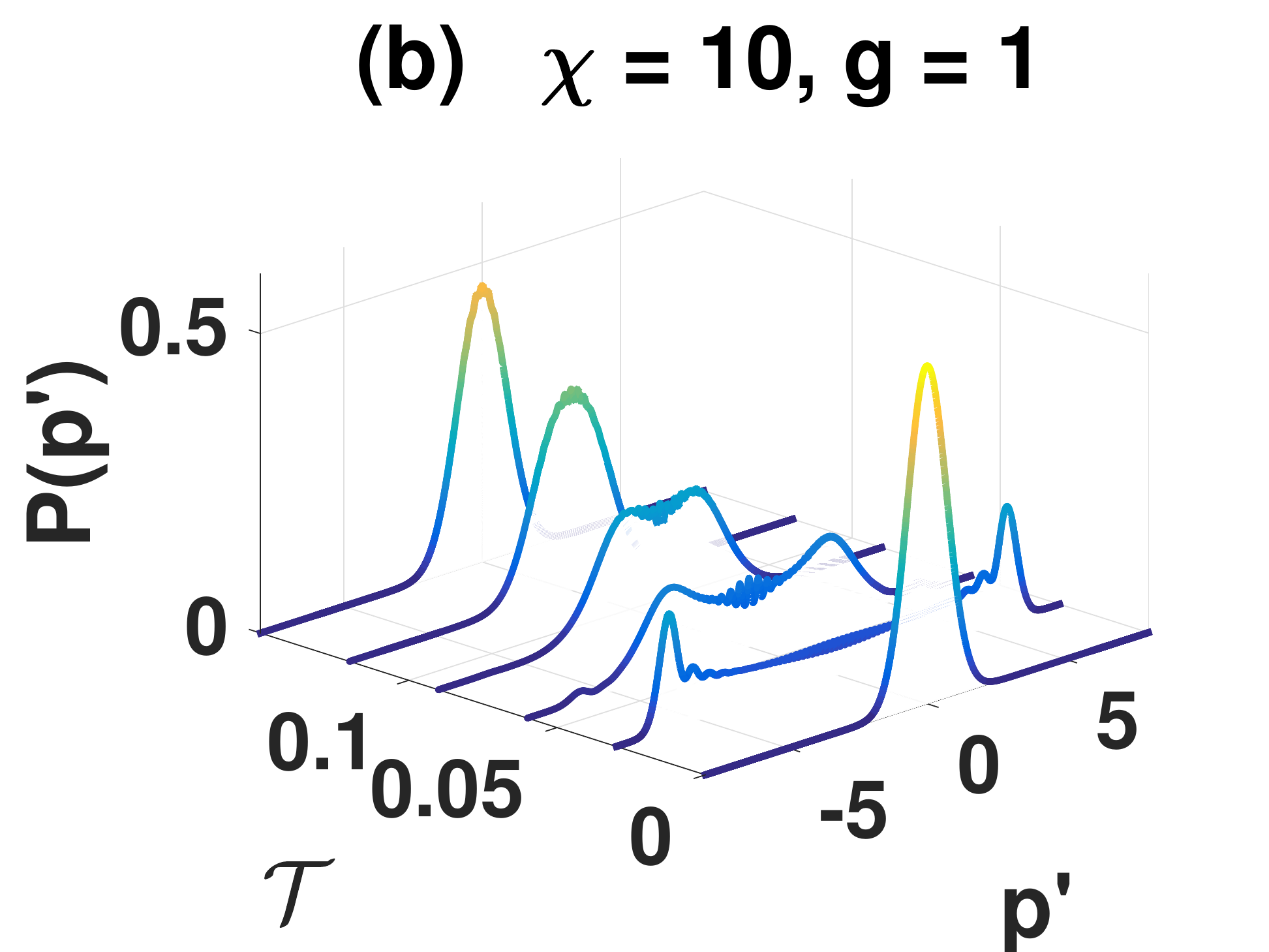}

\bigskip{}

\includegraphics[width=0.51\columnwidth]{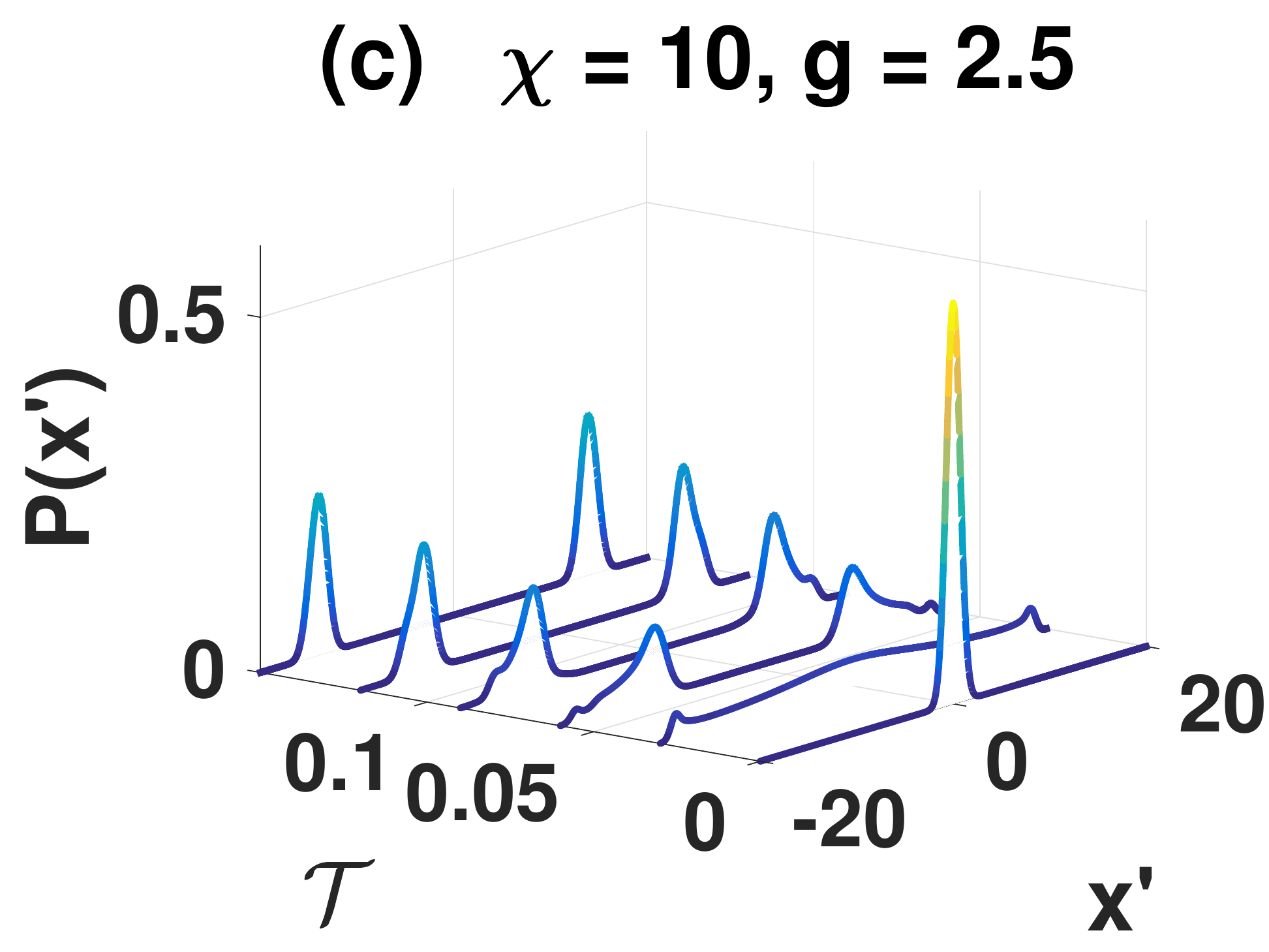}\includegraphics[width=0.51\columnwidth]{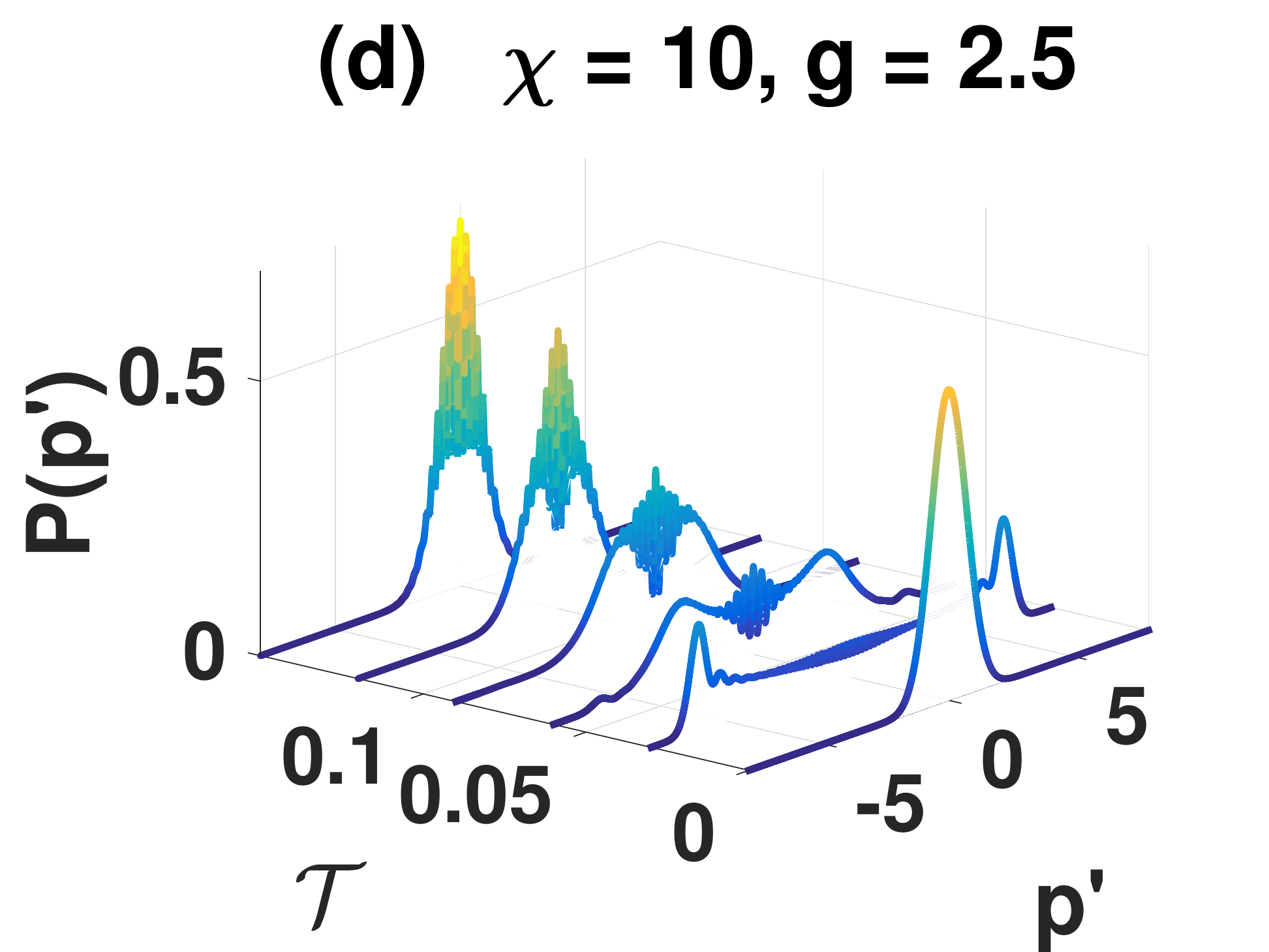}

\caption{Description as for Figure \ref{fig:lossy_kerr_quadrature_a05_r10}.
Here $|\alpha_{0}|=10$. The time range for all plots is $0-0.15$.
\label{fig:lossy_kerr_quadrature_a010_r10}}
\end{figure}

\begin{figure}[H]
\begin{centering}
\includegraphics[width=0.53\columnwidth]{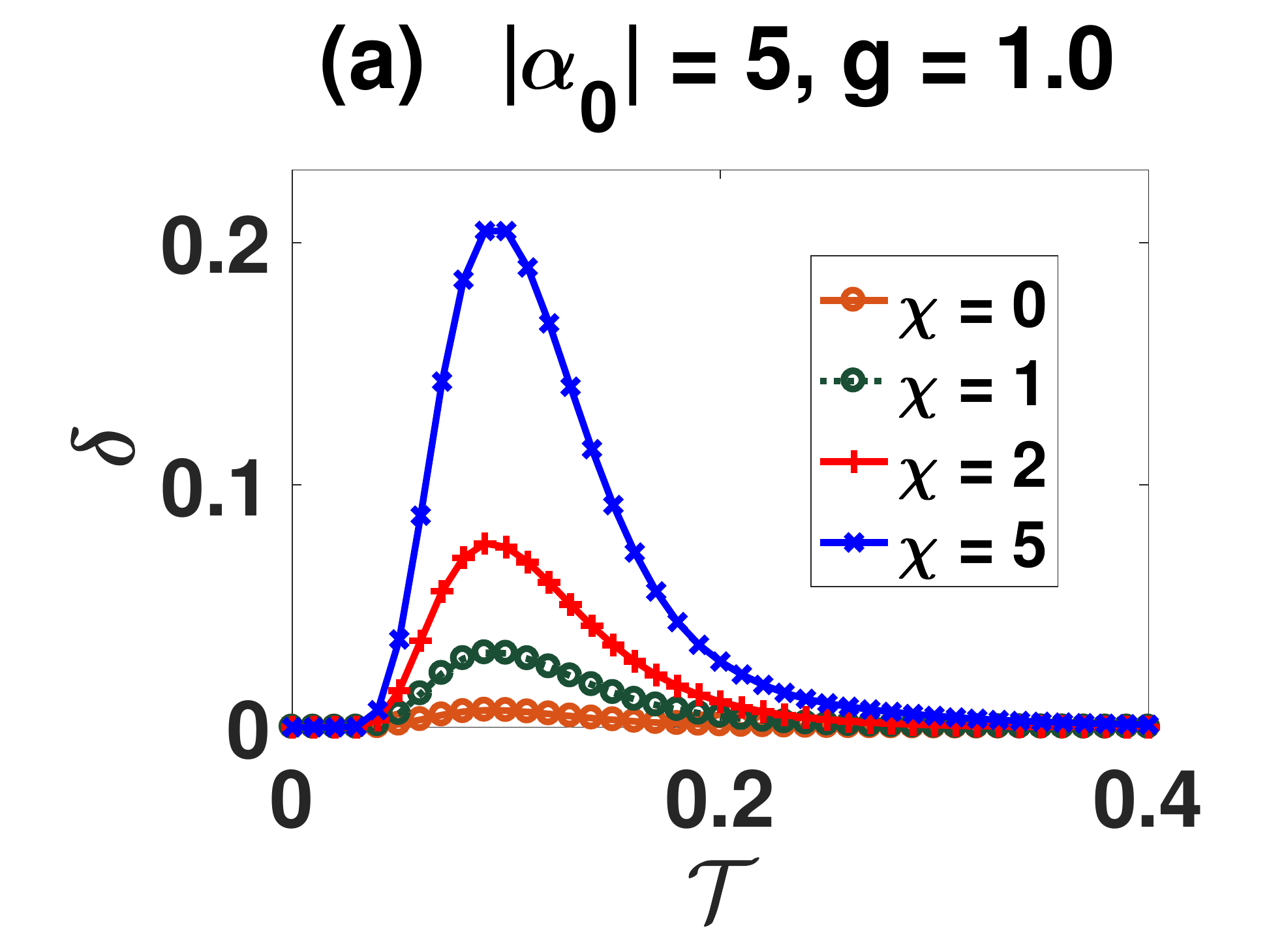}\includegraphics[width=0.53\columnwidth]{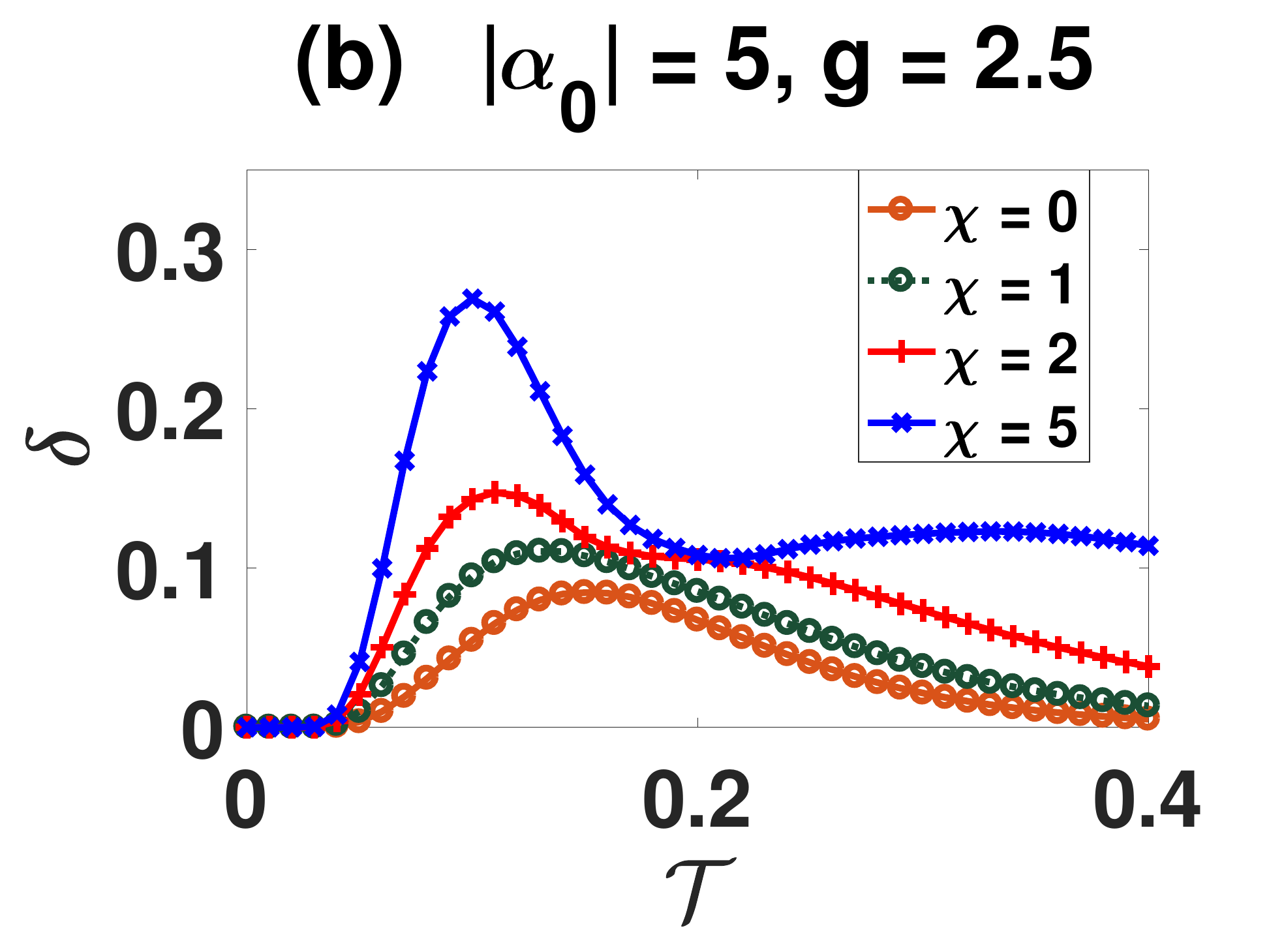}
\par\end{centering}
\caption{The evolution of the Wigner negativity with different $\chi$ values
for (a) $g=1$ and (b) $g=2.5$. In both cases, $\left|\alpha_{0}\right|=5$.
Note that a peak in the Wigner negativity does not imply the formation
of a cat-state (see main text). The verification of a cat-state can
only be drawn in conjunction with other cat-state signatures such
as in Fig. \ref{fig:lossy_kerr_quadrature_a05-r5}. \label{fig:lossy_kerr_ao5}\textcolor{blue}{}}
\end{figure}

\begin{figure}[H]
\begin{centering}
\includegraphics[width=0.53\columnwidth]{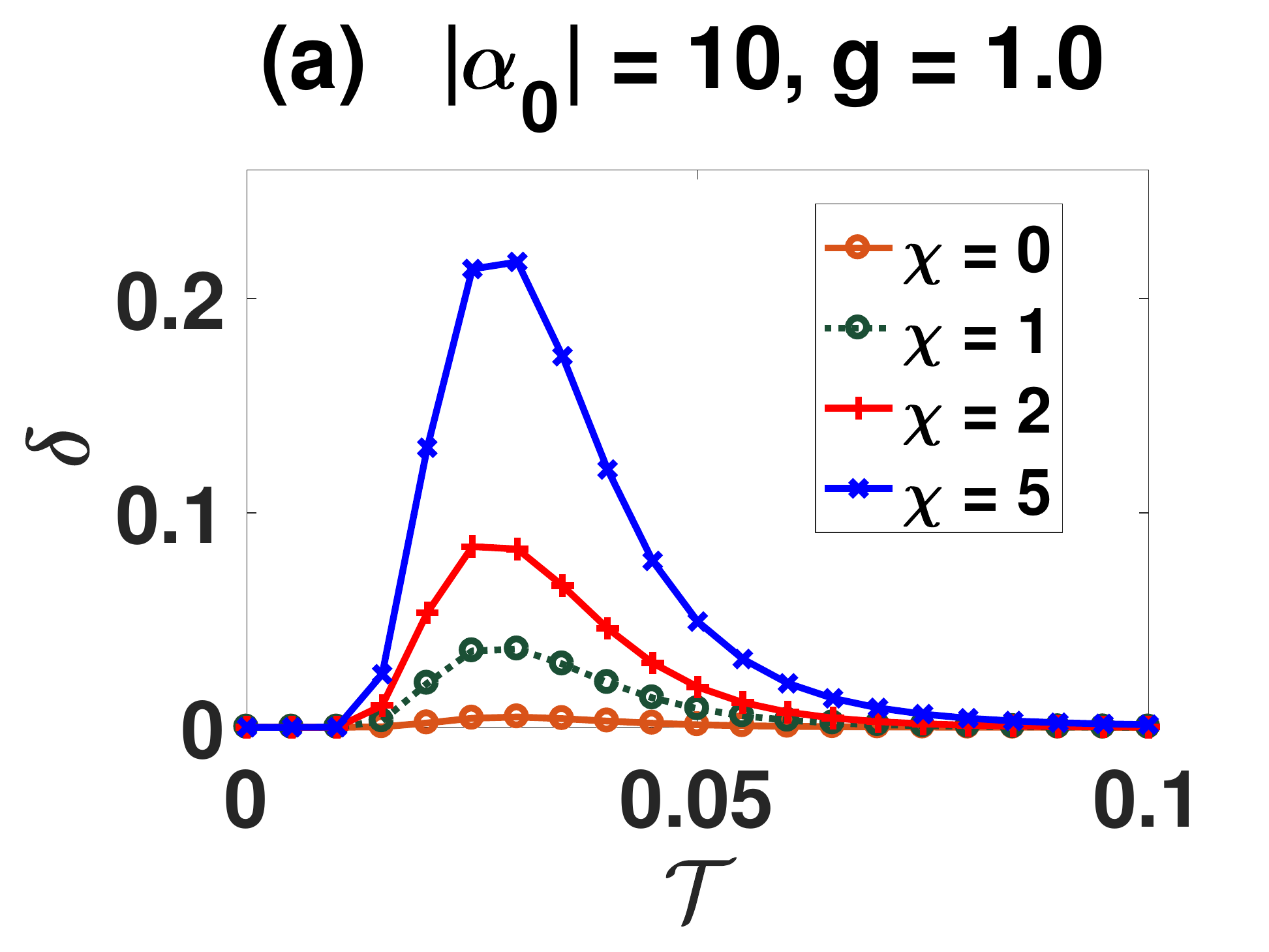}\includegraphics[width=0.53\columnwidth]{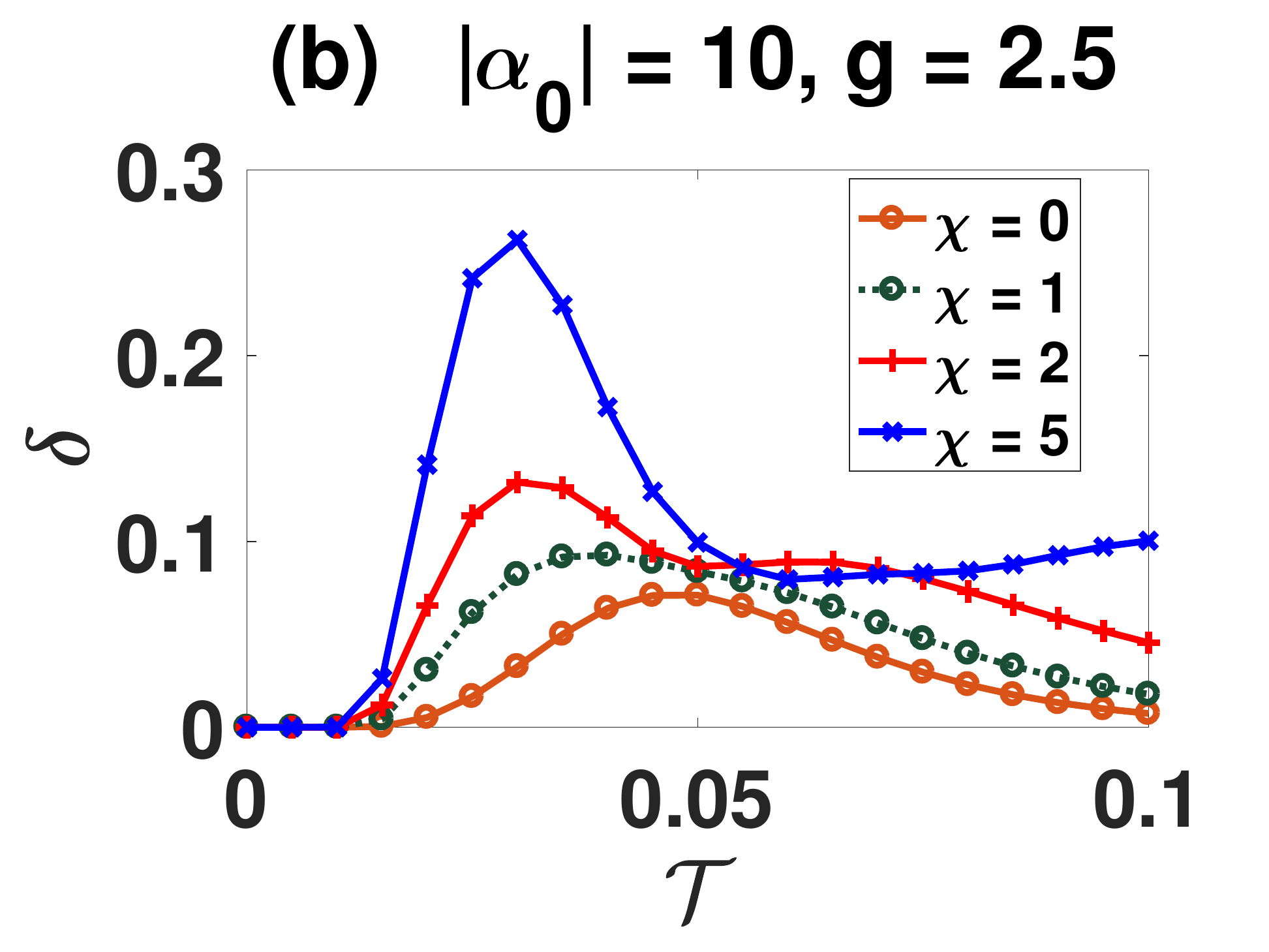}
\par\end{centering}
\caption{The evolution of the Wigner negativity with different $\chi$ values
for (a) $g=1$ and (b) $g=2.5$. In both cases, $\left|\alpha_{0}\right|=10$.
 \label{fig:lossy_kerr_ao10} \textcolor{red}{}}
\end{figure}

Another question to be answered is whether the presence of a Kerr
effect changes the threshold of $g$ required for a cat-state. We
find that $g>1$ is still required for the generation of a cat state.
The results for different $\chi$ and $|\alpha_{0}|$ are presented
in Figs. \ref{fig:lossy_kerr_quadrature_a05-r5}, \ref{fig:lossy_kerr_quadrature_a05_r10},
\ref{fig:lossy_kerr_quadrature_a010-r5}, and \ref{fig:lossy_kerr_quadrature_a010_r10}.
Figs. \ref{fig:lossy_kerr_quadrature_a05-r5} and \ref{fig:lossy_kerr_quadrature_a05_r10}
show the time evolution of the quadrature probability distributions
for $|\alpha_{0}|=5$ with $\chi=5$ and $\chi=10$, respectively.
The same quantities are plotted in Figs. \ref{fig:lossy_kerr_quadrature_a010-r5}
and \ref{fig:lossy_kerr_quadrature_a010_r10} for $|\alpha_{0}|=10$.
The emergence of interference fringes corresponds to $g\geq1$ even
in the presence of large Kerr strength. These results are confirmed
by the time evolution of the Wigner negativity as presented in Figs.
\ref{fig:lossy_kerr_ao5} and \ref{fig:lossy_kerr_ao10}. When $g$
is large enough to produce cat-like states, these figures also show
larger Wigner negativities with larger $\chi$ for the same $\alpha_{0}$
and $g$ values.

We emphasize the need to compute several cat-state signatures and
caution the use of any single signature alone to interpret the nonclassicality
of the physical state. For instance, the Wigner negativity is not
sufficient to infer the presence of a cat-state. The peak values of
Wigner negativity observed in Figs. \ref{fig:lossy_kerr_ao5} and
\ref{fig:lossy_kerr_ao10} for $\chi=5$ do not correspond to cat-states,
despite the large negativity values. These large negativities correspond
to nonclassical states that arise due to the large Kerr interaction
term, before the formation of cat-states. As previously discussed,
a cat-state is formed when two well-separated peaks in $P(x')$ are
observed and when interference fringes in the corresponding $P(p')$
distribution exist. This can be inferred from the quadrature probability
distributions or the Wigner function itself, but not directly from
the Wigner negativity.

We also see that with a finite $g$ (signal losses), as for the
earlier case without nonlinearity, the cat-state eventually decoheres
to a mixed state. More loss (lower $g$) gives a faster decay, for
fixed nonlinearity $\chi$ and $\alpha_{0}$. This is quantified in
the Table \ref{tab:larger kerr} which evaluates the Wigner negativity.
Also from the table, we see that for fixed $g$ and $\alpha_{0}$,
the cat-state decoheres faster for the larger $\chi$ value given
here.  

\begin{table}[H]
(a)%
\begin{tabular}{|c|c|c|c|c|}
\hline 
\multirow{2}{*}{$\chi$} & \multicolumn{3}{c|}{$\mathcal{T}_{\text{life}}$} & \multicolumn{1}{c|}{$t_{\text{life}}=\mathcal{T}_{\text{life}}/\left(\gamma g^{2}\sqrt{1+\chi^{2}}\right)$
$(\mu s)$}\tabularnewline
\cline{2-5} 
 & $g=1$ & $g=1.5$ & $g=2.5$ & $g=2.5$\tabularnewline
\hline 
5 & $0$ & $0$ & $0.68$ & $0.85$\tabularnewline
\hline 
10 & $0$ & $0$ & $1.25$ & $0.80$\tabularnewline
\hline 
\end{tabular}

\bigskip{}

(b)%
\begin{tabular}{|c|c|c|c|c|}
\hline 
\multirow{2}{*}{$\chi$} & \multicolumn{3}{c|}{$\mathcal{T}_{\text{life}}$} & \multicolumn{1}{c|}{$t_{\text{life}}=\mathcal{T}_{\text{life}}/\left(\gamma g^{2}\sqrt{1+\chi^{2}}\right)$
$(\mu s)$}\tabularnewline
\cline{2-5} 
 & $g=1$ & $g=1.5$ & $g=2.5$ & $g=2.5$\tabularnewline
\hline 
5 & $0$ & 0 & $0.177$ & $0.222$\tabularnewline
\hline 
10 & $0$ & 0 & $0.324$ & $0.206$\tabularnewline
\hline 
\end{tabular}

\caption{The cat-like state lifetime for different $\chi$ and $g$ values,
for (a) $|\alpha_{0}|=5$ and (b) $|\alpha_{0}|=10$. Here, $\gamma=2\pi\times3.98\text{kHz}$.
We comment that for $g=1.5$, the small value of negativity is not
associated with well-separated peaks in the distribution of $x'$
(Figs. \ref{fig:lossy_kerr_quadrature_a05-r5}, \ref{fig:lossy_kerr_quadrature_a05_r10},
\ref{fig:lossy_kerr_quadrature_a010-r5}, and \ref{fig:lossy_kerr_quadrature_a010_r10}).
Hence we do not claim these are cat-states. \label{tab:larger kerr}}

\end{table}

\section{Large transient cat\label{sec:Large-transient-cat}}

In this section, we investigate the feasibility of observing a transient
cat state using physical parameters that are achievable in an experiment
similar to the superconducting-cavity setup discussed in the previous
subsection. The effects of finite temperatures leading to thermal
noise are also included. We choose $g=2$ and $\left|\alpha_{0}\right|=20$,
which corresponds to a coherent amplitude of $20$. We focus on the
quadrature probability distribution as a cat-state signature.  In
order to achieve $g=2$ in an experiment, either the signal decay
rate has to be reduced or the nonlinear coupling strength has to be
enhanced, or both.

\begin{figure}[H]
\begin{centering}
\includegraphics[width=0.75\columnwidth]{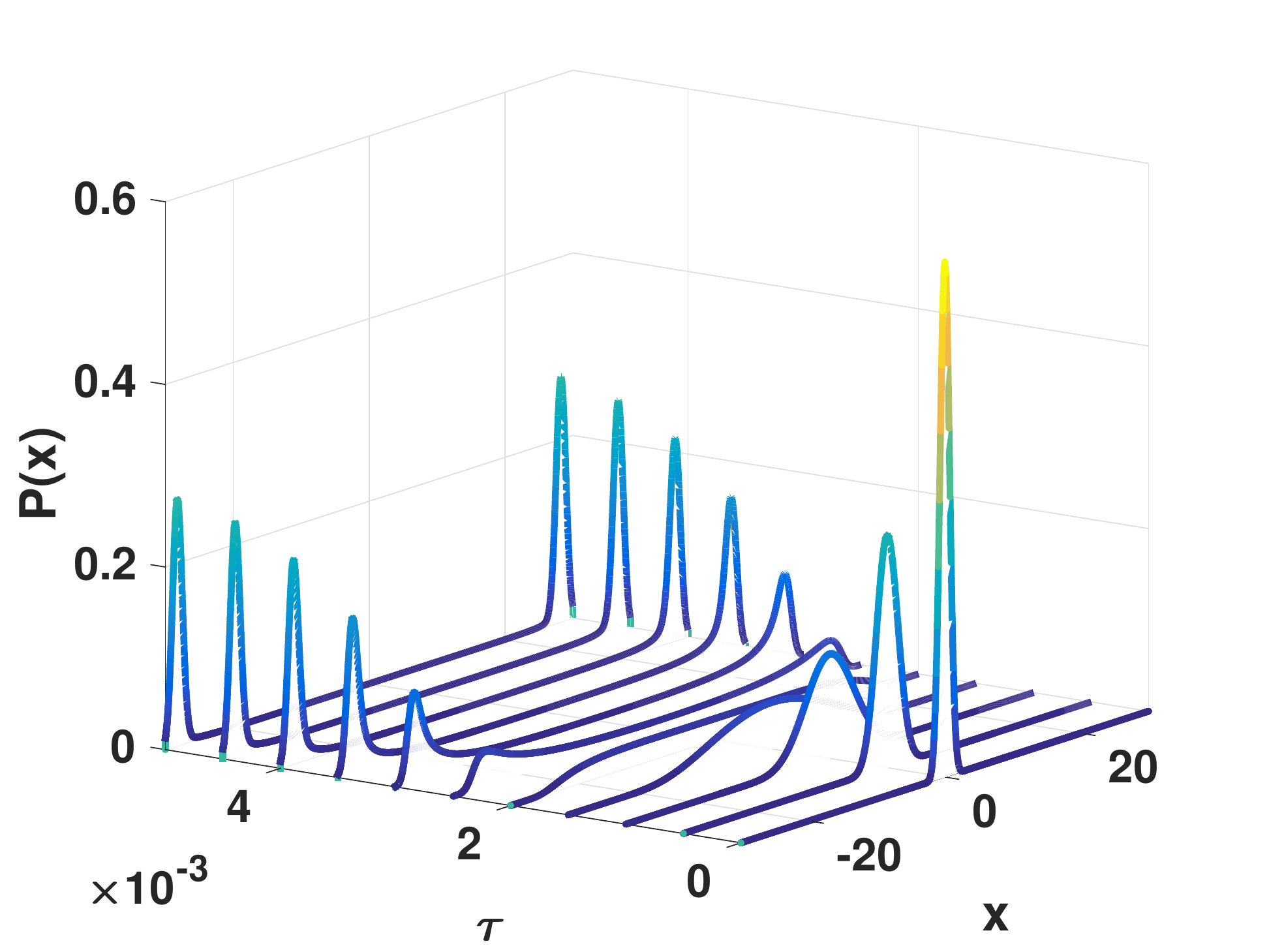}
\par\end{centering}
\begin{centering}
\includegraphics[width=0.75\columnwidth]{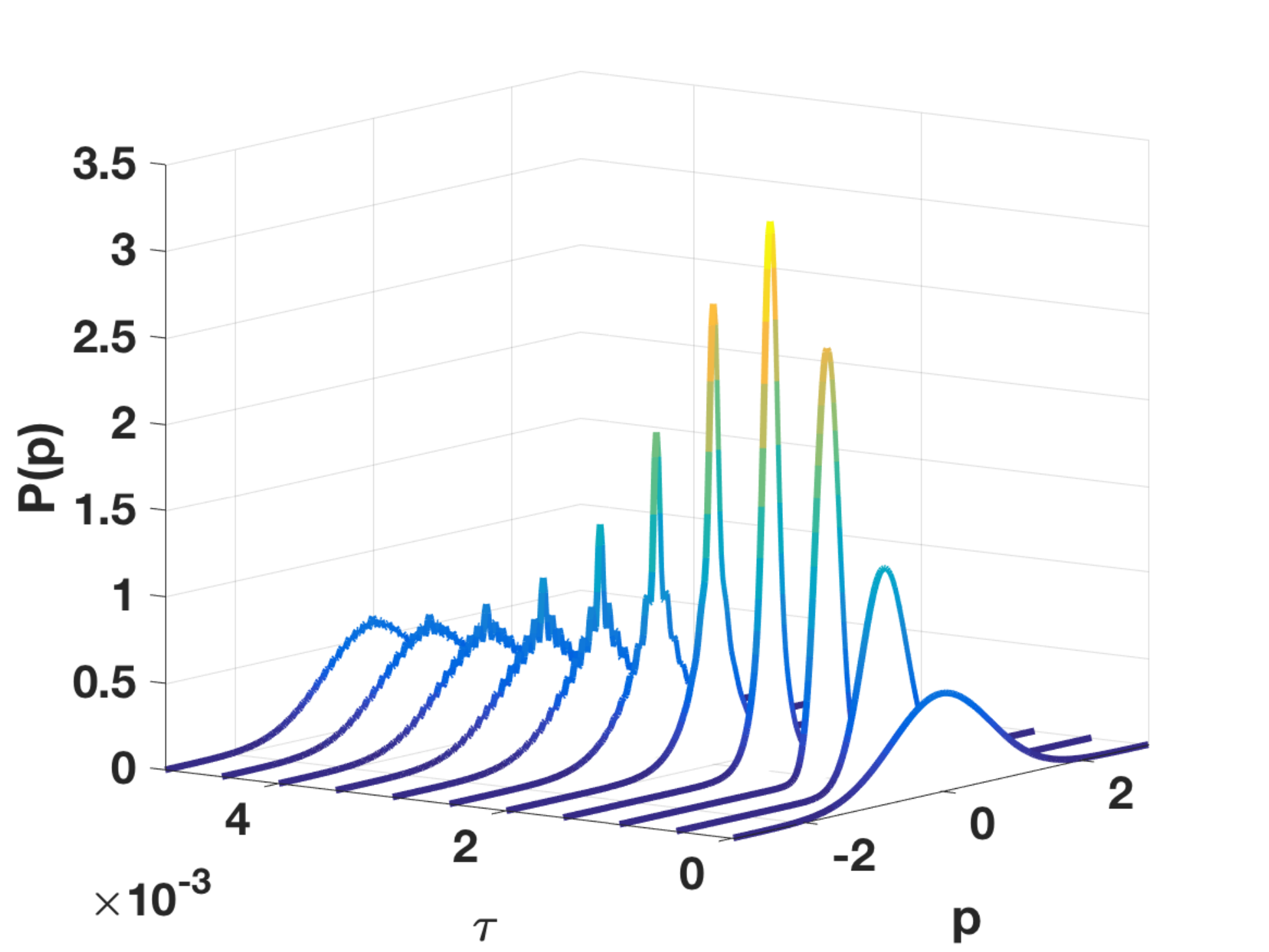}
\par\end{centering}
\caption{The evolution of the quadrature probability distributions in scaled
time $\tau=\gamma t$. Here, the parameters are $g=2$, $\left|\alpha_{0}\right|=20$
and $\chi=0$ at zero temperature. The time range for all plots is
$0-0.005$.\label{fig:large_alpha_P_p_manytimes}}
\end{figure}

\begin{figure}[H]
\begin{centering}
\includegraphics[width=0.75\columnwidth]{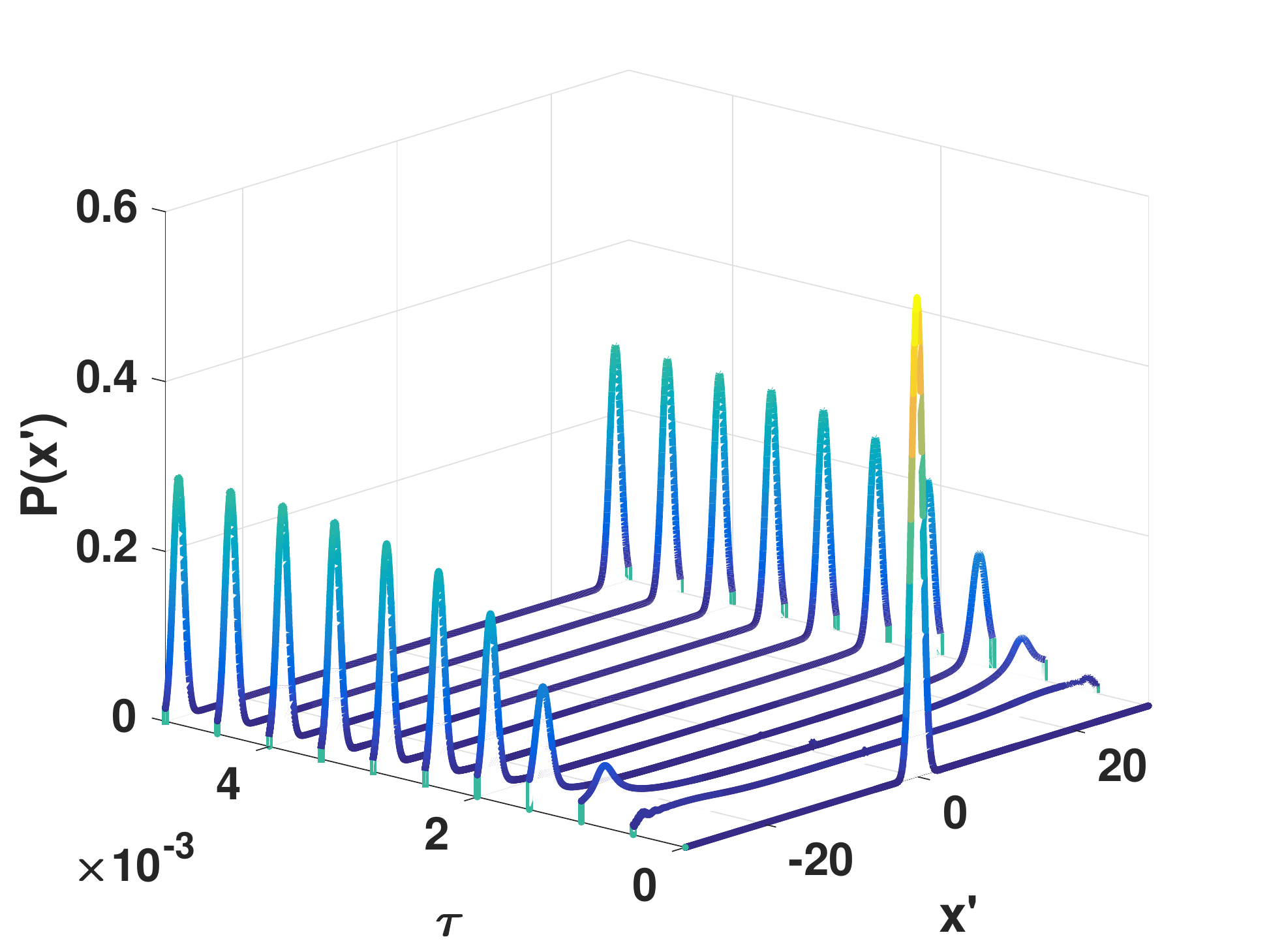}
\par\end{centering}
\begin{centering}
\includegraphics[width=0.75\columnwidth]{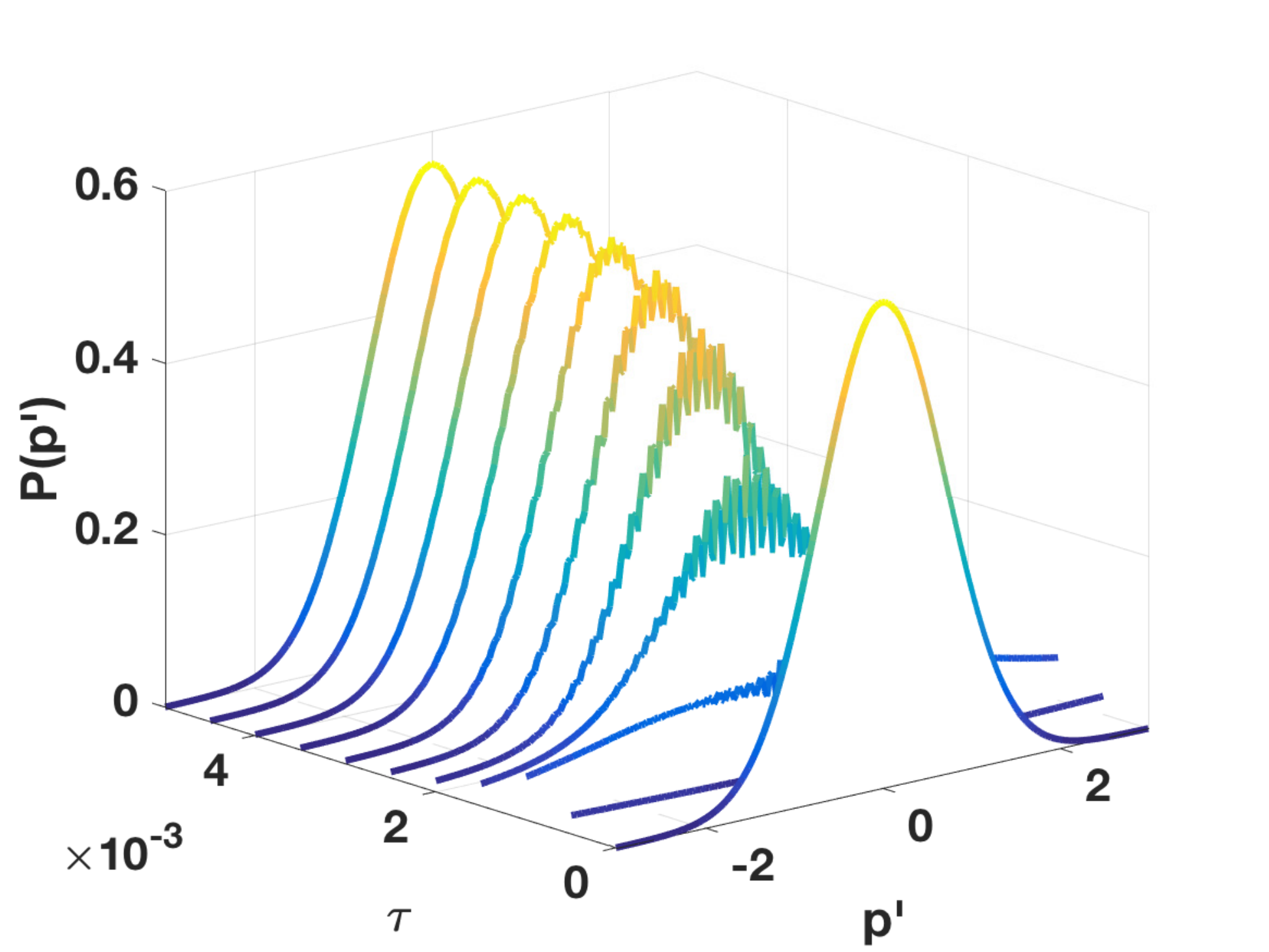}
\par\end{centering}
\caption{The evolution, in scaled time $\tau=\gamma t$, of the rotated quadrature
probability distributions $x'=x_{\phi}$ and $p'=x_{\phi+\pi/2}$
respectively. The angle $\phi$ is determined from the predicted complex
amplitude $\alpha_{0}=|\alpha_{0}|e^{i\phi}$ as given in Eq. (\ref{eq:defn-alpha-kerr}).
Here, the parameters are $g=2$, $\left|\alpha_{0}\right|=20$ and
$\chi=5$ at zero temperature. The time range for all plots is $0-0.005$.
\label{fig:large_alpha_P_p_manytimes-1}\textcolor{blue}{}}
\end{figure}

\begin{figure}[H]
\begin{centering}
\includegraphics[width=0.53\columnwidth]{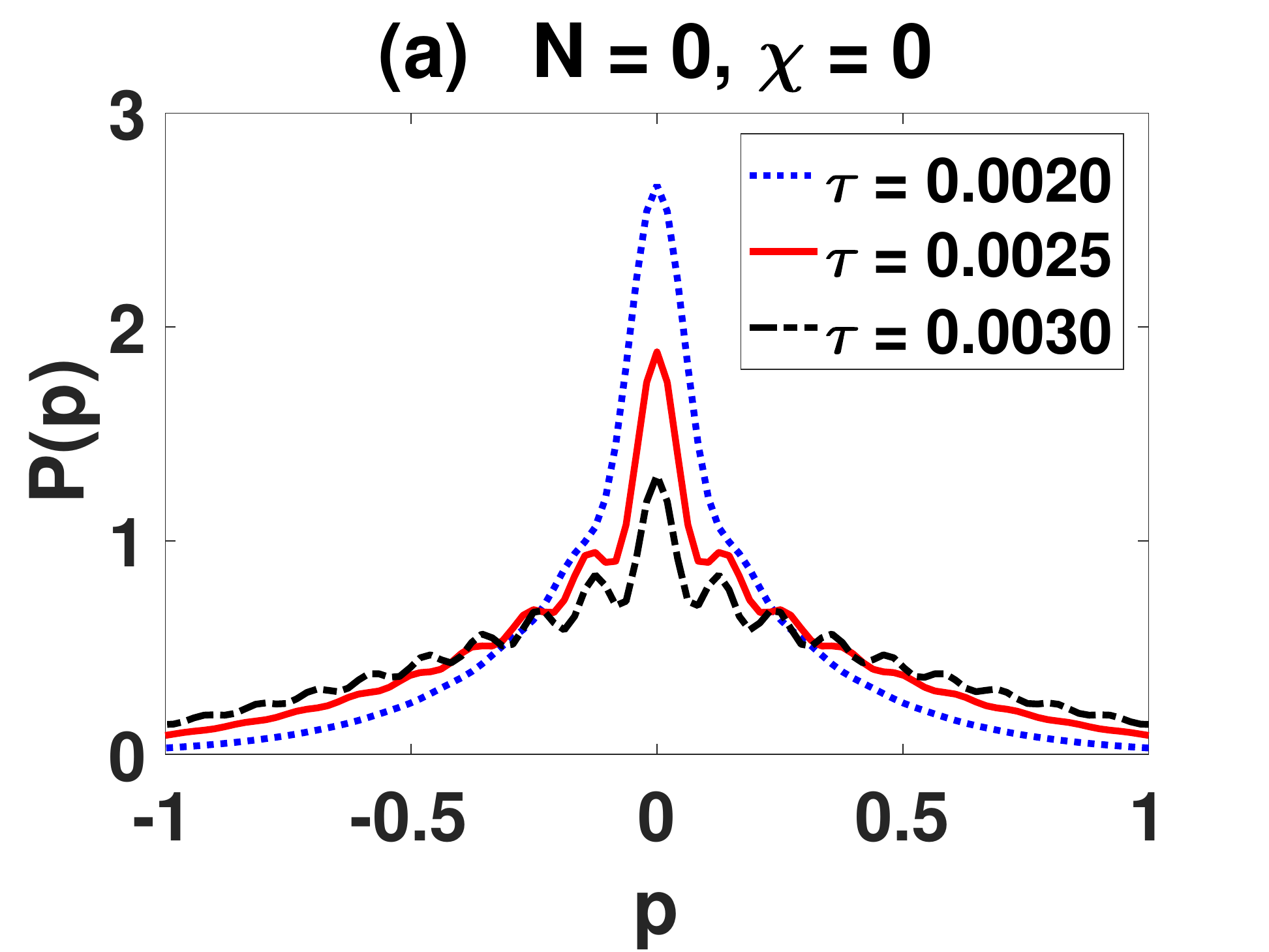}\includegraphics[width=0.53\columnwidth]{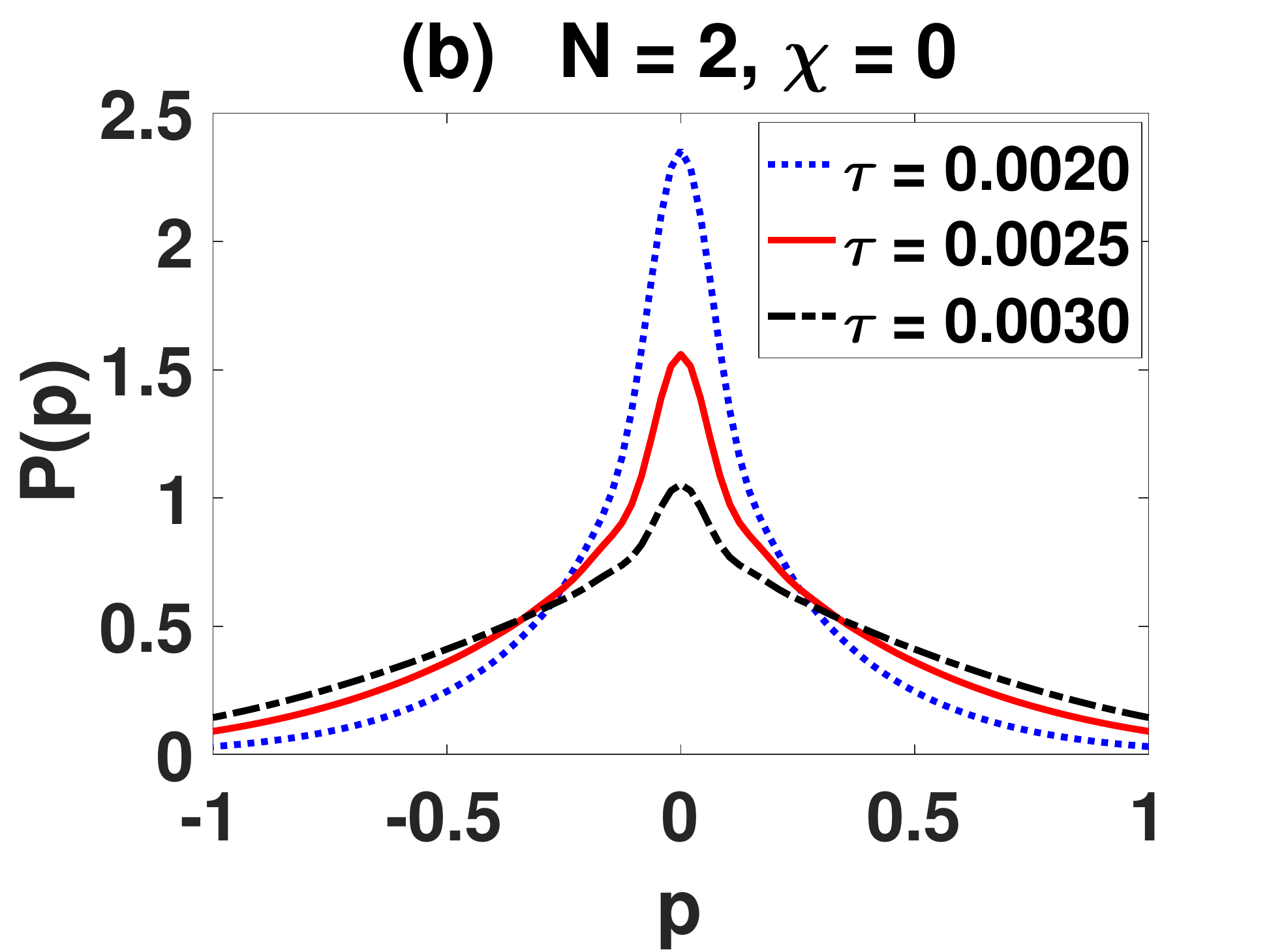}
\par\end{centering}
\bigskip{}

\begin{centering}
\includegraphics[width=0.53\columnwidth]{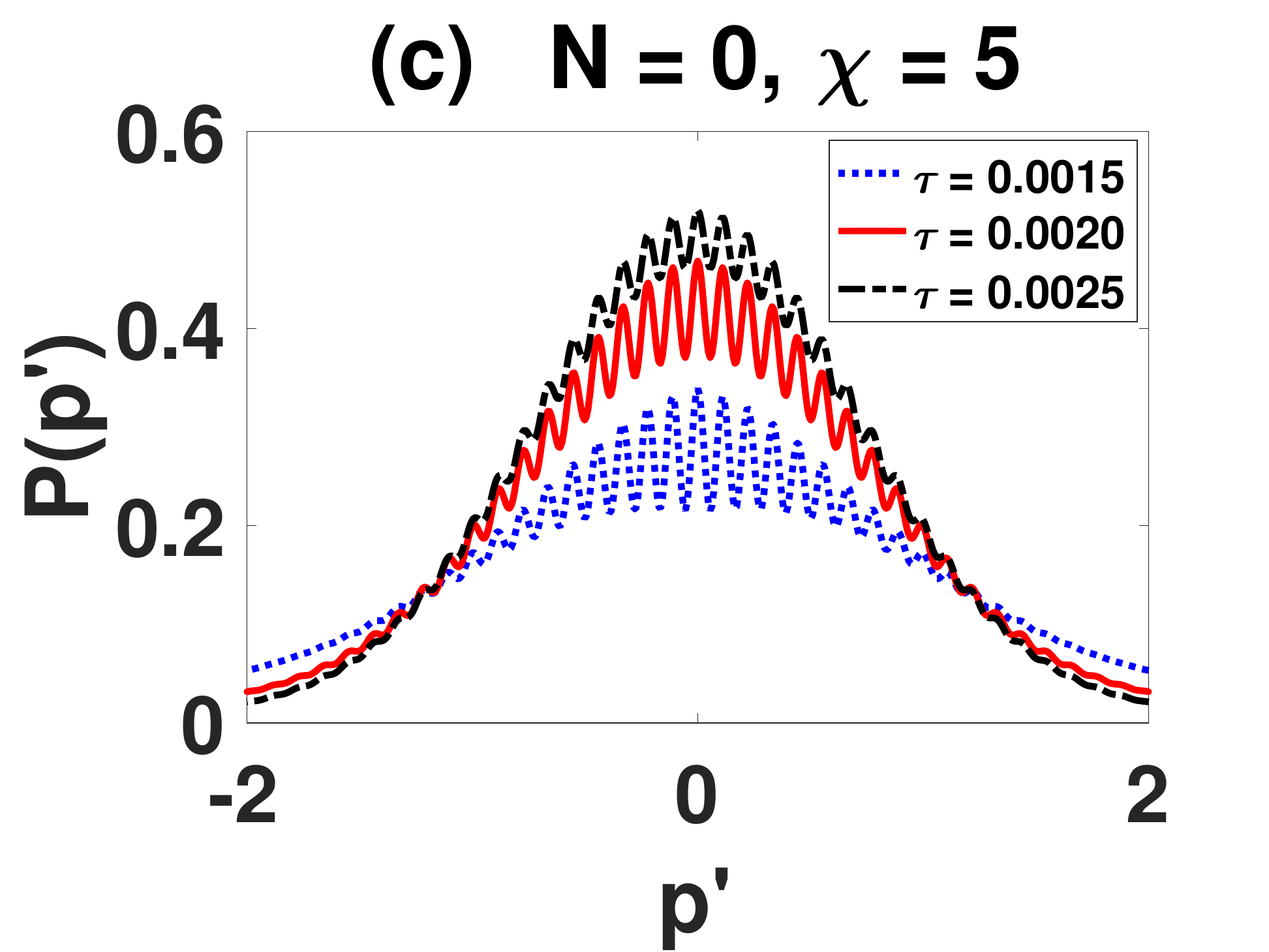}\includegraphics[width=0.53\columnwidth]{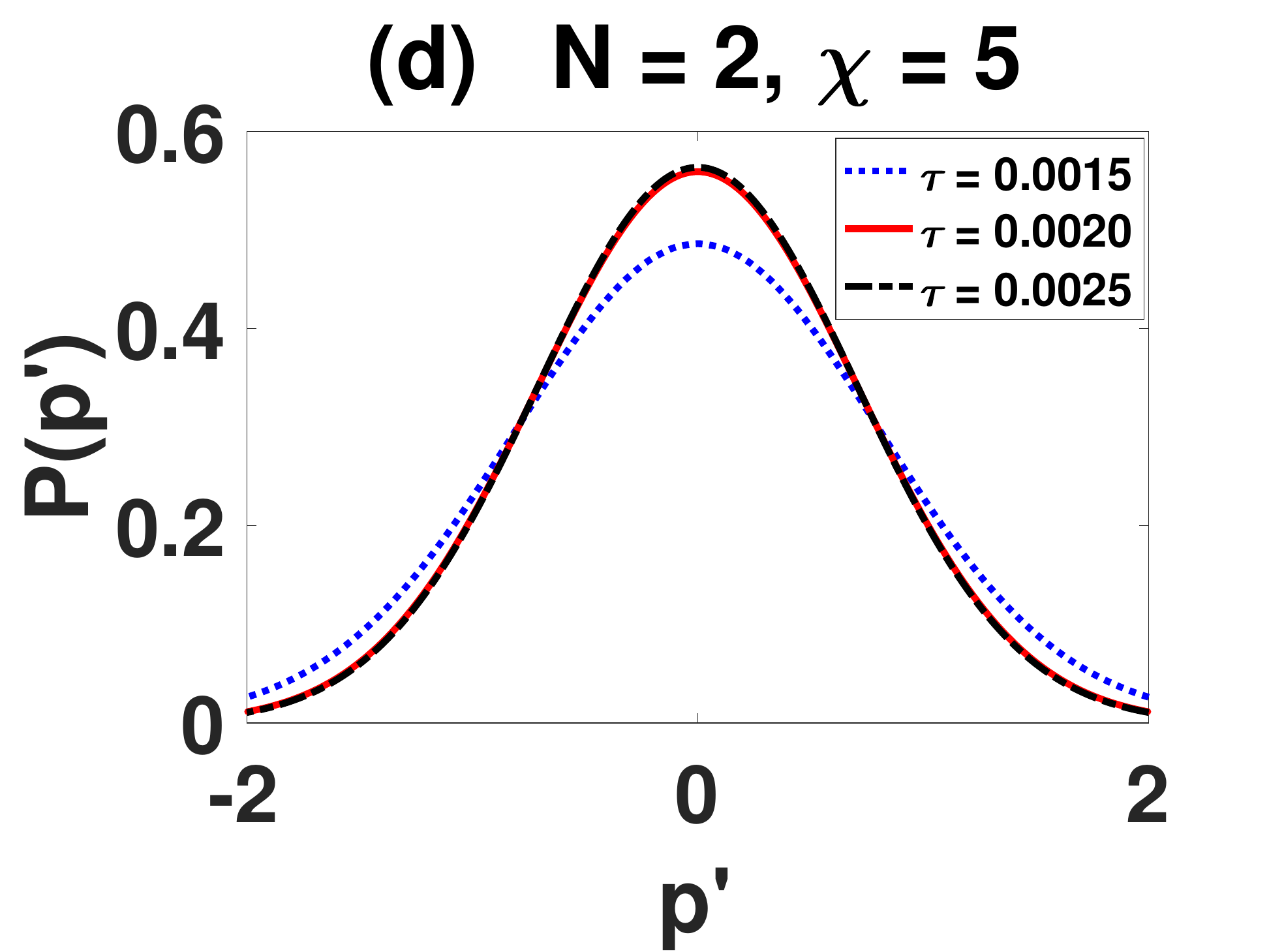}
\par\end{centering}
\caption{Snapshots of the quadrature probability distribution showing the interference
fringes for the case (top) without and (lower) with Kerr nonlinear
interaction. Here, the parameters $g=2$, $\left|\alpha_{0}\right|=20$
and $p'$ are the same as in Fig. \ref{fig:large_alpha_P_p_manytimes-1}.
The case with thermal noise is characterized by the mean thermal occupation
number $N$. \label{large_alpha_Pp_fewer_times}}
\end{figure}

We computed the evolution of the quadrature probability distributions
both with and without the Kerr nonlinear interaction at zero temperature.
The results are shown in Figs. \ref{fig:large_alpha_P_p_manytimes}
and \ref{fig:large_alpha_P_p_manytimes-1}. For the nonzero Kerr case,
it is the rotated quadrature probability distributions $P\left(x_{\phi}\right)$
and $P\left(x_{\phi+\pi/2}\right)$ that are plotted, where the angle
$\phi$ is determined from the predicted complex amplitude $\alpha_{0}=|\alpha_{0}|e^{i\phi}$
as given in Eq. (\ref{eq:defn-alpha-kerr}). From these figures, the
interference fringes appear sooner in the presence of Kerr nonlinear
interaction. This observation is confirmed in Fig. \ref{large_alpha_Pp_fewer_times},
where snapshots of these interference fringes in the quadrature probability
distributions are presented. We include plots with thermal noise of
$N$ thermal photons present in the reservoir. Even though we assume
an initial vacuum state, as in previous calculations, the reservoir
thermal noise causes a decoherence that destroys the cat-state.

The decoherence mechanism is known for this system. The single-photon
damping process switches the state between even and odd cat-states
with probabilities that scale with the single-photon damping rate,
and are further enhanced by the thermal noise. Eventually, the system
reaches a steady state where it is a mixture of the even and odd cat-states.
A detailed mathematical analysis of the decoherence process discussed
here can be found in Ref. \citep{carmichael2009statistical}.

It is appropriate to discuss a few points on the factors that might
limit the achievable cat-state amplitude. In the case without detuning
and Kerr-nonlinearity, the coherent state in the superposition has
an amplitude of $\sqrt{\lambda}/g$. Assuming all other cat-state
destroying parameters ($\gamma_{1},\,N$) remain the same, for larger
$g$, an even larger $\lambda$ is needed to obtain the same cat-state
amplitude, which can be hard to achieve.

There are also difficulties from the point of view of calculation.
This work uses the number-state basis expansion of the density operator
and the cutoff number scales roughly with the coherent amplitude as
$\left|\alpha_{0}\right|^{2}+\left|\alpha_{0}\right|$, where $\alpha_{0}$
is the coherent amplitude of the state. The super-operator that dictates
the time evolution of the density operator has a size of $n_{c}^{2}\times n_{c}^{2}$,
where $n_{c}$ is the cutoff number, and this quickly becomes problematic
even if the super-operator is represented as a sparse matrix. Also,
the cat-state signatures such as the Wigner function and its negativity
are almost not computable even with quadruple-precision computation.
Other methods such as the positive-P phase space representation are
available and are more suited for computations in this regime. However,
more sophisticated techniques \citep{Deuar_RPA2002} in phase space
methods have to be employed when the quantum noise is large ($g>1$).
Even though the Q-function is always positive and does not signify
nonclassicality when the state is a mixed state, it nevertheless has
the merit that its numerical computation is stable. Together with
other cat-state signatures, the Q-function can still serve as a good
nonclassicality indicator. Other cat-state signatures such as the
quadrature probability distributions can be computed to a very large
photon number cutoff (much larger than $500$, which is needed for
cat amplitude $\alpha_{0}>20$) though efficient algorithms such as
the Clenshaw algorithm for evaluating sums involving orthogonal polynomials
are required. 

\section{conclusion \label{sec:conclusion}}

It is known that a Schr\"odinger cat state is formed as the steady
state of a degenerate parametric oscillator, in the limit where single-photon
damping is zero and the initial condition is a vacuum state \citep{Gilles_PRA1994,Hach_PRA1994}.
In the same limit, under an additional nonlinear Kerr interaction,
the corresponding steady state is also a cat-state \citep{sunschrodinger}.
It is illuminating to study the dynamics in the lossless case, as
the interplay between the different nonlinear interactions affects
the cat-formation time, providing a better understanding of the physics
involved in the formation of a cat-state. In Sections IV.A, and VI.A,
we have examined this limit, showing in Section VI.A how the Kerr
nonlinearity can enhance the formation of the cat-state. In particular,
we examine the effect of the Kerr nonlinearity on the threshold value
of $g$, and illustrate how the formation time and lifetime of the
cat-state is affected by the Kerr nonlinearity in the zero temperature
limit. 

In practice, the cavity single-photon damping is very important. This
causes decoherence and eventually destroys the cat-state, as known
from previous exact steady-state results. In Section IV.B, we analyze
the effect of this using a parameter $g$ which gives the strength
of parametric nonlinearity relative to the single-photon signal decay
rate. A threshold value of $g$ is necessary for a cat-state to form.
When $g$ is large enough to form a cat-state, we find that a larger
$g$ will lead to a physical state with a larger Wigner negativity,
implying the formation of a more nonclassical state. However, the
more nonclassical the state is, the shorter is its lifetime.

We also examine the effects of detuning $\Delta$, in Section V. With
all other DPO parameters being equal, the presence of detuning rotates
the physical state in phase space, and does not affect the Wigner
negativity and purity of the state throughout the dynamics of the
system. The Kerr nonlinearity also rotates the state in phase space.
Unlike detuning, however, the Kerr interaction also changes the nonclassicality
of the state. This is examined in Section \ref{sec:Degenerate-parametric-oscillatio}. 

For large $\chi$, where the Kerr interaction strength is larger than
the parametric gain $g$, the Kerr term dominates the dynamics of
the system in the early stage. The larger the Kerr strength, the larger
the value of the corresponding Wigner negativity. As the two stable
states with equal amplitudes but opposite phases are gradually formed
due to the parametric term, the Wigner negativity decreases, before
increasing again as the cat-state is fully reached. A cat-state must
have two well-separated peaks along the phase space axis where the
two amplitudes lie, and hence the dynamical picture given here is
only clear when different cat-state signatures are computed and compared.
The Wigner negativity alone does not provide conclusive evidence of
a cat-state. Two distinct probability peaks and the presence of interference
fringes in the orthogonal quadrature are also necessary. This can
be seen in the quadrature probability distribution and Wigner function,
which confirms the macroscopic coherence between the two peaks.

With this physical picture established, we carried out a numerical
simulation in Section VI.B of a recent experiment of Leghtas et al.
\citep{Leghtas_Science2015}. While a nonclassical state is produced,
in agreement with experimental measurements of Wigner negativity,
it does not appear to be a fully mesoscopic cat-state. The coherent
peaks are not fully separated and the nonclassicality is relatively
weak. This is indicated by absence of significant interference fringes
in the quadrature probability distribution and relatively small Wigner
negativities. We nevertheless agree that this was an important experimental
step towards demonstrating a fully developed mesoscopic superposition
of two well-separated coherent states.

By exploring the parameter space, we find that $g>1$ is still required
for the cat-state generation, irrespective of the Kerr interaction
strength $\chi$. When $g$ is large enough for cat formation, for
a fixed coherent amplitude $|\alpha_{0}|$, a larger $\chi$ takes
a shorter time to form a cat-state and also has a larger Wigner negativity.
However, it has a shorter lifetime, as defined by the Wigner negativity
in Section \ref{sec:Results}.

The ability to compute the time evolution of the physical state allows
us to estimate the lifetime of a cat-state including thermal noise.
An example is given for large $|\alpha_{0}|$ in Section VII. To obtain
a large cat amplitude in the presence of thermal noise, which tends
to destroy the coherence of the cat-state, a large value of $g$ is
necessary. Alternatively, a system that has a lower temperature or
lower cavity decay rate is required. The engineering of the reservoir,
for instance, with squeezed states, as a means of noise reduction
is also possible and will be explored in a future publication.

\section*{Acknowledgements}

This work was performed on the OzSTAR national facility at Swinburne
University of Technology. OzSTAR is funded by Swinburne University
of Technology and the National Collaborative Research Infrastructure
Strategy (NCRIS). PDD and MDR thank the hospitality of the Weizmann
Institute of Science. This work was funded through Australian Research
Council Discovery Project Grants DP180102470 and DP190101480, and
through a grant from NTT Phi Laboratories. The research was performed
in part at Aspen Center for Physics, which is supported by National
Science Foundation grant PHY-1607611. 

\appendix

\section*{Appendix}

\section{Cat-state and signatures \label{sec:Cat signature}}

Here we summarize the cat-state signatures that verify the presence
of cat-states in the system. We focus on the simplest example, in
which we use these signatures to distinguish the difference between
a cat-state
\begin{align}
|\psi_{\text{cat}}\rangle & =\mathcal{N}_{\theta}\left(|\alpha_{0}\rangle+e^{i\theta}|-\alpha_{0}\rangle\right)\,\label{eq:cat-1}
\end{align}
which is a superposition of two coherent states $|\pm\alpha_{0}\rangle$
well-separated in phase space ($\mathcal{N}_{\theta}$ is a normalization
constant and $\theta$ a phase), and an arbitrary mixture of the two
coherent states given by the density operator
\begin{equation}
\rho_{\text{mix}}=P_{+}|\alpha_{0}\rangle\langle\alpha_{0}|+P_{-}|-\alpha_{0}\rangle\langle-\alpha_{0}|\,,\label{eq:mixture}
\end{equation}
where $P_{\pm}$ are probabilities and $P_{+}+P_{-}=1$.

The objective is to confirm that the system is \emph{not }in the coherent
state mixture (\ref{eq:mixture}). Thus, if we consider systems confined
to be in a mixture of the two coherent states, or in a mixture of
superpositions of the two coherent states, the exclusion of the mixture
(\ref{eq:mixture}) implies some type of \emph{cat-like} state, although
not necessarily a pure cat-state. For definiteness, we also require
that a cat-state have clear operational signatures of fringes or Wigner
negativity, as we explain below.

Realizing it is possible the system may be in a state of reduced purity,
the general confined density operator can be written with off-diagonal
terms as
\begin{eqnarray}
\rho & = & P_{11}|\alpha_{0}\rangle\langle\alpha_{0}|+P_{22}|-\alpha_{0}\rangle\langle-\alpha_{0}|\nonumber \\
 &  & +P_{12}|\alpha_{0}\rangle\langle-\alpha_{0}|+P_{21}|-\alpha_{0}\rangle\langle\alpha_{0}|\label{eq:impure-cat-mix}
\end{eqnarray}
This state can also be written in terms of the odd and even cat-states
as
\begin{align}
\rho & =p_{++}|\psi_{\text{even}}\rangle\langle\psi_{\text{even}}|+p_{--}|\psi_{\text{odd}}\rangle\langle\psi_{\text{odd}}|\nonumber \\
 & +p_{+-}|\psi_{\text{even}}\rangle\langle\psi_{\text{odd}}|+p_{-+}|\psi_{\text{odd}}\rangle\langle\psi_{\text{even}}|\,\label{eq:mix2}
\end{align}

We note that these ``impure cat-states'' may or may not give a result
that, for example, has interference fringes. As a result, it is an
open question whether such intermediate states are identifiable by
any of the criteria in common use. It is also possible that the system
cannot be represented in terms of the two coherent states alone, in
which case a broader class of mixtures needs to be excluded. Alternative
approaches to detecting mesoscopic coherence are discussed elsewhere
\citep{Bennett_PRA1996,Terhal_PRA2000,Brennen_QInfo_2003,Cavalcanti_PRL2006,Korsbakken_PRA2007,Cavalcanti_PRA2008,Frowis_NJP2012,Baumgratz_PRL2014,Sekatski_PRA2014,Frowis_OptsComm2015,Oudot_JOSAB2015,Frowis_PRL2016,Yadin_PRA2016,Frowis_RMP2018,Teh_PRA2018}
and include those based on uncertainty relations \citep{Cavalcanti_PRL2006,Cavalcanti_PRA2008,Teh_PRA2018,reid2019criteria}.

In this paper, we identify the cat-state using \emph{both} interference
fringes and negativity of the Wigner function. Where the distribution
for one quadrature phase amplitude ($X$) shows two well-separated
Gaussian peaks corresponding to the two coherent states, the observation
of interference fringes in the orthogonal quadrature ($P$) excludes
all models of the form of (\ref{eq:mixture}). This gives evidence
of a significant quantum coherence, which is one type of signature
of a Schr\"odinger cat-state.

However, if the associated Wigner function is observed to be positive,
then there exists a joint probability distribution $P(x,p)$ to correctly
describe the marginal probability distributions $P(x)$ and $P(p)$
for the results $x$ and $p$ of measurements $X$ and $P$. It is
then possible to construct two ``elements of reality'', the variables
$x$ and $p$, that directly and simultaneously predetermine the results
for $X$ and $P$. While these ``elements of reality'' $x$ and
$p$ do not describe \emph{quantum} states (being simultaneously precisely
defined \citep{reid2019criteria}), the system can nonetheless, with
respect to these variables, be interpreted as being in one \emph{or}
other of states corresponding to the Gaussian peaks in $X$. This
interpretation is not possible for the ideal cat-state (\ref{eq:cat-1})
which possesses a negative Wigner function. Thus, the observation
of interference fringes associated with a negative Wigner function
(consistent with that of the state (\ref{eq:cat-1})) gives strong
evidence of a cat-state.

\subsection{Interference fringes in the quadrature probability distribution}

One of the earliest proposed cat state signatures is the presence
of interference fringes in the quadrature probability distribution
\citep{PhysRevLett.57.13,Yurke_physicalB1988}. In order to understand
the origin of the interference fringes, consider an even cat-state
\begin{align}
|\psi_{\text{even}}\rangle & =\mathcal{N}_{+}\left(|\alpha_{0}\rangle+|-\alpha_{0}\rangle\right)\,\label{eq:even_cat}
\end{align}
Without losing generality, we assume that $\alpha_{0}$ is real,
and that $|\alpha_{0}|$ is large. The $x$-quadrature for this state
has two contributions from two well-separated phase points along $x$-axis.
The corresponding $x$-quadrature probability distribution has two
significant Gaussian distributions centered around these two phase
points along the $x$-axis. This gives us justification to assume
the system is either a superposition, or a mixture, as in (\ref{eq:impure-cat-mix}).

To exclude the statistical mixture (\ref{eq:mixture}), one measures
the orthogonal quadrature $p$. For a cat-state (\ref{eq:cat-1}),
the probability amplitudes for these two possible contributions $|\pm\alpha_{0}\rangle$
have to be summed, and hence there will be interference fringes in
the $p$-quadrature probability distribution for this cat state. These
fringes cannot arise for the system given by the classical mixture
(\ref{eq:mixture}) which is therefore excluded if fringes are observed.
If we consider the coherent-state manifold, with $\alpha_{0}\geq2.5$
to allow for distinct Gaussian distributions, the onset of fringes
implies failure of the mixture (\ref{eq:mixture}), so that $P_{12}$
and $P_{21}$ defined by eq. (\ref{eq:impure-cat-mix}) must be nonzero.

\begin{figure}[H]
\begin{centering}
\includegraphics[width=0.6\columnwidth]{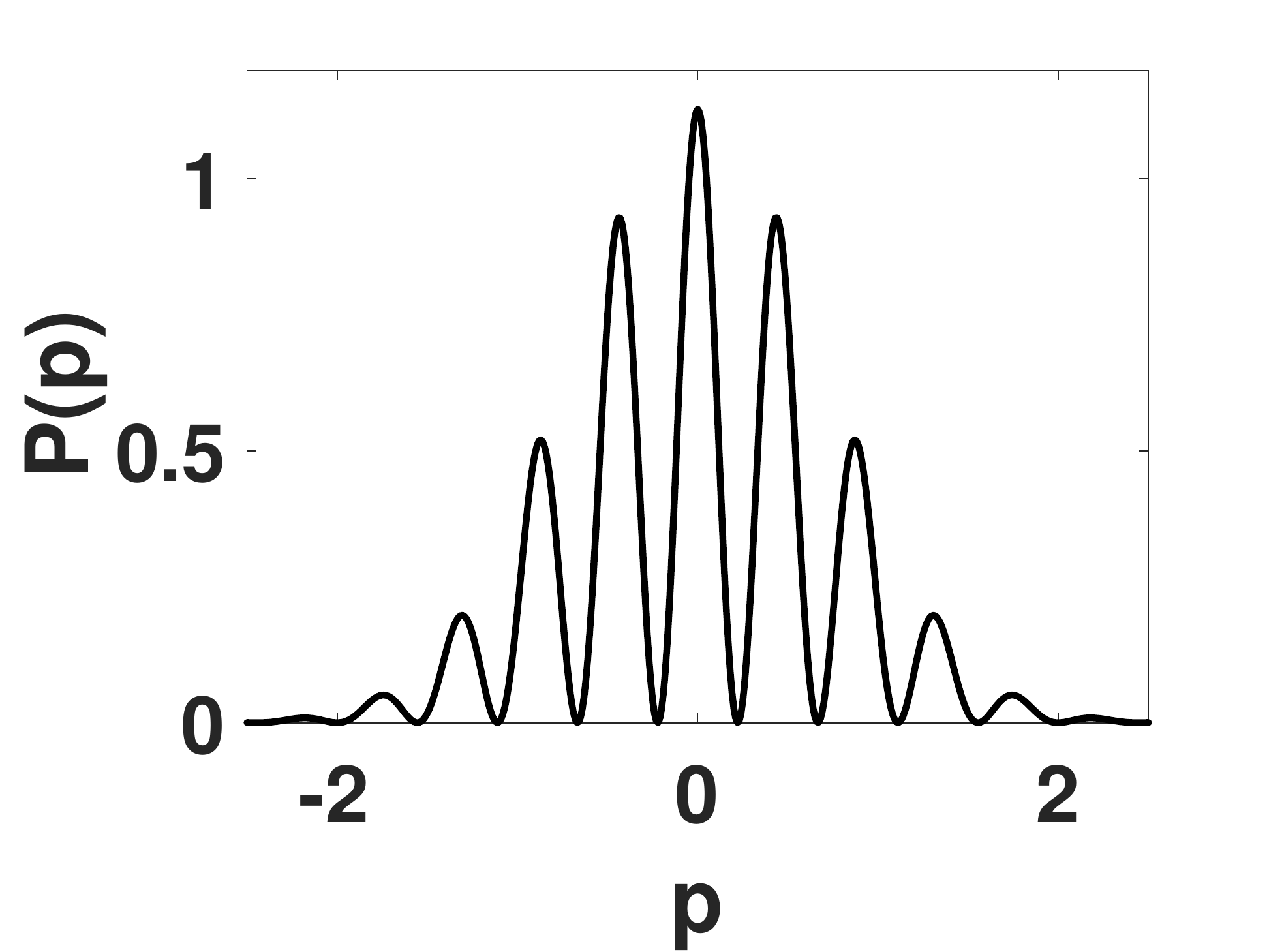}
\par\end{centering}
\caption{The $p$-quadrature probability distribution for a pure, even cat-state
with $\alpha_{0}=5$.\label{fig:Pp_cat_ao5}}
\end{figure}

More generally, a cat-state may be in a manifold of superposition
states spanned by two coherent states $\left\{ |\alpha_{0}\rangle,|-\alpha_{0}\rangle\right\} $,
where $\alpha_{0}$ is a complex number and these two coherent states
can have any phase relation between them. Therefore, we define a general
rotated quadrature operator $x_{\theta}=\left(e^{-i\theta}a+e^{i\theta}a^{\dagger}\right)/\sqrt{2}$.
The $x_{\theta}$-quadrature probability distribution can be computed
from a density operator $\rho$ which is expanded in the number state
basis. The probability distribution $P\left(x_{\theta}\right)$ is
then
\begin{align}
\langle x_{\theta}|\rho|x_{\theta}\rangle & =\langle x_{\theta}|\left(\sum_{n,m}\rho_{nm}|n\rangle\langle m|\right)|x_{\theta}\rangle\nonumber \\
 & =\sum_{n,m}\rho_{nm}\langle x_{\theta}|n\rangle\langle m|x_{\theta}\rangle\,,\label{eq:prob_dist}
\end{align}
where
\begin{align}
\langle x_{\theta}|n\rangle & =\frac{e^{-i\theta n}}{\sqrt{2^{n}n!\sqrt{\pi}}}e^{-\frac{x_{\theta}^{2}}{2}}H_{n}\left(x_{\theta}\right)\,.\label{eq:x_theta_n}
\end{align}
Here, $H_{n}\left(x\right)$ is the Hermite polynomial. In particular,
for $\theta=0$, $x_{\theta=0}=x$ and for $\theta=\pi/2$, $x_{\theta=\pi/2}=p$,
and their inner products with a number state are given by 
\begin{align}
\langle x|n\rangle & =\frac{1}{\sqrt{2^{n}n!\sqrt{\pi}}}e^{-\frac{x^{2}}{2}}H_{n}\left(x\right)\label{eq:z_n}\\
\langle p|n\rangle & =\frac{\left(-i\right)^{n}}{\sqrt{2^{n}n!\sqrt{\pi}}}e^{-\frac{p^{2}}{2}}H_{n}\left(p\right)\label{eq:p_n}
\end{align}
respectively. For an even cat-state with real-valued coherent amplitudes,
$\alpha_{0}$, the $p$-quadrature probability distribution is given
by \citep{Yurke_physicalB1988,Teh_PRA2018}
\begin{align}
P\left(p\right) & =\frac{1}{\sqrt{\pi}}\mathcal{N}_{+}^{2}\left\{ 2\text{exp}\left(-p^{2}\right)\left[1+\cos\left(2\sqrt{2}p\alpha_{0}\right)\right]\right\} \,,\label{eq:cat_Pp}
\end{align}
For comparison purposes, we plot $P\left(p\right)$ for $\alpha_{0}=5$
in Fig. \ref{fig:Pp_cat_ao5} using Eq. (\ref{eq:cat_Pp}).

In this work, a number state cutoff of up to $500$ is used. There
is a floating point number overflowing issue in the numerical computation
of Eqs. (\ref{eq:z_n}) and (\ref{eq:p_n}), which arises from the
evaluation of the Hermite polynomials. This issue is overcome by using
a Matlab function \citep{rodney} that employs logarithmic manipulation.
Moreover, this Matlab function is based on the Clenshaw algorithm
\citep{clenshaw1955note,press2007numerical} that computes orthogonal
polynomials more efficiently and accurately \citep{smith1965algorithm}
than either naively computing the summations involved or other methods
using the Hermite polynomials recurrence relation such as the Forsythe
method \citep{Forsythe_SIAM1957}.

\subsection{Photon-number probability distribution}

Yet another aspect where the quantum superposition of a cat state
is manifested is in the photon-number probability distribution. The
even cat-state Eq. (\ref{eq:even_cat})
\begin{align}
|\psi_{\text{even}}\rangle & =\mathcal{N_{+}}\left(|\alpha_{0}\rangle+|-\alpha_{0}\rangle\right)\nonumber \\
 & =\mathcal{N_{+}}e^{-\frac{\left|\alpha_{0}\right|^{2}}{2}}\sum_{n}^{\infty}\frac{1}{\sqrt{n!}}\left[\alpha_{0}^{n}+\left(-\alpha_{0}\right)^{n}\right]|n\rangle\,,\label{eq:even_cat_number_state}
\end{align}
contains an even number of photons. Similarly, an odd cat-state $|\psi_{\text{odd}}\rangle=\mathcal{N_{-}}\left(|\alpha_{0}\rangle-|-\alpha_{0}\rangle\right)$
has an odd number of photons. For a classical mixture of $|\alpha_{0}\rangle$
and $|-\alpha_{0}\rangle$, the photon number probability distribution
is nonzero for both even and odd numbers of photons. Hence, \emph{assuming}
we are in the manifold of the superpositions of the two coherent states
(or their mixtures), the photon-number probability distribution reveals
both the nonclassicality of a cat-state and its phase relation. It
is possible that the system is in a superposition of both the even
and odd cat-states, and the photon number probability distribution
does not distinguish between this state and a classical mixture of
the even and odd cat-states. With that, we also computed the purity
of the state given by\textit{\emph{}}
\begin{align}
\mathcal{P} & {\normalcolor =}\text{Tr}\left(\rho^{2}\right)\,.\label{eq:purity}
\end{align}

\subsection{Phase-space distributions}

It is also useful to consider phase-space distributions that can
determine the entire quantum states, and display quantum features.
In particular, we compute the Husimi Q and Wigner functions. 

\subsubsection{Wigner function and its negativity}

The Wigner function gives us the joint probability distribution of
the real and imaginary parts of the coherent amplitude of the quantum
state, which allows the deduction of the form of a cat state. The
Wigner function for a density operator in a Fock state for a finite
particle number is given by \citep{Cahill_PhysRev1969,PATHAK2014117}
\begin{align}
W\left(\alpha,\alpha^{*}\right) & =\sum_{n}^{N_{c}}\rho_{nn}X_{nn}+2\text{Re}\left(\sum_{m=1}^{N_{c}}\sum_{n=0}^{m-1}\rho_{nm}X_{nm}\right)\,,\label{eq:wigner_function}
\end{align}
where $n<m$, $\rho_{nm}$ is the matrix element of the density operator
$\rho$ and $X_{nm}$ is \citep{Cahill_PhysRev1969,PATHAK2014117}
\begin{align}
X_{nm} & =\frac{2\left(-1\right)^{n}}{\pi}\sqrt{\frac{n!}{m!}}e^{-2\left|\alpha\right|^{2}}\left(2\alpha\right)^{m-n}L_{n}^{m-n}\left(4\left|\alpha\right|^{2}\right)\,.\label{eq:X_nm}
\end{align}
Here, $L_{b}^{a}\left(x\right)$ is the associated Laguerre polynomial.
For large cutoff photon numbers $N_{c}$, the direct evaluation of
the Wigner function in Eq. (\ref{eq:wigner_function}) leads to numerical
instabilities. These issues can be overcome by rewriting the expression
in Eq. (\ref{eq:wigner_function}) as
\begin{align}
W\left(\alpha,\alpha^{*}\right) & =\sum_{n}^{N_{c}}\rho_{nn}X_{nn}+2\text{Re}\left(e^{-2\left|\alpha\right|^{2}}\sum_{l=1}^{N_{c}}c_{l}\left(2\alpha\right)^{l}\right)\,,\label{eq:Wigner_new_expression}
\end{align}
where
\begin{align}
c_{l} & =\sum_{n=0}^{N_{c}-l}\rho_{n,l+n}\frac{2\left(-1\right)^{n}}{\pi}\sqrt{\frac{n!}{\left(l+n\right)!}}L_{n}^{l}\left(4\left|\alpha\right|^{2}\right)\,.\label{eq:c_l}
\end{align}
The first term in Eq. (\ref{eq:Wigner_new_expression}) involving
the sum of Laguerre polynomials is evaluated using the Clenshaw algorithm
\citep{clenshaw1955note}. For the second term, the same algorithm
is used to compute $c_{l}$ which contains the sum of associated Laguerre
polynomials. Then the sum of polynomials $2\alpha$ is computed using
the Horner's method for polynomial evaluation. We note that for $\alpha$
that has a large amplitude, large numerical errors are found and these
methods cease to work. 

In experiments, state tomography has to be carried out. It has been
proposed by Lutterbach and Davidovich \citep{Lutterbach_PRL1997}
that measurements of the photon number parity amounts to the determination
of a Wigner function. This is based on the fact that a Wigner function
is the expectation value of the number parity operator $\hat{\Pi}=exp\left(i\pi\hat{n}\right)$,
where $\hat{n}$ is the number operator, for a physical state that
is displaced by a coherent amplitude $\alpha$. Explicitly, it is
given as follows \citep{Cahill_PhysRev1969}:
\begin{align}
W\left(\alpha\right) & =\frac{2}{\pi}\text{Tr}\left(\hat{D}\left(-\alpha\right)\rho\hat{D}\left(\alpha\right)\hat{\Pi}\right)\,,\label{eq:Wigner_experiment}
\end{align}
where $\hat{D}\left(\alpha\right)$ is a displacement operator. This
method has been used to determine the Wigner function in experiments
\citep{Bertet_PRL2002,Vlastakis_Science2013,Sun_Nature2014,Wang_Science2016,Liu_Scienceadv2017,Wang_PRL2017}.
\begin{figure}[H]
\begin{centering}
\includegraphics[width=0.6\columnwidth]{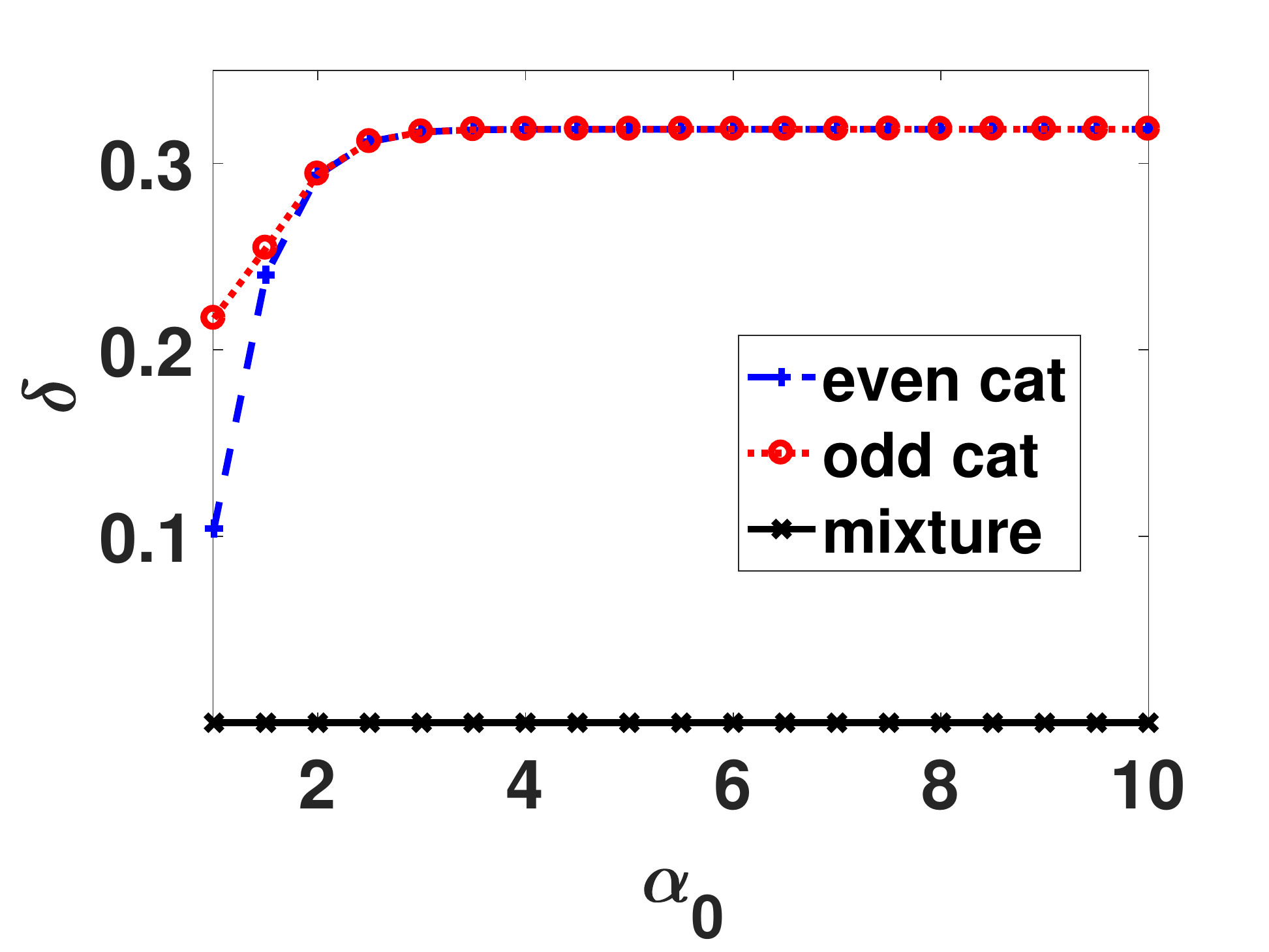}
\par\end{centering}
\caption{The Wigner negativity $\delta$ as a function of the coherent amplitude
$\alpha_{0}$, for both the ideal cat-state and the mixture Eq. (\ref{eq:mixture}).\label{fig:lossless_delta}}
\end{figure}

The negativity of the Wigner function can be quantified as \citep{Kenfack_2004}
\begin{align}
\delta & =\frac{1}{2}\intop\left[\left|W\left(\alpha,\alpha^{*}\right)\right|-W\left(\alpha,\alpha^{*}\right)\right]d^{2}\alpha\,.\label{eq:wigner_negativity}
\end{align}
A positive-valued Wigner function gives $\delta=0$ while $\delta$
is nonzero in the presence of any negative values in a Wigner function.
The Wigner functions $W_{+}$ and $W_{-}$ for the even cat-state
and odd cat-state respectively are given by \citep{Teh_PRA2018}
\begin{eqnarray}
W_{\pm}\left(\alpha,\alpha^{*}\right) & = & \frac{2}{\pi}\mathcal{N}_{\pm}^{2}\left\{ exp\left[-2\left(\alpha^{*}-\alpha_{0}^{*}\right)\left(\alpha-\alpha_{0}\right)\right]\right.\nonumber \\
 &  & +exp\left[-2\left(\alpha^{*}+\alpha_{0}^{*}\right)\left(\alpha+\alpha_{0}\right)\right]\nonumber \\
 &  & \pm\langle\alpha_{0}|-\alpha_{0}\rangle exp\left[-2\left(\alpha^{*}-\alpha_{0}^{*}\right)\left(\alpha+\alpha_{0}\right)\right]\nonumber \\
 &  & \left.\pm\langle-\alpha_{0}|\alpha_{0}\rangle exp\left[-2\left(\alpha^{*}+\alpha_{0}^{*}\right)\left(\alpha-\alpha_{0}\right)\right]\right\} \,.\nonumber \\
\label{eq:Wigner_cat_state}
\end{eqnarray}
which give negative values. For a mixture $\rho$ of Eq. (\ref{eq:mixture})
which has purity given by $\mathcal{P}=P_{+}^{2}+P_{-}^{2}+P_{+}P_{-}\text{e}^{-2\left|\alpha_{0}\right|^{2}}$,
the Wigner function is
\begin{align}
W_{\text{mix}}\left(\alpha,\alpha^{*}\right) & =\frac{1}{\pi}\left\{ exp\left[-2\left(\alpha^{*}-\alpha_{0}^{*}\right)\left(\alpha-\alpha_{0}\right)\right]\right.\nonumber \\
 & +exp\left[-2\left(\alpha^{*}+\alpha_{0}^{*}\right)\left(\alpha+\alpha_{0}\right)\right]\label{eq:Wigner_mix}
\end{align}
and does not admit any negative values. Fig. \ref{fig:lossless_delta}
plots the value of $\delta$, using Eq. (\ref{eq:wigner_negativity}),
for the cat-states $|\psi_{\text{even}}\rangle$ and $|\psi_{\text{odd}}\rangle$
versus $\alpha_{0}$, and for the mixture $\rho$ of Eq. (\ref{eq:mixture})
with $P_{+}=P_{-}=1/2$, which has purity given by $\mathcal{P}=\left(2+e^{-4\left|\alpha_{0}\right|^{2}}\right)/4$.
Hence, the magnitude of Wigner function negativity $\delta$ quantitatively
captures the nonclassicality of the quantum state. This is because
the mixture Eq. (\ref{eq:mixture}) has a non-negative Wigner function.
Hence, if we assume the system is constrained to the manifold of superpositions
of the two coherent state (or their mixtures), the negativity is a
signature of a cat-state. We note however that more generally, the
negativity does not always imply a cat-state, due to the possible
presence of microscopic superpositions.

Numerically, the computation of the Wigner negativity in Eq. (\ref{eq:wigner_negativity})
requires schemes of numerical integration that have errors as a finite
grid size is used. In this work, a trapezoidal numerical integration
as well as the Gauss-Lobatto numerical integration are used. With
the same grid size, the Gauss-Lobatto method is known to be much more
accurate than the trapezoidal numerical integration. The Wigner negativities
computed using both of these methods agree up to four significant
figures, indicating that the grid size chosen is fine enough and the
Wigner negativities computed have small grid size errors.

\subsubsection{Husimi Q function}

The Husimi Q function is defined by $Q\left(\alpha,\alpha^{*}\right)=\langle\alpha|\rho|\alpha\rangle/\pi$.
The expression of a Q function for a density operator in the number
state basis is given by
\begin{align}
Q\left(\alpha,\alpha^{*}\right) & =\frac{1}{\pi}\langle\alpha|\rho|\alpha\rangle\nonumber \\
 & =\frac{1}{\pi}\langle\alpha|\left(\sum_{n,m}\rho_{nm}|n\rangle\langle m|\right)|\alpha\rangle\nonumber \\
 & =\sum_{n,m}\rho_{nm}\frac{\left(\alpha^{*}\right)^{n}\alpha^{m}}{\pi\sqrt{n!m!}}exp\left(-\left|\alpha\right|^{2}\right)\,.\label{eq:q-function}
\end{align}
Unlike the Wigner function which admits negative values and is used
as an indicator of nonclassicality, the Husimi Q function is always
positive. However, it has been shown by L\"{u}tkenhaus and Barnett
\citep{Lutkenhaus_PRA1995} that a highly nonclassical state will
have zeros in the Q function, where the corresponding Wigner function
at these zero points have equal positive and negative contributions.
Also, in the case where the calculation of the Wigner function is
too numerically intensive to be computed, the Q function can serve
as a phase-space visualization guide that complements other cat state
signatures.

\bibliographystyle{apsrev4-1}
\bibliography{DPO_cat_final}

\end{document}